\documentclass[11pt]{article}
\usepackage[left=0.75in,right=0.75in,top=0.75in,bottom=1.0in]{geometry}
\usepackage{amsmath,amsfonts,amssymb,bm}
\usepackage{mathtools}
\usepackage{graphicx}
\usepackage{booktabs}
\usepackage{float}
\usepackage{subcaption} 
\usepackage{placeins}
\usepackage[numbers,sort&compress]{natbib}
\usepackage{etoolbox}

\AtBeginEnvironment{thebibliography}{%
  \small                             
  \setlength{\parskip}{0pt}%
  \setlength{\itemsep}{2pt plus 0.3ex}%
  \setlength{\parsep}{0pt}%
}
\usepackage[hidelinks]{hyperref}

\usepackage{authblk}
\title{Spectrum and Physics-Informed Neural Networks (SaPINNs) for Input–State–Parameter Estimation in Dynamic Systems subjected to Natural Hazards-Induced Excitation}
\author[1]{Antonina Kosikova\thanks{Corresponding author. Email: ak4967@columbia.edu}}
\author[1]{Apostolos Psaros}
\author[1]{Andrew Smyth}
\affil[1]{Department of Civil Engineering and Engineering Mechanics, Columbia University, New York, NY 10027, USA}

\date{} 
\begin{document}
\maketitle
\newcommand{\keywords}[1]{%
  \vspace{0.75\baselineskip}%
  \noindent\textbf{Keywords:} #1%
}
\begin{abstract}
System identification under unknown external excitation is an inherently ill-posed problem, typically requiring additional knowledge or simplifying assumptions to enable reliable state and parameter estimation. The difficulty of the problem is further amplified in structural systems subjected to natural hazards such as earthquakes or windstorms, where responses are often highly transient, nonlinear, and spatially distributed. To address this challenge, we introduce Spectrum and Physics-Informed Neural Networks (SaPINNs)  for efficient input–state–parameter estimation in systems under complex excitations characteristic of natural hazards. The proposed model enhances the neural network with governing physics of the system dynamics and incorporates spectral information of natural hazards by using empirically derived spectra as priors on the unknown excitations.  This integration improves inference of unmeasured inputs, system states, and parameters without imposing restrictive assumptions on their dynamics. The performance of the proposed framework is demonstrated through comparative studies on both linear and nonlinear systems under various types of excitation, including the El Centro earthquake, where the seismic spectrum is assumed to be not precisely known.  To account for predictive uncertainty, the proposed architecture is embedded within a Deep Ensemble (DEns) networks architecture, providing distributions over possible solutions.  The results demonstrate that the proposed approach outperforms conventional PINNs, as the incorporation of spectral information introduces an inductive bias that guides the network more effectively through the solution space and enhances its capability to recover physically consistent state and parameter estimates with realistic uncertainty levels.
\end{abstract}

\section{Introduction}
In recent years, the increasing frequency and severity of natural hazards such as earthquakes, hurricanes, and tsunamis have posed significant risks to civil infrastructure systems. These events subject structures to complex dynamic excitations, challenging traditional modeling and simulation techniques. Conventional approaches, such as the finite element method (FEM) and Bayesian system identification techniques, often demand extensive computational power and high-fidelity data, resources that are not always available in real-world scenarios. Consequently, there has been a growing interest in advanced computational frameworks that integrate physics-based principles and data-driven strategies, promising more efficient and scalable solutions. Alongside these developments, Structural Health Monitoring (SHM) has benefited from new technologies such as non-destructive evaluation tools, sensor networks, and artificial intelligence (AI)-based algorithms, which  greatly enhanced the ability to detect early-stage damage and assess structural system integrity in real-time \cite{vlachas2022coupling, https://doi.org/10.1002/stc.290}.  In practical applications, the assessment of a dynamic system’s operational state is still predominantly carried out using Operational Modal Analysis (OMA), or  output-only system identification \cite{azam2015dual}. However, one of the main challenges in output-only estimation is distinguishing the system's natural frequencies from those of the excitation force, especially when the forcing spectrum overlaps with the structural modes. An alternative approach involves employing a reduced-order numerical model subjected to a known input force, where model parameters are iteratively adjusted to match the predicted response to measured structural behavior.  While this approach can provide valuable insights into system states, such analysis becomes infeasible when the excitation force is not precisely known.  To address this challenge,  both probabilistic and deterministic methods in system identification have evolved. Probabilistic approaches, in particular, have gained prominence due to their ability to incorporate uncertainties arising from environmental variability, sensor noise, and operational conditions. Notable examples include Augmented Kalman Filter (AKF) \cite{lourens2012augmented,dertimanis2019input, VETTORI2023109654}, the Extended Kalman Filter (EKF) with smoothing for joint input–state–parameter estimation  \cite{maes2019tracking},  as well as the integration of a time-varying autoregressive (TAR) model with the Unscented Kalman Filter (UKF) for input force identification and FEM-based model calibration \cite{castiglione2020auto}, among others.  However, although widely adopted, Kalman filter–based methods require careful tuning of the process and measurement noise covariance matrices and are susceptible to long-term drift in the estimated states and input forces—particularly when the system model or noise characteristics are not well known. Moreover, in AKF formulations, the unknown input force is typically modeled as a random walk process. While this assumption may be adequate in generic settings, such as ambient excitation,  it is often unrealistic for a more complex loading scenarios including intermittent, or band-limited loads, commonly encountered in wind, traffic, or seismic environments. \\
With the development of ever increasing potential of the machine learning (ML) techniques, various types of models have  been introduced for system identification, capable of capturing nonlinear and partially observed dynamics. In the context of unknown input force estimation, Gaussian Process Latent Force Models (GP-LFMs) offer a regression-based framework that reconstructs unknown input by modeling it as latent process, parameterized through kernel functions encoding temporal correlation \cite{alvarez2009latent}. These models have been applied in combination with Kalman filtering for joint input-state prediction in linear systems \cite{nayek2019gaussian}, and further extended for the input-state-parameter estimation problem \cite{rogers2020application}.  Further, several studies have demonstrated the potential of fully data-driven surrogate models for capturing complex input–output relationships in dynamic systems, particularly in cases where deterministic, physics-based approaches are either analytically intractable or computationally prohibitive. A hybrid architecture combining a long short-term memory (LSTM) network with an autoencoder has been employed to train a recurrent neural network directly on system outputs, enabling the extraction of a low-dimensional representation of the latent dynamics without requiring explicit knowledge of the governing equations or direct access to input force measurements \cite{vlachas2022multiscale,simpson2021machine}. By leveraging the memory capabilities of LSTM networks, such models can effectively encode temporal dependencies in structural response data, enabling the reconstruction of complex, nonstationary input force histories without relying on explicit physical modeling. However, these purely data-driven approaches often lack the interpretability and physical consistency required for reliable system identification. In particular, they do not enforce governing equations or system modeling constraints, making it difficult to associate inferred quantities with physically meaningful parameters (e.g., stiffness, damping, or input force frequencies). In this context, Physics-Informed Neural Networks (PINNs) have emerged as a more robust alternative, combining the flexibility of neural networks with the capability to embed physical laws directly into the learning process \cite{raissi2019physics}. The PINNs have been coupled with the EKF for joint state-parameter estimation, yielding reliable parameter estimation results under the assumption of known excitation forces \cite{liu2024neural}.  Further,  Bayesian (B-PINNs) were proposed for solving forward and inverse nonlinear problems, demonstrating improved predictive accuracy over standard PINNs, particularly in scenarios characterized by high signal-to-noise ratio \cite{yang2021b}.  Another study considered augmentation of the PINNs with sparse‐regression techniques for identification of the discrepancy terms in nonlinear dynamics using the modal characteristics of the monitored system \cite{lai2021structural}.  This framework was further validated on an experimental dataset from a miniature cable-stayed bridge \cite{lai2022neural}.  A more recent study \cite{Haywood-Alexander2025} demonstrated that PINNs can be effectively applied for joint state-parameter estimation in sparse sensing scenarios, while also capturing modeling uncertainties. In the context of input force identification, PINNs have been applied to estimate slowly varying load using both parallel and sequential network architectures, where the authors advocate for the use of displacement measurements to jointly infer the applied load and the underlying system parameters
 \cite{buildings13030650}.  Although several studies demonstrated accurate state and parameter recovery, they generally assume a known excitation history or focus primarily on ambient excitations, limiting their applicability in scenarios involving unknown nonstationary external forces.  The focus of this work is to introduce Spectrum- and Physics-Informed Neural Networks (SaPINNs), a hybrid framework that integrates the spectral characteristics of the input forces into the PINN formulation to enhance both interpretability and computational efficiency in the joint input–state–parameter estimation problem for systems subjected to natural hazard-induced excitations.\\
The organization of this paper is as follows. Section \ref{motiv} outlines the motivation for the proposed framework.  Section \ref{section.PINNs} provides an overview of the classical PINN framework and highlights the limitations that motivate the proposed extension. Section \ref{section.4} presents the SaPINN formulation, including theoretical developments, observability analysis, and uncertainty quantification. Section \ref{examples} illustrates the performance of the proposed framework through numerical examples on both linear and nonlinear systems subjected to sinusoidal, wind, and seismic excitations, including the El Centro earthquake. Finally, Section \ref{conclusion} provides discussion of the results, limitations of the approach, and potential applications.
\section{Motivation}\label{motiv}
The annual frequency of natural hazards has been steadily increasing worldwide  \cite{alimonti2024number}, and with the intensification of these events, the risk of significant damage to civil infrastructure also rises.  Natural hazards, such as earthquakes and strong winds, induce substantial responses in civil engineering structures. Under extreme events, these forces can cause significant or even catastrophic damage. In milder conditions, however, they induce strong excitations that activate majority of the structural  modes, providing a richer basis for system identification than free vibration alone. Extreme loading events can induce nonlinear behaviors, such as stiffness degradation, hysteretic damping, and localized damage, which are typically dormant under operational or low-intensity excitations, thus the damage mechanisms such as cracking, yielding, or joint slippage remain latent until triggered by high-stress demands.
Analyzing the system’s performance under high-intensity excitations yields a more comprehensive and realistic characterization of its in-service behavior, enabling the refinement of numerical models, validation of design assumptions, and reevaluation of safety margins. However, system identification under natural hazard-induced excitations presents a series of complex challenges stemming from the nonlinear, transient, and often unpredictable nature of extreme loading events.  These events typically generate short-duration, high-frequency excitations, complicating analysis and invalidating stationarity assumptions.  Moreover, it is common and often unavoidable, that the exact time-history of the excitation is unknown. In seismic applications, ground motions are typically recorded at regional stations or free-field locations rather than directly at the structure’s base. Local site conditions, soil–structure interaction, and foundation dynamics can significantly alter the effective input, complicating the accurate representation of the applied force at the monitored system.  Similarly, wind and tornado-induced pressures are inherently transient, spatially non-uniform, and rarely captured with sufficient resolution during real events. Further, sensor deployments are often sparse, leading to limited observability and increased difficulty in reconstructing full system dynamics. 
Despite these challenges, a key advantage in this context is the availability of established spectral characterizations and empirical models of well-studied natural hazard events. These include standardized ground motion spectrum in earthquake engineering, as well as canonical models such as Kaimal or Davenport spectrum for wind-induced loads. Such information can be integrated into the prediction models to guide the learning of physically plausible excitation profiles, enhancing identifiability and reducing the solution space of the inverse problem. To incorporate these empirical models into the system identification framework, the SaPINNs are proposed as an extension of classical PINNs architecture. In this formulation, spectral information and empirical models of excitation profiles are explicitly embedded into the training process, introducing spectrum- and physics-consistent regularization that effectively constrains the solution space. This integration enhances the fidelity of system identification and improves the model’s predictive performance under complex and uncertain loading scenarios typical of natural hazard-induced events, as is be shown in the subsequent sections.
\section{Overview of PINNs}\label{section.PINNs}
Physics-based modeling represent a class of deep learning frameworks that incorporate governing physical laws directly into the neural network  training. By embedding the governing equations into the loss function, the admissible solution space is restricted to functions consistent with the underlying physics. This is accomplished by jointly minimizing the discrepancy between model predictions and observed data, along with the residuals of the governing equations, guiding the network to learn solutions that adhere to both the observations and physical laws within a defined tolerance. The general form of the PINNs loss function used during training comprises both physics-based and data-driven components as:
\begin{equation}
\mathcal{L}(\Theta) 
= \frac{\lambda_p}{N_p} \sum_{i=1}^{N_p} 
\left\|  f(t_i) -\mathcal{N}_\Theta[y_\Theta](t_i) \right\|_2^2
+\frac{\lambda_D}{N_D} \sum_{j=1}^{N_D} 
\left\|  y_j^{\text{obs}}-y_\Theta(t_j)  \right\|_2^2,
\end{equation}
where $\left\{y^{\mathrm{obs}}\right\}$ is the observed data,  $y_\Theta $ is a neural network approximator of a target function $y(\boldsymbol{t})$, parameterized by a parameter set $\Theta$—which includes both neural network and unknown physical parameters—and dependent on the input information $\boldsymbol{t}$ .  The  $N_{\Theta}$ denotes the differential operator acting on the solution $y_{\Theta}$,  and $f(\boldsymbol{t})$  represent the input in the governing equation. The scaling factors {${\lambda_p, \lambda_D}$} are used to balance the contributions of the terms in the loss function. These factors are typically selected based on the fidelity of the available data and its relative impact on the total loss \cite{psaros2022meta}.  An important feature of PINNs is automatic differentiation (AD) that enables efficient computation of function derivatives defined over computational graphs. As opposed to symbolic or numerical differentiation AD allows the network to evaluate the residuals from the differential operator $N_{\Theta}$ directly from the network outputs, allowing for the enforcement of physics governing equations through the loss function. In inverse problems, this operator encodes the latent dynamics of the system through its dependence on unknown parameters $\Theta$. When combined with observed data, it constrains the network to learn latent system dynamics that are consistent with both the measurements and the governing physical laws. The classical model of a linear time-invariant (LTI) system is typically represented by the following equation: 
\begin{equation}
    M\ddot{\boldsymbol{x}}(t) + C\dot{\boldsymbol{x}}(t) + K{\boldsymbol{x}}(t) = U(t),
\end{equation}
where $M$ is the mass matrix, $C$ is the damping coefficient matrix, $K $ is the stiffness matrix. Here,  the  $x(t)$ denotes the displacement, $\dot{x}(t)$ and $\ddot{x}(t)$ are the velocity and acceleration, respectively, and $U(t)$ is the external force applied to the system for the time of the measurements $t$. Consider the problem, where the objective is to infer a set of unknown structural parameters denoted by $\boldsymbol {\theta}$, along with unknown system states $\boldsymbol{z}(t)=[{\boldsymbol{x}(t)}, \dot{{\boldsymbol{x}}}(t)]^T$. The input–output relationship of the dynamic system can be represented using a state-space formulation:
\begin{align}
    \dot{\boldsymbol{z}} (t)&= A (\boldsymbol {\theta}){\boldsymbol{z}}(t)+ BU(t) \\
    {\boldsymbol{y}}(t) &= H{ \dot{\boldsymbol{z}}}(t) + \boldsymbol{\eta}(t),
\end{align}
where $A(\boldsymbol {\theta})$, $B$, are the system matrices derived from the system's differential equation, with $\boldsymbol {\theta}$ being the set of unknown, considered time-invariant, structural parameters, typically taken as stiffness and/or damping ratios at the degrees-of -freedom (DoF) of interest; $H$ is the observation matrix,  $\boldsymbol{y}(t)$ is the system output, considered as measured acceleration, and $\boldsymbol{\eta}(t)$ represents measurement noise in the observations.  To incorporate the system equations as constraints within the PINN formulation, the observed data can be treated as the dynamic system output, while the unknowns in the model include both the structural parameters $\boldsymbol {\theta}$ and the neural network parameters $ \boldsymbol {\xi}$ denoted collectively within the parameter set $\Theta=[\boldsymbol {\theta}, \boldsymbol {\xi}]$. Using as the noisy acceleration  $\boldsymbol{y}^{obs}(t)$ as the observed data, and treating the corresponding velocity and displacement vectors as latent processes, the loss function can be formulated as the following: 
\begin{equation}
\mathcal{L}(\boldsymbol{\Theta})_{PINN} = \frac{\lambda_P}{N_P} \sum_{i=1}^{N_P}
\left\| 
\dot{\mathbf{z}}_{\boldsymbol{\xi}}(t_i)
- A(\boldsymbol{\theta}) \mathbf{z}_{\boldsymbol{\xi}}(t_i)
- BU(t_i)
\right\|_2^2
+
\frac{\lambda_D}{N_D} \sum_{j=1}^{N_D} 
\left\| \mathbf{y}^{\mathrm{obs}}(t_j)-H \dot{\mathbf{z}}_{\boldsymbol{\xi}}(t_j)  \right\|_2^2.
\end{equation}
In this equation the excitation force is considered a known quantity,  $\mathbf{z}_{\boldsymbol{\xi}}$ is a network approximation of the latent state dynamics,  and the time of the measurements $t$ is taken as the input for the training. 
To enable simultaneous reconstruction of the excitation force, unknown states, and system parameters, the input force $U(t)$ can also be considered a latent process to be approximated through the network parameters as $U_{\boldsymbol{\xi}}(t)$.  However, inferring latent states, excitation force, and system parameters from acceleration-only data using PINNs is inherently ill-posed problem, prone to identifiability issues, and requires careful modeling choices. Unless the loss function is augmented by additional known quantities, the problem remains ill-posed, leading to non-unique or physically inconsistent solutions. In particular, the network may converge to spurious combinations of latent states, input forces, and system parameters that reproduce the observed data but are inconsistent with the actual system.  Moreover, classic PINNs often struggle with capturing high-frequency components of dynamic signals, particularly when the system is subjected to broadband or impulsive loads \cite{mustajab2024physics}.  
\vspace{1em} 
\section{Spectrum and Physics-Informed Neural Network }\label{section.4}
\subsection{Methodology}\label{section.4.1}
\begin{figure}
    \centering
      \includegraphics[scale=0.42]{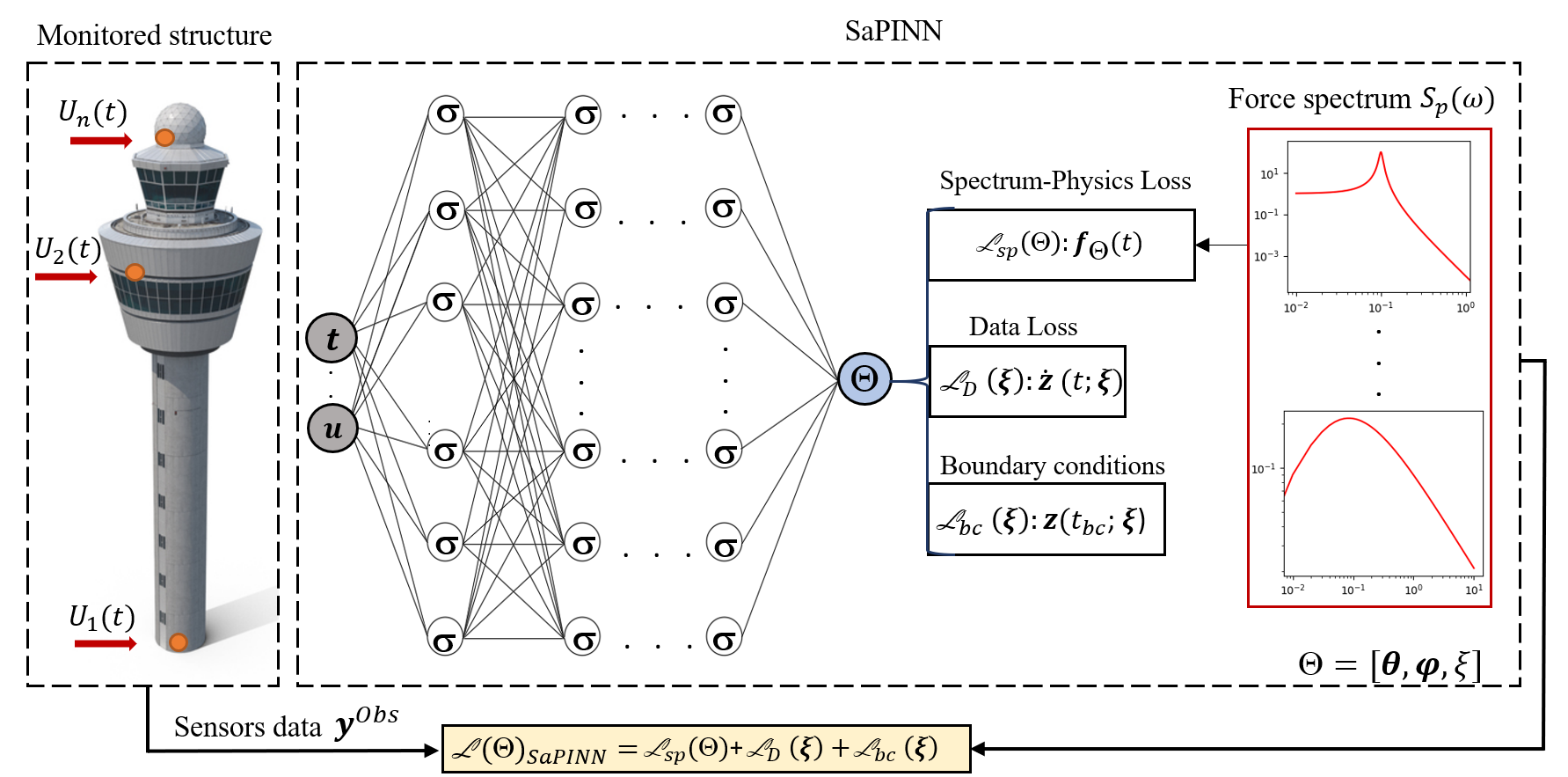}
        \caption{Schematic of the proposed framework for system identification under natural hazard-induced excitations. The monitored structure is subjected to unknown excitation forces \( U_1(t), U_2(t), \ldots, U_n(t) \), while system responses $\mathbf{y}^{\mathrm{obs}}$ are recorded at monitored locations. The loss function \( \mathcal{L}_{\text{SaPINN}} \) consists of the spectrum-physics residual \( \mathcal{L}_{\text{sp}} \), data loss \( \mathcal{L}_D \), and boundary condition loss \( \mathcal{L}_{\text{bc}} \). Monitoring data from a structure are incorporated into the loss function, which evaluates the model's predictive performance. The input layer consist of the time of the measurements $t$ and any other available information from the monitored structure denoted here as $u$. The output from the training is the set of identified structural parameters $\theta$, the excitation force phase angles $\phi$, and the neural network parameters $\xi$. } 
        \label{fig:Fig1}
          \end{figure}

To guide the PINN toward a physically realistic solution space in the absence of measured input forces, the assumed spectral characteristics of the natural hazard can be incorporated as an additional constraint.  In this study, system identification is carried out in the time domain, while the force spectrum is introduced through the Spectral Representation Method (SRM) \cite{DEODATIS1996149}. By specifying a target power spectral density (PSD) representative of the excitation type, the method enables the generation of plausible realizations of a stationary stochastic process $p(t) $ through a Fourier-type expansion of cosines derived from the given spectrum $S_p$ as:
\begin{equation}
p(t)= \sum_{i=1}^{N_{\omega}}\sqrt{2\, S_p(\omega_i)\, \Delta \omega}\,\cos\!\bigl(\omega_i t+\phi_i\bigr),
\end{equation}
where $S_p (\omega_i)$ is the PSD at frequency $\omega_i$, $\Delta \omega$ is the frequency increment, $\phi_i$ is the phase associated with $\omega_i$, and $N_{\omega}$ is the number of frequencies considered for the approximation.  The SRM considers randomness to be incorporated in the uniformly distributed phase angles  $\phi\sim  [0, 2\pi]$, which are unique for each realization. Therefore, to identify a realization of the stochastic process that corresponds to the true excitation, the phases associated with each frequency component of the input force must be estimated so that the resulting signal is consistent with the measured system output. Assuming the excitation spectrum $S_p(\omega)$ is known for a specific extreme event, the corresponding excitation at a DoF of interest can be modeled with:
\begin{equation}
U_{ \phi}(t) = h(t) \, p(t; \phi),
\label{eq:uu}
\end{equation}
where $h(t)$ is a time-dependent modulation (envelope) function specific to the hazard, and $p(t; \phi)$ is a realization of the stationary stochastic process from spectrum $S_p$, parameterized by a set of phase angles $\phi \in \mathbb{R}^{N_{\omega}}$.  It is important to emphasize that in this work we assume the functional form of the modulating function is either known a priori or can be inferred from regional analyses conducted during or after the occurrence of a natural hazard event. Previous studies have demonstrated that it is possible to approximate the parameters of the modulating function by analyzing the observed structural response, as the cumulative energy content and its temporal evolution are correlated with the growth and decay characteristics of the time-dependent modulating function \cite{ gustfactor,solari2016thunderstorm, rezaeian2010stochastic, STAFFORD20091123}.  Further, depending on the empirical characterization of the event, the modulation can be implicitly embedded within a time-varying Evolutionary (EPSD).  In such cases, the nonstationary nature of the excitation is captured directly through the spectrum, which governs the envelope of the excitation processes \cite{RONCALLO2022104978}. \\
Building on these assumptions, here we show why SaPINNs offer a more robust formulation than conventional PINNs for input–state–parameter estimation. Consider an input force on a fixed time horizon $[0,T]$ characterized by an envelope $h (t):[0,T]$ and hazard–specific amplitudes $a_i \coloneqq \sqrt{2\,S_p(\omega_i)\,\Delta\omega}$ with discrete frequencies $\{\omega_i\}_{i=1}^{N_\omega}$, the admissible excitation set can be represented as: 
\begin{equation}
\mathcal{A} \;\coloneqq\; \biggl\{ U_\phi(t) \;=\; h(t)\sum_{i=1}^{N_\omega} a_i \cos\!\bigl(\omega_i t + \phi_i\bigr) \;:\; \phi \in [0,2\pi)^{N_\omega} \biggr\}.
\label{eq:admissible_set}
\end{equation}
The $\mathcal{A}$ set implies that the input is parameterized by the phase vector $\phi \in \mathbb{R}^{N_{\omega}}$, rather than by a generic trajectory $U(t)\in L^2(0,T)$ as in unconstrained PINNs, yielding a lower-dimensional search space. The PINNs approach result in the force parameterization $U_{\boldsymbol{\xi}}(t) \in \mathbb{R}^{N_{\boldsymbol{\xi}} }$, where $N_{\boldsymbol{\xi}}$ is the number of network parameters, which requires optimizing a high-dimensional set of $\boldsymbol{\xi}$, which not only increases computational cost but also introduces instability and ill-conditioning in the optimization process, thereby amplifying uncertainty in the estimation of latent states and system parameters. Let $\mathcal{G}_\Theta$ denote the LTI (or weakly nonlinear) dynamics operator mapping input $U$ to the noiseless system output $y$ as follows:
\begin{equation}
\mathcal{G}_\Theta:\; U \mapsto y \quad \text{with}\quad M\ddot x + C(\theta)\dot x + K(\theta)x = U, \;\; y = H\ddot x.
\end{equation}
With measured system response $\mathbf{y} $ at $N^{obs}$ DoF, the constrained inverse problem becomes:
\begin{equation}
\min_{\boldsymbol{\Theta}}\; \frac{\lambda_D}{N_D}  \sum_{i=1}^{N_D}\!\bigl\|\mathbf{y}(t_i)-H\,\dot{\mathbf{z}}_{\boldsymbol{\xi}}(t_i)\bigr\|_2^2
\;+\;\frac{\lambda_{sp}}{N_{sp}}\!\sum_{j=1}^{N_{sp}}\!\bigl\|\mathbf{y}(t_j)-H\,f_{\boldsymbol{\Theta}}(t_j)\bigr\|_2^2
\quad \text{s.t.}\;\; U \,=\, U_{\boldsymbol{\phi}} \in \mathcal{A},
\label{eq:constrained}
\end{equation}
where $ \lambda_D, \lambda_{\text{sp}}$ are the data and spectrum-physics loss weight factors, $\mathbf{z}_{\boldsymbol{\xi}}$ are the latent states generated by the network, and $f_{\boldsymbol{\Theta}}(t)$ is the spectrum and physics-informed function of the system output, defined for the brevity of notation as:
\begin{equation}
f_{\boldsymbol{\Theta}}(t) =
A(\boldsymbol{\theta}) \, \mathbf{z}_{\boldsymbol{\xi}}(t)
 -\boldsymbol{ B} \, U_\phi(t)).
\end{equation}
In the proposed framework the phases $\phi$ are treated as additional parameters to be learned during the model training, resulting in a reformulated set of trainable variables $\Theta = \left[ \boldsymbol{\theta}, \, \boldsymbol{\xi}, \, \phi \right]$.
As shown in the above equations, the SaPINNs implement structured input constraints rather than an unconstrained $U_{\boldsymbol{\xi}}(t)$ penalized by the physics and data-based losses. This formulation also leads to a more stable solution as the input force gradients with respect to the phase angles are explicit and bounded for backpropagation through the spectrum–physics residual:
\begin{equation}
\frac{\partial U_\phi(t)}{\partial \phi_i} \;=\; -\,h(t)\,a_i\,\sin\!\bigl(\omega_i t + \phi_i\bigr), \qquad
\bigl\|\partial U_\phi/\partial \phi_i\bigr\|_{L^2 (0,T)} \le \|h\|_{L^2 (0,T)} a_i.
\label{eq:phase_grad}
\end{equation}
Further,  it is important to  to consider the problem if the spectrum is not known precisely. Suppose the true input $U^\star\notin\mathcal{A}$, and let the $L^2$–projection onto $\mathcal{A}$ be $\Pi_{\mathcal{A}}U^\star=U_{\phi^\star}$ with the projection error denoted as:
\begin{equation}
\varepsilon_U \;\coloneqq\; \bigl\| U^\star - \Pi_{\mathcal{A}}U^\star \bigr\|_{L^2(0,T)} 
 \;=\; \min_{U\in\mathcal{A}} \|U^\star-U_{\hat\phi}\|_{L^2(0,T)}.
\end{equation}
For an LTI system, the input–output map $\mathcal{G}_\Theta$ satisfies $\|\mathcal{G}_\Theta\|_{L^2\to L^2}\le \|\mathcal{H}(j\omega;\boldsymbol{\theta})\|_{H_\infty}$ , where $\mathcal{H}(j\omega;\boldsymbol{\theta})$ is the transfer function from input $U$ to measured noiseless output $y$ , hence for the learned $U_{\hat\phi}\in\mathcal{A}$, the gained energy in the response from the excitation satisfies:
\begin{equation}
\bigl\| y_{\boldsymbol{\theta}}(U^\star)- y_{\boldsymbol{\theta}}\bigl(U_{\hat\phi}\bigr) \bigr\|_{L^2(0,T)} \;\le\; \|\mathcal{H}(j\omega; \boldsymbol{\theta})\|_{H_\infty}\,\varepsilon_U,
\end{equation}
with  $\|\mathcal{H}(j\omega; \boldsymbol{\theta})\|_{H_\infty}$ being the norm representing the worst-case (maximum) gain over all frequencies conditional on the set of structural parameters $\boldsymbol{\theta}$.  Thus, any mismatch between the empirical spectrum and the true excitation (including envelope misfit, frequency truncation, or discretization) induces a bounded error proportional to $\varepsilon_U$,  scaled by the system energy gain. In practice, this can appear as nonzero residuals, frequency-localized misfit, and larger uncertainty bounds in the estimated system states and parameters (as demonstrated in section \ref{section.elcentro}). 
Since the observed system output is typically contaminated by measurement noise, to account for its effect the observations can be modeled as:
\begin{equation}
y^{\mathrm{obs}}(t) \;=\; y_{\boldsymbol{\theta}^\star}(U^\star)(t) \;+\; \eta(t).
\end{equation}
where $y_{\boldsymbol{\theta}^\star}(U^\star)$ is the noise-free output generated by the true system parameters $\boldsymbol{\theta}^\star$ under the true excitation $U^\star$, and $\eta(t)$ denotes the measurement noise (assumed Gaussian zero-mean). Consequently, the prediction error with respect to the observations reflects not only spectral misspecification but also model modeling error and measurement noise:
\begin{equation}
\begin{aligned}
\bigl\|y^{\mathrm{obs}}-y_{\boldsymbol{\hat\theta}}(U_{\hat\phi})\bigr\|_{L^2(0,T)}
&\le \underbrace{\bigl\|y_{\theta^\star}(U^\star)-y_{\theta^\star}(U_{\hat\phi})\bigr\|_{L^2(0,T)}}_{\text{spectrum error}}
\;+\; \underbrace{\bigl\|\bigl(\mathcal{G}_{\Theta^\star}\bigr)U_{\hat\phi}-\mathcal{G}_{\hat\Theta}\bigr\|_{L^2(0,T)}}_{\text{modeling error}}
\;+\; \underbrace{\|\eta\|_{L^2(0,T)}}_{\text{measurement noise}} \\[3pt]
\end{aligned}
\end{equation}
The inequality shows that even with a perfect spectral fit ($\varepsilon_U=0$), residuals will remain due to measurement noise and model mismatch due to the difference between structural parameters $\hat{\boldsymbol{\theta}}\neq \boldsymbol{\theta}^\star$. \\
Now, let $Y_{\mathrm{PINN}}(\boldsymbol{\theta}) := \{\, y_{\boldsymbol{\theta}}(U_{\boldsymbol{\xi}}) : \boldsymbol{\xi}\in\mathbb{R}^{N_\xi}\,\}$ denote the outputs reachable by PINN under model parameters $\boldsymbol{\theta}$.
With noisy $y^{\mathrm{obs}} $and PINN-generated input force surrogate $U_{\boldsymbol{\xi}}$, the residual decomposes into three contributions: the modeling error, arising from the discrepancy between the true and estimated dynamics evaluated on the surrogate input; the input force mismatch, reflecting the deviation of the surrogate input from the true excitation; and the measurement noise:

\begin{equation}
\begin{aligned}
r_{\mathrm{PINN}}(\boldsymbol{\xi};\hat{\boldsymbol{\theta}})
&= {y}^{\mathrm{obs}} - y_{\hat{\boldsymbol{\theta}}}(U_{\boldsymbol{\xi}})
= \underbrace{\bigl(\mathcal{G}_{\boldsymbol{\theta}^\star}-\mathcal{G}_{\hat{\boldsymbol{\theta}}}\bigr)U_{\boldsymbol{\xi}}}_{\text{modeling error}}
\;+\; \underbrace{\mathcal{G}_{\boldsymbol{\theta}^\star}\!\big(U^\star - U_{\boldsymbol{\xi}}\big)}_{\text{input error}}
\;+\; \eta.
\end{aligned}
\end{equation}
As $U_{\boldsymbol{\xi}}$ is only partially constrained by the physics loss,  the optimizer can reduce the residual by letting the input-mismatch term cancel the modeling error term, yielding a small $\|r_{\mathrm{PINN}}(\hat{\boldsymbol{\theta}})\|_{L^2}$ even when $\hat{\boldsymbol{\theta}}\neq\boldsymbol{\theta}^\star$ .
In contrast, by restricting inputs $U_{\phi}\in\mathcal A$, SaPINNs (i) reduce the degrees of freedom available to "absorb” the modeling error $\delta_{\mathrm{model}}$, and (ii) yield less biased $\hat{\boldsymbol{\theta}}$ under local identifiability (discussed in the section \ref{Observ}), since the optimizer will not be able to trade system parameters error for the parameters describing the input force. Therefore, by constraining the excitation to a hazard-consistent spectral family, SaPINNs reduce the dimensionality of the input space and prevent the optimizer from compensating for modeling errors through input surrogates. As a result, the proposed framework provides more stable training, bounded error propagation under spectral misspecification, and less biased parameter estimates than conventional PINNs, even when the residuals are not minimized to the same extent. \\
Using the proposed SaPINN formulation, the total loss function combines spectrum–physics, data, and boundary/initial conditions:
\begin{equation}
\mathcal{L}(\Theta)_{SaPINN} = \frac{\lambda_{sp}}{N_{sp}} \sum_{i=1}^{N_{sp}}
\left\| 
\mathbf{y}^{\mathrm{obs}}(t_i)
- Hf_{\Theta}(t_i)
\right\|_2^2
+
\frac{\lambda_D}{N_D} \sum_{j=1}^{N_D} 
\left\| \mathbf{y}^{\mathrm{obs}}(t_j)-H\dot{\mathbf{z}}_{\boldsymbol{\xi}}(t_j)  \right\|_2^2
+
\frac{\lambda_{bc}}{N_{bc}} \sum_{k=1}^{N_{bc}}\left\|\mathbf{z}_{bc}(t_k) -\mathbf{z}_{\boldsymbol{\xi}}(t_k)   \right\|_2^2.
 \label{eq:sappins_loss}
\end{equation}
Here, $\mathbf{z}_{\mathrm{bc}}$ encodes any known boundary or initial conditions, and $(\lambda_{sp},\lambda_D,\lambda_{bc})$ weight the relative contributions (with normalization by sample counts $N_{sp},N_D,N_{bc}$). In this formulation, we intentionally rely on the observed data \( \mathbf{y}^{\text{obs}} \) for the spectrum and physics residual to improve convergence toward the observed system response. The states derivative \( \dot{\mathbf{z}}_\xi(t) \) are computed using AD, enabling the enforcement of dynamic consistency through the governing equations and the approximation of latent states through the network parameterized mapping.  The structure of a loss function depends on the type and quality of data available from the monitored structure, as well as the number and nature of physical constraints incorporated into the training process. In this work, we consider a setting in which the only available dynamic measurements are noisy accelerations, and the boundary conditions are specified through initial state information. The loss function is defined as a weighted sum of Mean Squared Errors (MSE), normalized by the dataset size. The solution is obtained by minimizing a composite loss function that incorporates multiple components, each designed to reflect the specific observability and requirements of the problem. As such, the formulation of the loss function is inherently problem-dependent and central to ensuring stability and accuracy of the inverse problem. An overview of the proposed framework for the input-state-parameter estimation using SaPINNs is presented in Fig.\ref{fig:Fig1}.

The proposed architecture employs a fully connected multilayer feedforward neural network (FNN) with periodic activation functions. Specifically, for implicit neural representations, a sine activation function is employed in each layer, defined as: 
\begin{equation}
\boldsymbol{\sigma}_i=sin(\textbf{W}_it_i+\textbf{b}_i),
\end{equation}
where the function $\boldsymbol{ \sigma_i }$ corresponds to the \( i^{\text{th}} \) layer of the network, \( t_i \in \mathbb{R}^{N} \) is the input to the layer, \( \mathbf{W}_i \in \mathbb{R}^{L \times N} \) is the weight matrix, and \( \mathbf{b}_i \in \mathbb{R}^{L} \) is the bias vector, with \( L \) denoting the number of neurons in the \( i^{\text{th}} \) layer and $N$ represents the dimensionality of the input vector. This choice of activation is motivated by Fourier feature neural representations, which have been shown to enhance model performance in modeling solutions with oscillatory or periodic characteristics  \cite{sitzmann2020implicit}. The sine nonlinearity is applied element-wise to the affine transformation, enabling the network to represent functions with dominant periodic structure. 
\subsection{ Observability: Ensuring a Well-Posed Inverse Problem }\label{Observ}
In any parameter estimation problem, including those approached using machine learning techniques, it is essential to ensure that the system is observable and the parameters are uniquely identifiable. The system identifiability becomes especially important in SHM scenarios where the excitation force is unknown and must be inferred alongside the system states and parameters. If the system is unobservable, multiple combinations of excitation forces and parameter values may produce indistinguishable outputs, thereby undermining confidence in the inferred quantities. Ensuring uniqueness of the solution is therefore critical: if the mapping from parameters and excitations to observations is not injective or well-defined, even a perfectly trained model cannot recover the true physical parameters unambiguously. Therefore, a prior analysis of the model formulation is essential when dealing with limited knowledge about the system, to ensure that the problem is well-posed. 
The methodology proposed herein for solving the joint input-state-parameter estimation problem assumes an observable system, where only acceleration measurements are available for the inference. However, to determine how many unknown quantities can be uniquely identified from these measurements, a formal observability analysis of the problem must be addressed first. To assess whether the model satisfies the necessary conditions for the observability of dynamic system states, the Observability Rank Condition (ORC) \cite{observability} can be applied. The ORC helps determine whether the states of a system can be uniquely identified given a set of measured quantities, and it is particularly useful for systems in which the output measurements are directly influenced by the input forces. In the case of a system with an unknown input force, the ORC method extends classical observability analysis by constructing an augmented observability matrix that accounts for direct feedthrough (DF) effects \cite{MAES2019378}. Given that the proposed framework considers structural response data under natural hazard event, where the system output is continuously affected by external forces, it is appropriate to adopt such an observability analysis to determine the identifiability of latent states and parameters. \\
Consider a single-degree-of-freedom (SDoF) system excited by a sinusoidal force $u(t)=A \sin(\omega_0 t)$. The system mass $m$ , the response $y(t)=\ddot{x}(t)$, and the spectral form of the excitation force are assumed to be known; while the velocity, displacement, excitation force in time domain, and the damping and stiffness parameters, denoted by $\theta_1,$ and $\theta_2$, respectively, are treated as unknown variables. Assuming the system parameters remain time-invariant over the measurement period, an extended observability matrix can be constructed to capture the sensitivity of the system output with respect to the unknowns. Based on the identity of harmonic functions, the sinusoidal excitation can be equivalently expressed as a phase-shifted cosine, which can be rewritten according to the SaPINN proposed formulation as:
\begin{equation}
u(t) = h(t) \, p(t;\phi)=A\cos (\omega_0 t+\phi).
\end{equation}
In this special case, the modulating function $h(t)=A$ is constant, and $p(t; \phi)$  represents the spectral realization of the excitation process, parameterized by a single phase angle $\phi = -\frac{\pi}{2}$. The governing physics-spectrum equation for the problem is then: 
\begin{equation}
m\ddot{x}+\theta_1 \dot{x}+\theta_2 x=A\cos (\omega_0 t+\phi).
\end{equation}
To assess the system’s observability, an augmented state vector of the system $\mathbf{\boldsymbol{z_a}}(t)$ is constructed as: 
\begin{equation}
\mathbf{\boldsymbol{z}}_a(t)
=
\begin{bmatrix}
x_1 \\
x_2 \\
x_3 \\
x_4 \\
x_5
\end{bmatrix}
=
\begin{bmatrix}
x(t) \\
\dot{x}(t) \\
\theta_1 \\
\theta_2 \\
\phi
\end{bmatrix},
\end{equation}
where the unknown parameters $\theta_1, \theta_2$ are treated as state variables, and the functional form of the input force is considered to be known but parameterized by unknown parameter $\phi$, which is also treated as a time-invariant state variable.  Assuming the mass $m=1$ and the measured output is the acceleration, the system response can be expressed as a function of the state variables in augmented notation:
\begin{equation}
y(t)=\ddot{x}(t)=-\,x_3\,x_2 \;-\; x_4\,x_1 \;+\; A \cos(\omega_0 t+x_5) .
\end{equation}
Denoting the system output as a function of the augmented state vector $y(t) = g_s(\mathbf{\boldsymbol{z}}_a(t))$, where the function \( g_s \) maps the augmented state variables directly to the measured acceleration, the system dynamics can be represented with:
\begin{equation}
\dot{\mathbf{\boldsymbol{z}}}_a(t) = \mathbf{f_s}(\mathbf{\boldsymbol{z}}_a(t)),
\end{equation}
which describes how the augmented state vector \( \boldsymbol{z}_a(t) \) evolves over time. This formulation encapsulates both the physical dynamics and the time-invariant parameters treated as states within the problem. Following the established notation, the observability matrix based on ORC-DF formulation can be constructed as:
\begin{equation}
\mathcal{O}(\mathbf{\boldsymbol{z}_a})
\;=\;
\begin{bmatrix}
\nabla\!\bigl(\mathcal{L}_{\mathbf{f}}^{0}\,g_s(\mathbf{\boldsymbol{z}_a})\bigr)
\\[6pt]
\nabla\!\bigl(\mathcal{L}_{\mathbf{f}}^{1}\,g_s(\mathbf{\boldsymbol{z}_a})\bigr)
\\[6pt]
\vdots\\[2pt]
\nabla\!\bigl(\mathcal{L}_{\mathbf{f}}^{l-1}\,g_s(\mathbf{\boldsymbol{z}_a})\bigr)
\end{bmatrix},
\end{equation}
where $\displaystyle \mathcal{L}_{\mathbf{f}}^{0}\,g_s(\mathbf{z}_a) \;=\; g_s(\mathbf{z}_a(t))$ is the zero-order Lie derivative, and $l$ is the dimension of the augmented state vector $\mathbf{z_a}$.  The higher-order Lie derivatives are defined recursively as: 
\begin{equation}
\mathcal{L}_{\mathbf{f}}^{k}\, g_s(\boldsymbol{z}_a)
=
\nabla \bigl( \mathcal{L}_{\mathbf{f}}^{k-1}\, g_s(\boldsymbol{z}_a) \bigr)
\cdot
\mathbf{f}_s(\boldsymbol{z}_a).
\quad k= 1, \dots, l-1
\end{equation}
The $\nabla(\cdot)$   denotes the gradient with respect to the vector $\mathbf{z}_a$, and the derivatives provide a generalization of how the output evolves along the trajectories defined by the nonlinear vector field $\mathbf{f_s}$. The rank of the observability matrix  $\mathcal{O}(\mathbf{\boldsymbol{z}_a})$ indicates the number of linearly independent parameters that can be uniquely identified based on the available data and the system equations. Since the problem is formulated in the time domain with continuous excitation, the non-degeneracy condition $x_i \not\equiv 0,\quad \forall\, i = 1, \dots, l$  is naturally satisfied, and the rank of the observability matrix for the system is $\operatorname{rank}(\mathcal{O}) = l = 5$. This indicates that all states and parameters in the augmented state vector are locally observable, provided a sufficiently rich and long acceleration dataset. A fully constructed observability matrix is presented in Appendix A, and an additional analysis for the case where the amplitude of the excitation force is considered unknown is provided in Appendix B. \\
As the structural systems subjected to natural hazard-induced loading exhibit more complex dynamics than those excited by a purely sinusoidal function, a further analysis of the observability condition is necessary.  Using previously established Eq. \ref{eq:uu} , the simple case of the sinusoidal excitation can be extended to a more complex scenario, where the force is expressed as: 
\begin{equation}
u(t)= h(t)\sum_{i=1}^{N_{\omega}}\sqrt{2 S_p(\omega_i) \Delta \omega}\cos(\omega_i t+\phi_i).
\end{equation}
Here, we assume the $S_p (\omega)$  and  $h(t)$ to be known quantities, as these terms are available from the empirical models for the specific natural hazard event.  The augmented state vector is therefore expanded to include $N_{\omega}$ phase angles to model the unknown hazard-induced excitation:
\begin{equation}
\mathbf{\boldsymbol{z_a}}(t)
=
\begin{bmatrix}
x_1 \\
x_2 \\
x_3 \\
x_4 \\
x_5\\
\vdots\\[2pt]
x_l
\end{bmatrix}
=
\begin{bmatrix}
x(t) \\
\dot{x}(t) \\
\theta_1 k\\
\theta_2c \\
\phi_1\\
\vdots\\[2pt]
\phi_{N_\omega}
\end{bmatrix},
\end{equation}
where the size of the augmented state vector increases to $l=N_{\omega}+4$. The formulation of the observability matrix remains unchanged; however, a greater number of  Lie derivatives is required to verify observability of the problem. Extreme hazard excitations naturally introduce a more complex forcing function compared to pure sinusoidal input, however, within the SRM formulation, the phase angles describing the stochastic process remain linearly independent over any non-degenerate time interval. The frequencies $\omega_i$ are distinct for each cosine,  which means that changes in $\phi_i$ affect the output in a way that is independent of changes in $\phi_j$ for $i \neq j$. This ensures that, despite the increase in the number of phases needed to describe the excitation force, the observability matrix remains full-rank and invertible, satisfying the ORC-DF criteria and preserving the identifiability of the augmented state vector. This “global” recovery of the augmented state from the observed system response is a practical consequence of local observability in conjunction with persistent excitation.  As a result, it provides sufficient independent information to uniquely recover the augmented state vector.   Since the outputs are measured over a time interval, the system exhibits behavior consistent with practical or interval-wise global observability, even though the ORC remains a local criterion. 
Further, as the problem is proposed to be solved using the SaPINNs, the additional constraints imposed by the loss function, such as boundary conditions on the states and priors on the phase angles (i.e. $\phi \in [0, 2\pi]$) help to ensure that all parameters in the augmented state vector are uniquely identifiable within the proposed framework.

\subsection{Uncertainty quantification}
Although classic PINNs offer a compelling framework for integrating data with physics-based modeling, they are not inherently probabilistic. As a result, both the standard PINN and by extension the SaPINN architecture yield point estimates of the solution, without providing any measure of uncertainty or confidence in their predictions. However, several methods exist for incorporating uncertainty into PINN-based frameworks. Hidden physics models couple PINNs with a Gaussian process surrogates to capture model discrepancies, enabling both model and data uncertainty quantification \cite{raissi2018hidden}. Bayesian PINNs address uncertainty by treating neural network weights as random variables with priors and performing posterior inference using variational or sampling methods, enabling estimation of predictive distributions over the unknowns \cite{yang2021b}. Dropout-based approaches, originally developed as a regularization technique, can be reinterpreted as approximate Bayesian inference, yielding a coarse but tractable surrogate for deep Gaussian processes; although they offer lower computational cost and have been applied successfully to large-scale datasets, they are generally less accurate than Bayesian PINNs or Hidden Physics Models \cite{gal2016dropout}. The above mentioned methods offer principled frameworks for uncertainty quantification, however, they often struggle to scale to large architectures or long training horizons and may fail to capture multimodal posteriors, which are critical for representing plausible solutions in complex systems with sparse measurements. In contrast, Deep Ensembles (DEns) offer a simple yet effective approach: multiple networks are trained independently with different initializations, producing diverse solutions that collectively approximate the predictive distribution over the latent processes and unknown variables \cite{psaros2023uncertainty}.  This method has three key advantages in the SaPINN setting: (i) they scale efficiently to high-dimensional inverse problems, as models can be trained in parallel without altering the architecture; (ii) they reduce sensitivity to optimization difficulties by exploring diverse local minima through different initializations; and (iii) they naturally capture multimodal posterior structure, which is critical for ill-posed problem of input–state–parameter estimation.  Accordingly,  in this study we adopt the DEns to address uncertainty estimation in the predicted input, latent states, and system parameters. By training multiple instances of the SaPINN, the resulting ensemble approximates an empirical posterior distribution over the inferred quantities. Compared to a single network, the ensemble provides mean and variance of the predictions and offers a straightforward measure of epistemic uncertainty \cite{NIPS2017_9ef2ed4b}. By considering $q \in Q$  models of the network to be trained, each $q^{th}$ model yields its own prediction of the target output:
\begin{equation}
\hat \Theta^{(q)} = \arg \min_{\Theta^q} \, \mathcal{L}(\Theta^q), \quad \text{for } q = 1, \dots, Q
\end{equation}
where \( \Theta^{(q)} \) are the randomly initialized parameters of the \( q^{\text{th}} \) network in the ensemble. Following determination of $\hat \Theta^{(q)}$, the predictions from each $q$ model are then aggregated to estimate both the mean of the inferred variables and processes, as well as the associated uncertainty.  Specifically, for a given input $t$, the predictive mean and variance of the excitation force are estimated under the assumption of a normal distribution: 

\begin{equation}
\mathbb{E}[U(t)] 
\approx \frac{1}{Q} \,  \, h(t) \sum_{q=1}^{Q} p(t; \boldsymbol{\hat{\phi}}^{(q)}),
\end{equation}
\begin{equation}
\mathrm{Var}[U(t)] \approx \frac{1}{Q - 1} \sum_{q=1}^{Q} \left( h(t)\, p(t; \boldsymbol{\hat{\phi}}^{(q)}) - \mathbb{E}[U(t)] \right)^2.
\end{equation}

In this approach, uncertainty in the predicted solution is characterized by the variability in parameter estimates obtained from multiple independently trained models within the ensemble. 
\section{Illustrative examples}\label{examples}
In this section, the performance of the proposed architecture is evaluated on both linear and nonlinear dynamic systems subjected to unknown excitation time histories. The capabilities of classic PINNs and SaPINNs are compared in terms of input force reconstruction, latent states, and system parameters estimation. The examples begin with a toy problem in which a SDoF system is excited by a sinusoidal force applied at the center of mass. Subsequently, the analysis is extended to more realistic scenarios involving stochastic wind loads from a thunderstorm and seismic excitations. This progression allows for a direct comparison of the framework performance under simple harmonic excitation and complex non-stationary dynamic loading conditions. In this study, the networks are trained using the Adam optimizer \cite{kingma2014adam}, which adaptively adjusts the learning rates of the parameters based on estimates of the first and second moments of the gradients.  The learning rate for the optimizer was taken as $1 \times 10^{-3}$ for each full-batch training step. In the presented examples, the available observations are limited to noise-contaminated accelerations, with the system states and stiffness parameters treated as unknown variables,  except for the toy example, where the excitation amplitude is also considered unknown. The sampling rate for the measurements in all examples was set to 100 Hz. The scaling factors for the system parameters were initialized at 1.2, while the true values were set to 1. Within the proposed SaPINNs architecture, for each ensemble,  the phase angles ${\phi}$  are initialized using a random uniform distribution over a constrained domain $[0,2\pi]$.  To ensure consistent basis for comparison, both PINNs and SaPINNs employed identical network architectures (same number of layers and nodes), and the boundary conditions specified as initial state values were included in the loss formulation of each model. 
\subsection{Linear systems}\label{ex.linear}
\subsubsection{Toy problem: SDoF system under Sinusoidal Excitation }
\begin{figure}[H]
    \centering
    \begin{subfigure}[b]{\textwidth}
        \centering
        \begin{minipage}[b]{0.32\textwidth}
            \centering
            \includegraphics[width=\textwidth]{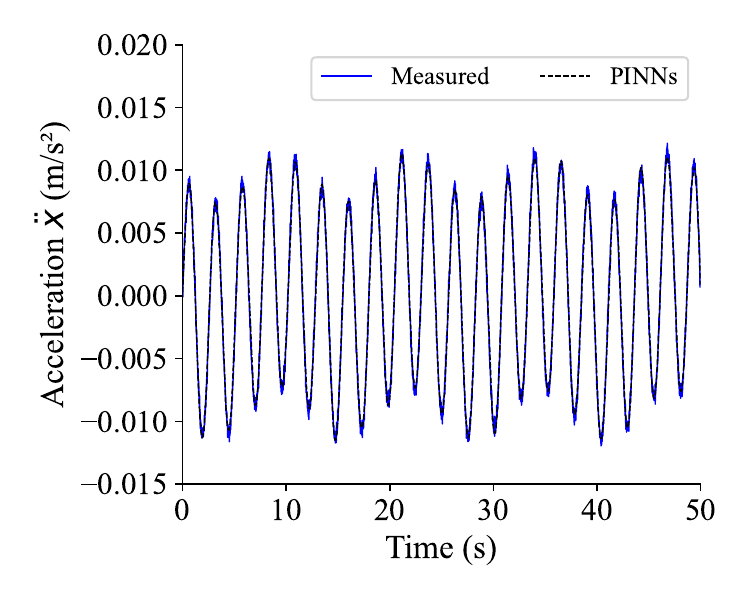}
        \end{minipage}
        \hfill
        \begin{minipage}[b]{0.32\textwidth}
            \centering
            \includegraphics[width=\textwidth]{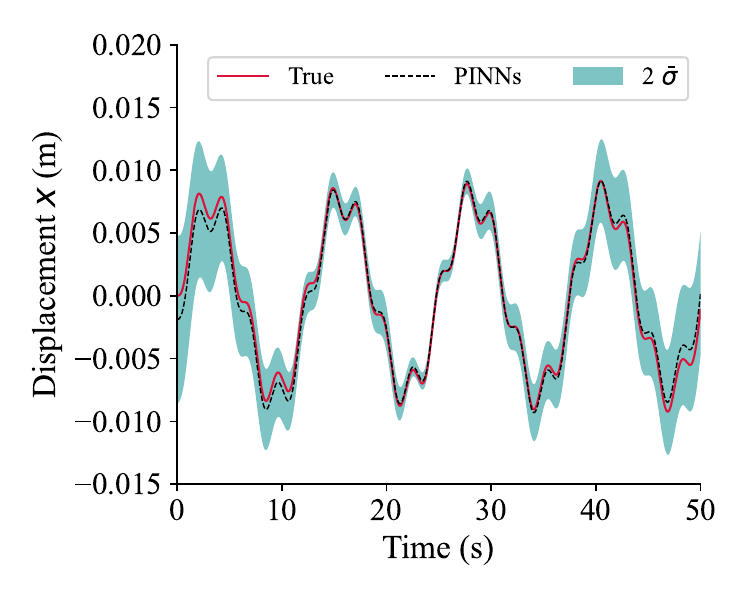}
        \end{minipage}
        \hfill
        \begin{minipage}[b]{0.32\textwidth}
            \centering
            \includegraphics[width=\textwidth]{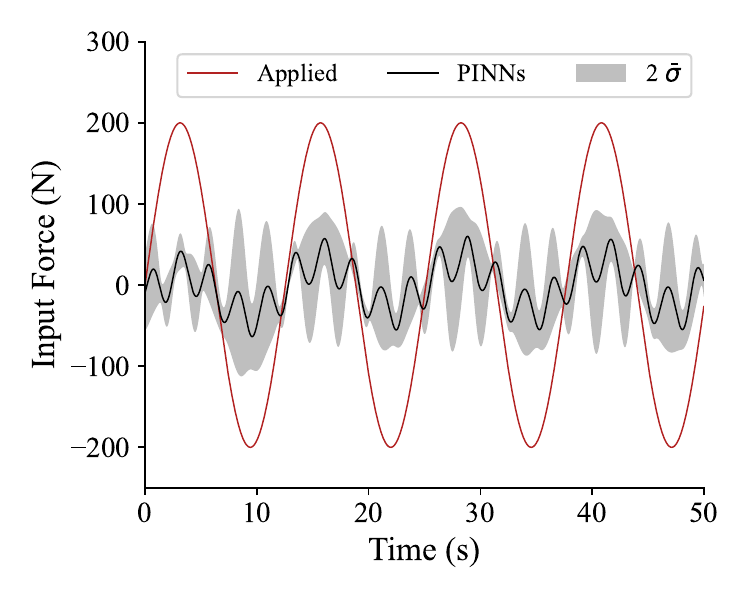}
        \end{minipage}
         \vspace{-2ex}
         \captionsetup{justification=centering}
         \caption*{(a)}
    \end{subfigure}

    \vspace{-1ex}
    \begin{subfigure}[b]{\textwidth}
        \centering
        \begin{minipage}[b]{0.32\textwidth}
            \centering
            \includegraphics[width=\textwidth]{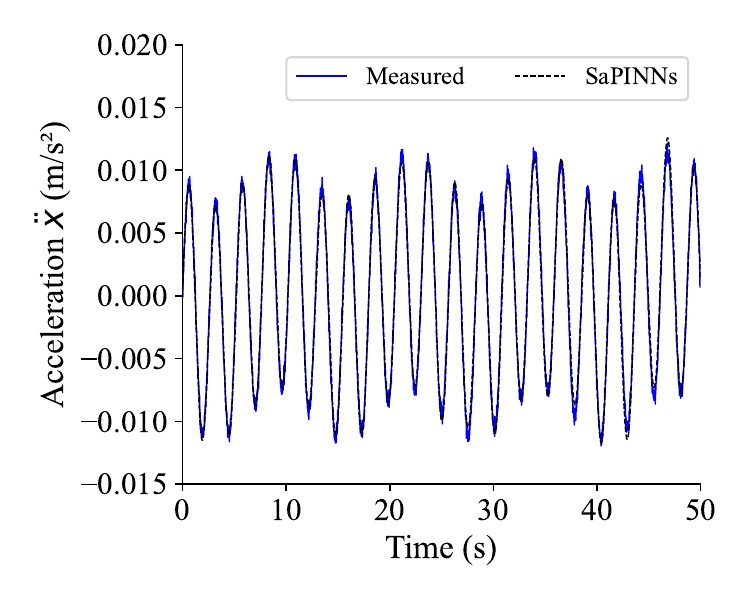}
        \end{minipage}
        \hfill
        \begin{minipage}[b]{0.32\textwidth}
            \centering
            \includegraphics[width=\textwidth]{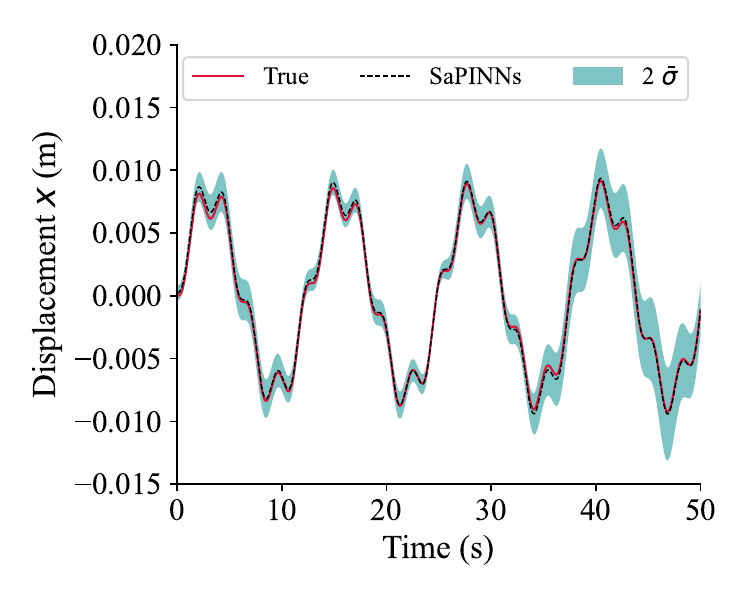}
        \end{minipage}
        \hfill
        \begin{minipage}[b]{0.32\textwidth}
            \centering
            \includegraphics[width=\textwidth]{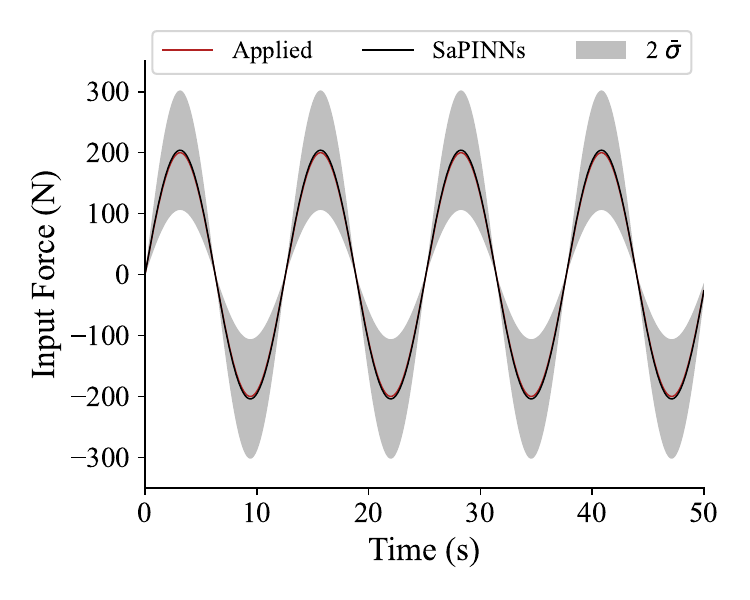}
        \end{minipage}
         \vspace{-2ex}
        \captionsetup{justification=centering}
        \caption*{(b)}
    \end{subfigure}
    \vspace{-2ex}
     \caption{System response used for the training, predicted displacements, and reconstructed input force from (a) PINNs and (b) SaPINNs ensemble estimation. Each group shows the same quantities in corresponding order.}
    \label{fig:Fig2}
\end{figure}

\begin{figure}[!htbp]
    \centering
    \begin{subfigure}[b]{0.32\textwidth}
        \includegraphics[width=\textwidth]{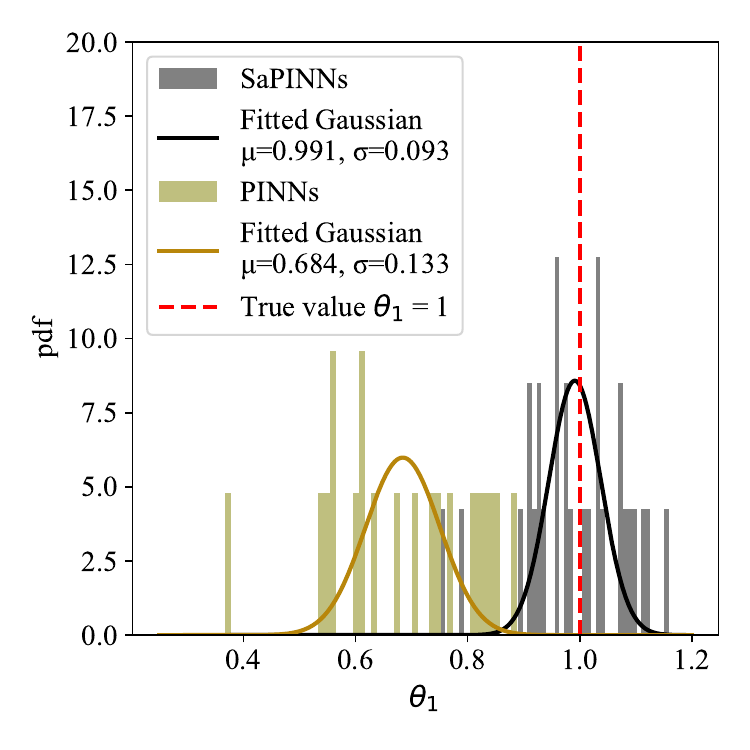}
        \label{fig:plot1}
    \end{subfigure}
    \begin{subfigure}[b]{0.32\textwidth}
        \includegraphics[width=\textwidth]{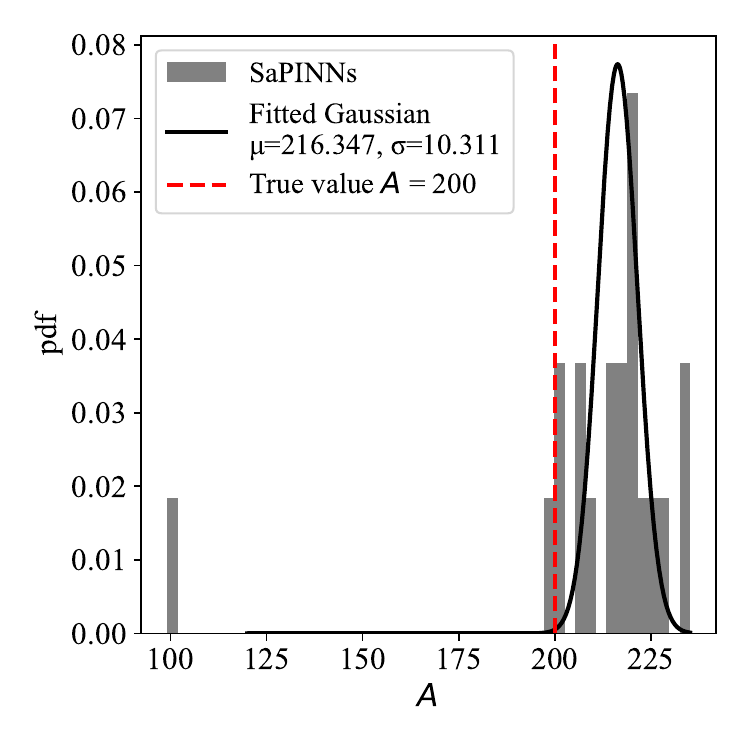}
        \label{fig:plot2}
    \end{subfigure}
    \begin{subfigure}[b]{0.32\textwidth}
        \includegraphics[width=\textwidth]{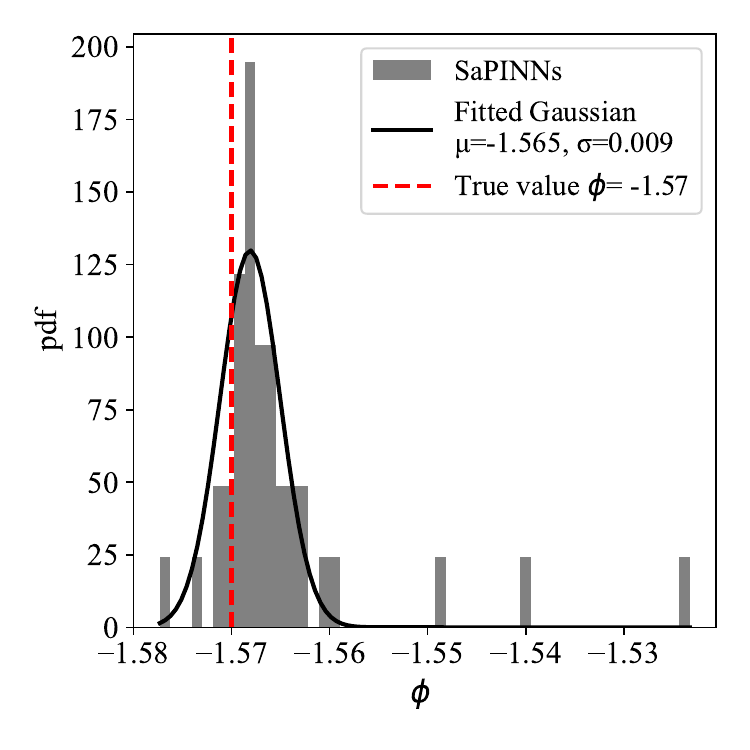}
        \label{fig:plot3}
    \end{subfigure}
    \vspace{-4ex} 
    \caption{SDoF stiffness scaling factor estimates from ensemble of PINNs and SaPINNs, inferred force amplitude $A$ (initialized at 300 N), and phase angle estimates from SaPINNs. }
    \label{fig:Fig3}
\end{figure}

\begin{figure}[H]
    \centering
    \includegraphics[width=0.4\textwidth]{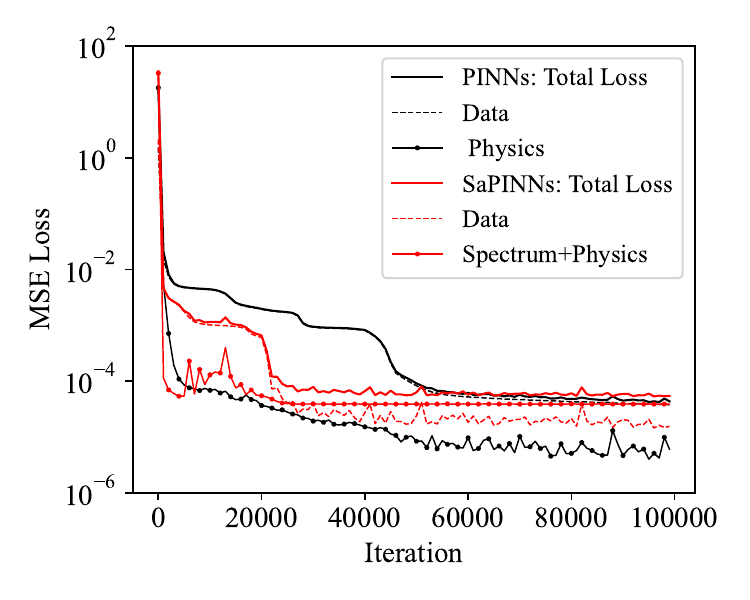} 
    \vspace{-3ex} 
    \captionsetup{justification=centering}
    \caption{PINNs and SaPINNs loss functions for the toy problem, showing the first component of the ensemble.}
    \label{fig:Fig4}
\end{figure}

Consider a dynamic system subjected to a pure sinusoidal excitation. The system is modeled with parameters: mass \( m = 4500 \, \text{kg} \), stiffness \( k = 27 \, \text{kN/m} \), and viscous damping coefficient \( c = 0.245 \, \text{Ns/m} \). For illustrative purposes no modeling error was considered in this example. 
The external excitation is defined as: 
\[
U(t) = A \sin(\omega t),
\]
where \( A = 200 \,  \text{N} \), and the angular frequency of excitation \( \omega=0.5 \, \text{rad/s} \). To emulate realistic monitoring conditions, the system response is observed via a single-channel acceleration measurement:
\begin{equation}
    y^{obs}(t) = \ddot{x}(t) + \eta(t),
\end{equation}
where $\eta(t)$ represents additive Gaussian white noise, simulated here with an amplitude equal to 15\% of the root-mean-square (RMS) of the acceleration signal.  These noisy acceleration measurements serve as training data for the networks. The measurements are generated for a 50-second observation window, providing sufficient temporal resolution for system identification. In this simplified scenario of a single-frequency sinusoidal excitation, the physics-spectrum residual is defined as: 
\begin{equation}
\mathcal{L}(\boldsymbol{\Theta})_{sp}= \sum_{i=1}^{N_{sp}}
\left\| 
 y^{obs}(t_i)-\frac{1}m(A\cos (\omega t_i+\phi)-c\dot{x}_\xi(t_i)-\theta_1k  x_\xi(t_i))
\right\|_2^2 .
\end{equation}
Here $\theta_1$ is considered as an unknown scaling factor to the nominal value of the stiffness $k$, and both the amplitude $A$ and phase $\phi$ are considered unknown quantities in this example.  Correspondingly, the PINNs physics loss is formulated as the following:
\begin{equation}
\mathcal{L}(\boldsymbol{\Theta})_{p}= \sum_{i=1}^{N_{p}}
\left\| 
y^{obs}(t_i)-\frac{1}m(U_\xi(t_i)-c\dot{x}_\xi(t_i)-\theta_1k  x_\xi(t_i))
\right\|_2^2 .
\end{equation}
Since the SDoF system with sinusoidal excitation represents one of the simplest case scenarios,  the estimation is performed using an ensemble of 20 parallel networks, each consisting of two hidden layers with 20 nodes per layer.  The results for the input force reconstruction and the estimated unknown displacements are presented in Fig. \ref{fig:Fig2}, while the system and force parameters obtained by each model are demonstrated in Fig. \ref{fig:Fig3}. 
The ensembles' results demonstrate that the SaPINNs architecture significantly outperforms the baseline PINNs in estimating the excitation force and system parameters. While both models are able to recover the displacement response with reasonable accuracy, the standard PINNs fail to reconstruct the input force, yielding predictions that deviate substantially from the true excitation. On the other hand,  the parameterized SaPINNs were able to infer the phase angle of the input force with high precision and, although, the estimated amplitude deviated by approximately 8\%, the overall force reconstruction was consistent with the true excitation. Additionally, the SaPINNs model accurately estimated the stiffness scaling factor $\theta_1$,  producing values close to the ground truth, whereas the PINNs produced a broadly distributed set of estimates centered farther away, indicating failure to converge to a well-defined or physically meaningful solution.  The discrepancy between the models performance is further highlighted by the evolution of their respective loss functions, as shown in Fig. \ref{fig:Fig4}. 
Although the physics-based loss in the PINNs achieved a lower numerical value compared to the spectrum-physics loss in the SaPINNs, this does not translate into more accurate parameter estimates. In fact, the physics loss in PINNs exhibits noticeable fluctuations, indicating instability in training and potential overfitting to the observed data, while the SaPINNs loss rapidly decreases and stabilizes after $\sim$ 20,000 iterations,  suggesting convergence to an optimal solution. These results highlight that spectral information provides a stronger inductive bias, enabling recovery of physically consistent parameters, and underscore a key limitation of PINNs: low training loss alone does not guarantee meaningful solutions.
\vspace{1em}
\subsubsection{SDoF system subjected to thunderstorm wind excitation}\label{sec.thunder}
\begin{figure}[H]
    \centering
    \begin{subfigure}[b]{0.36\textwidth}
        \centering
        \includegraphics[width=\textwidth]{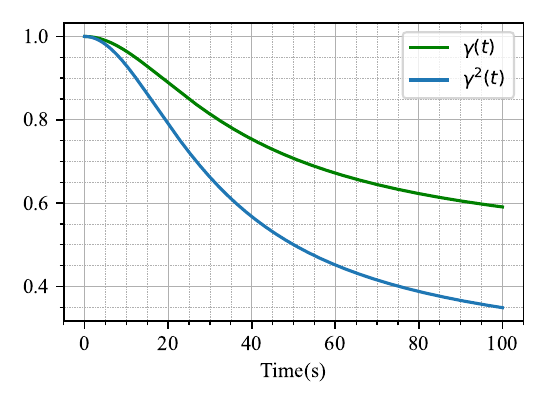}
        \captionsetup{justification=centering}
        \caption{} 
        \label{subfig:Th_mod}
    \end{subfigure}
    \begin{subfigure}[b]{0.4\textwidth}
        \centering
        \includegraphics[width=\textwidth]{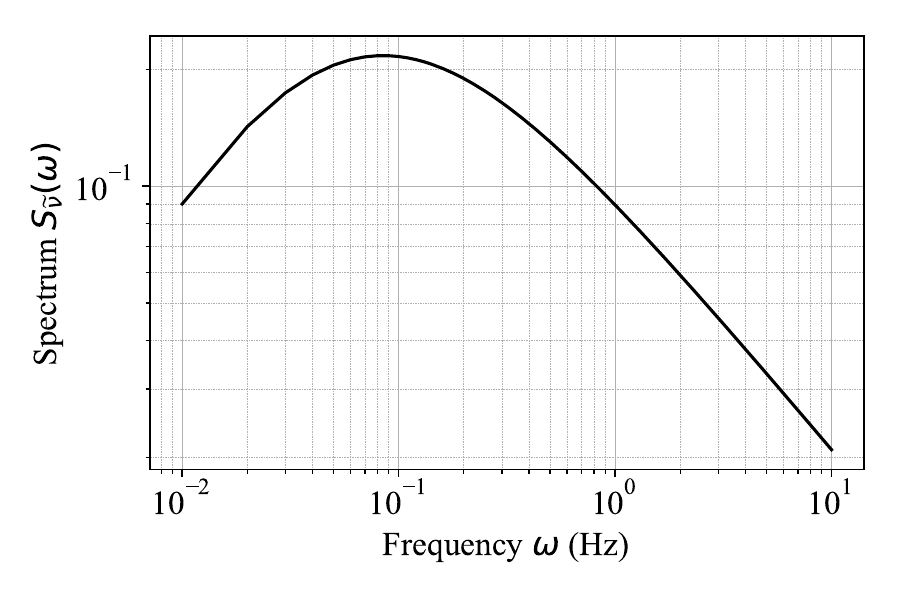}
        \captionsetup{justification=centering}
        \caption{} 
        \label{subfig:Th_spec}
    \end{subfigure}
    \captionsetup{justification=justified}
    \caption{(a) Modulating function plotted for different powers $\gamma (t), \gamma (t)^2$,  showing decaying behavior over time. (b) Turbulence fluctuations spectrum \( S_\nu (\omega) \), illustrating the frequency distribution of turbulence energy. }
    \label{fig:my_label}
\end{figure}

To simulate a more realistic loading condition, we consider the same SDoF system as described in the previous section and apply an excitation force representative of wind loading induced by a thunderstorm.  Thunderstorms present a unique set of challenges and conditions that differ significantly from those encountered during normal wind events.  During a thunderstorm, wind speeds and directions can change rapidly, resulting in highly non-stationary excitation profiles. The proposed SaPINNs framework accounts for these conditions by embedding relevant atmospheric fluid dynamics, derived from extensive studies of thunderstorms,  alongside the system's governing equations of motion.  While the Von Kármán and Kaimal spectra are standard for boundary layer winds, thunderstorm winds are more transient and intense, prompting the use of alternative, empirically derived models. This study employs an EPSD model to describe the non-stationary and turbulent nature of wind fluctuations induced by thunderstorms \cite{SOLARI200173}. Within this framework, the wind-induced force is a non-stationary process with time-modulated amplitude, and the turbulent fluctuations are modeled as a Gaussian process defined in the frequency domain. Using this formulation, the aerodynamic force $U(t)$ caused by thunderstorm winds is characterized by a wind velocity $v(t)$, which can be decomposed into a slowly varying mean component and superimposed turbulent fluctuations. Assuming negligible spatial coherence effects in the wind field, the along-wind aerodynamic force acting on a structure can be expressed as: 
\begin{equation}
U(t)=\frac{1}{2} \rho v^2(t) A_{eff} C_D,
\end{equation}
where $\rho $ is the air density,  $v(t)$ is the instantaneous wind velocity, $A_{eff}$ is the effective area of the structure exposed to the wind, and $C_D $ is the drag coefficient. The value of $C_D$ depends on the geometry and aerodynamic characteristics of the structure and can be influenced by the wind angle of attack. However, it is reasonable to assume the coefficient to be known, based on prior aerodynamic characterization of a specific structure under representative wind loading conditions \cite{doi:10.1061/JSDEAG.0003465}. The wind velocity $v(t)$ can be modeled as a uniformly modulated process \cite{roncallo2020evolutionary}:
\begin{equation}
v(t)=\bar{v}(z_h) \gamma(t)[1+\bar{I}_v \widetilde{v}(t)],
\end{equation}
where $\bar{v}$ is the reference mean wind speed at the height $z_h$ relative to the altitude of the DoF on which the force is acting,   $\bar{I}_v$ is the mean value of the turbulence intensity and $\tilde{v}(t)$ is the zero-mean fluctuating component. It is worth highlighting that the turbulence intensity is constant; therefore, the mean wind velocity and turbulent fluctuations are modulated by the same function  $\gamma(t)$, which captures the slowly varying trend of the wind during a thunderstorm, described as:
\begin{equation}
\gamma(t)=\frac{1-\gamma^*}{\left[1+\left(\frac{t}{T_\gamma}\right)^2\right]^{\frac{1}{2}}}+\gamma^*.
\end{equation}
In this formulation  $\gamma^*$ is a measure of the intensity of the mean wind velocity and $T_\gamma$ is a duration of the thunderstorm peak, which can vary between different thunderstorms, but considered constant for a single thunderstorm event. Based on previous studies,  the values $\gamma^*=0.45$ and $T_\gamma=26.45 \mathrm{~s}$, extracted from 129 full-scale thunderstorm records \cite{roncallo2020evolutionary}, were adopted in this work to provide an effective representation of typical wind speed trends. The mean wind velocity can be estimated from weather station in the geographical area of the monitored structure using the logarithmic wind profile:
\begin{equation}
\bar{v}(z_h)=\frac{ u^*}{K} \ln (\frac{z_h}{z_0}),
\end{equation}
where $z_0$ is a surface roughness length,  $K$ is the von Karman constant, and $u^*$ is the shear velocity of the flow. The fluctuating component, representing turbulence in the wind velocity is modeled as: 
\begin{equation}
\widetilde{v}(t)=\gamma(t)^2\widetilde{\nu}.
\end{equation}
Here,  $\widetilde{\nu}$ is a reduced turbulent fluctuation, generated as a stationary realization $p(t;\phi)$ from a baseline one-sided PSD: 
\begin{equation}
S_{\widetilde{\nu}}(\omega)=\frac{6.868 \omega L_w / \bar{v}}{\left[1+10.302 \omega L_w / \bar{v}\right]^{5 / 3}},
\end{equation}
where $L_w / \bar{w}$ represents the turbulence length scale, taken as $1.72 \mathrm{~s}$ as per thunderstorm time histories \cite{RONCALLO2022104978}.  The modulating function used in this study is presented in Fig.\ref{subfig:Th_mod}, and Fig. \ref{subfig:Th_spec} illustrates the one-sided turbulence spectrum. 


For the simulation of synthetic measurements,  the following parameters were used: $A_{eff}=8$ $ m^2$, $C_D$=1,  $\bar{v}(z_h) =10$ $m/s$, and the turbulence intensity was taken as $\bar{I}_v =0.2$.  The modeling error is introduced to the problem by misspecification of the system mass by 5 \%, and the only observation of the system dynamics is available though a noisy acceleration $y^{obs}(t)$. The unknown system parameters are considered as scaling factors applied to the nominal stiffness and damping coefficients, denoted by $\theta_1$ and $\theta_2$, respectively.  Since the example presented here involves a simple linear SDoF system, the SaPINNs spectrum-physics residual is defined explicitly as:
\begin{equation}
\mathcal{L}(\boldsymbol{\Theta})_{sp}= \sum_{i=1}^{N_{sp}}
\left\| 
y^{obs}(t_i)-\frac{1}m(h(t)p(t_i;\mathbf{\phi})-\theta_1k  x_ {\boldsymbol {\xi}}(t_i)-\theta_2\dot{x}_ {\boldsymbol {\xi}}(t_i))
\right\|_2^2,
\end{equation}
where
\begin{equation}
h(t)\, p(t;\phi) = \frac{1}{2} \rho A_{\text{eff}} C_D 
\left[
\bar{v}(z_h)\, \gamma(t) 
\left( 
1 + \bar{I}_v\, \gamma(t)^2 \sum_{i=1}^{N_\omega} \sqrt{2 S_{\widetilde{\nu}}(\omega)\Delta \omega} \cos(\omega_i t + \phi_i) 
\right)
\right]^2.
\end{equation}
 
\begin{figure}[H]
    \centering
    \begin{subfigure}[b]{0.49\textwidth}
        \centering
        \includegraphics[width=\textwidth]{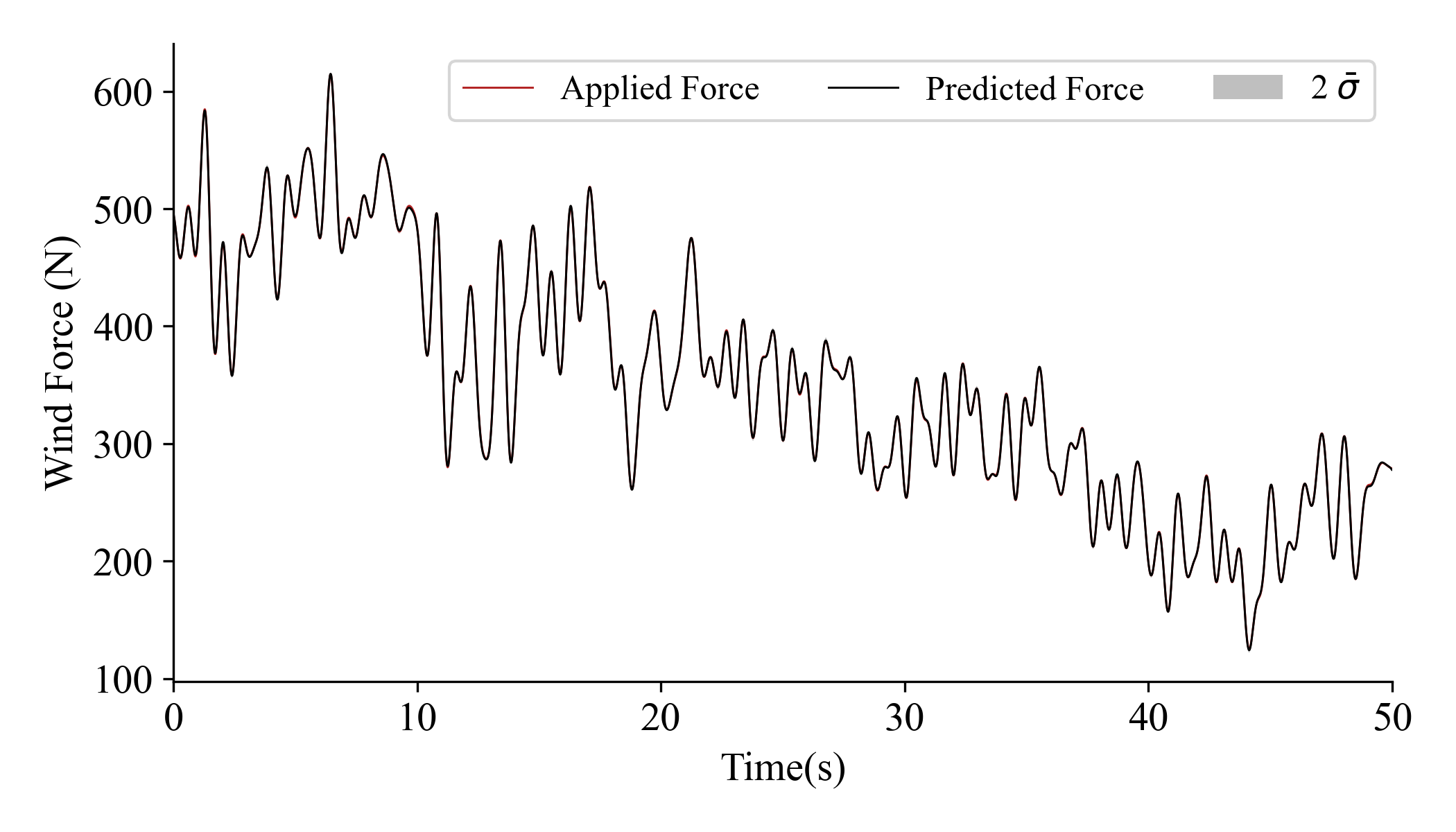}
        \captionsetup{justification=centering}
        \caption{}
        \label{fig:plot1}
    \end{subfigure}
    \begin{subfigure}[b]{0.49\textwidth}
        \centering
        \includegraphics[width=\textwidth]{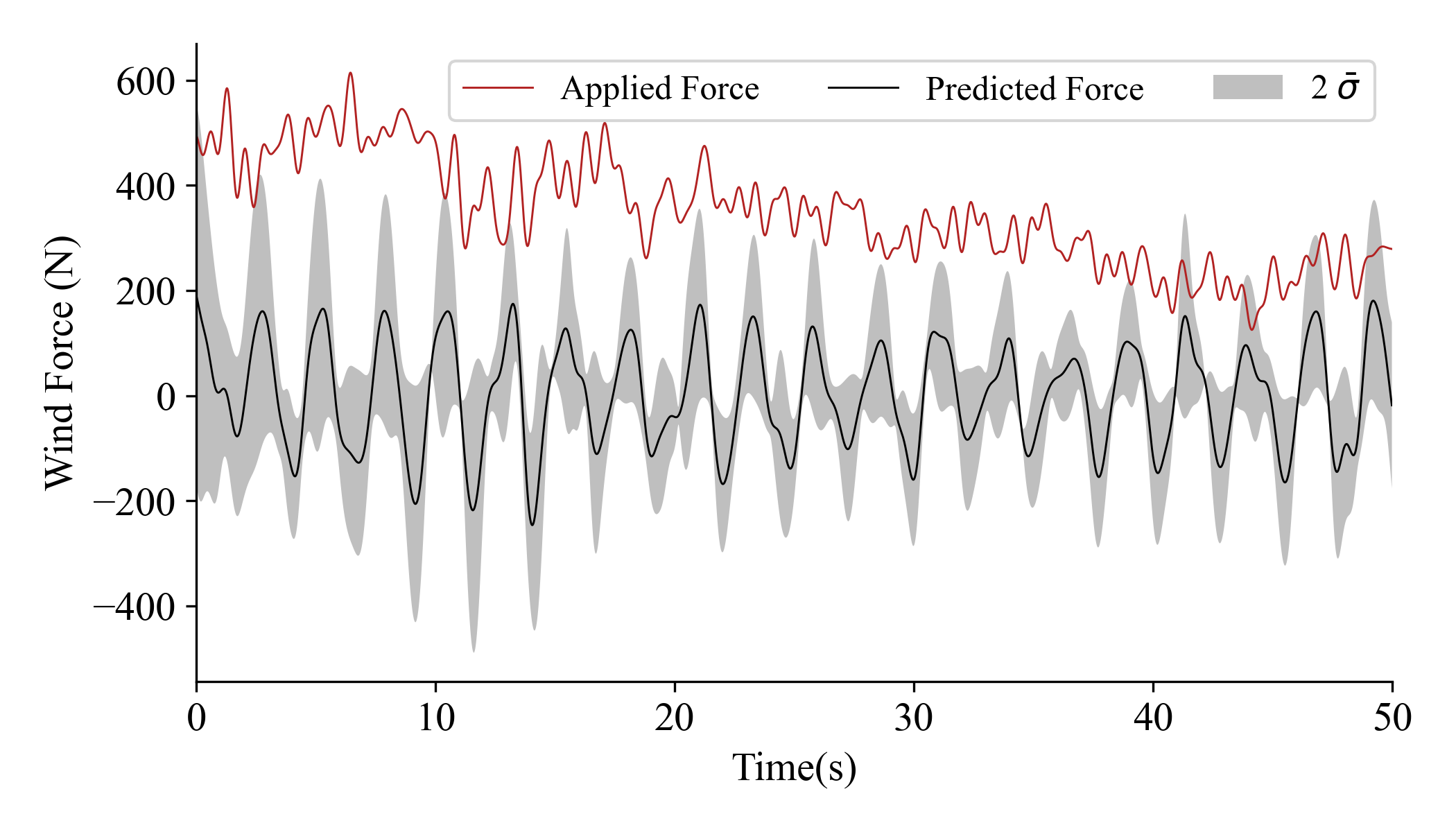}
        \captionsetup{justification=centering}
        \caption{}
        \label{fig:plot2}
    \end{subfigure}
     \vspace{0em} 
    \begin{subfigure}[b]{0.49\textwidth}
        \centering
        \includegraphics[width=\textwidth]{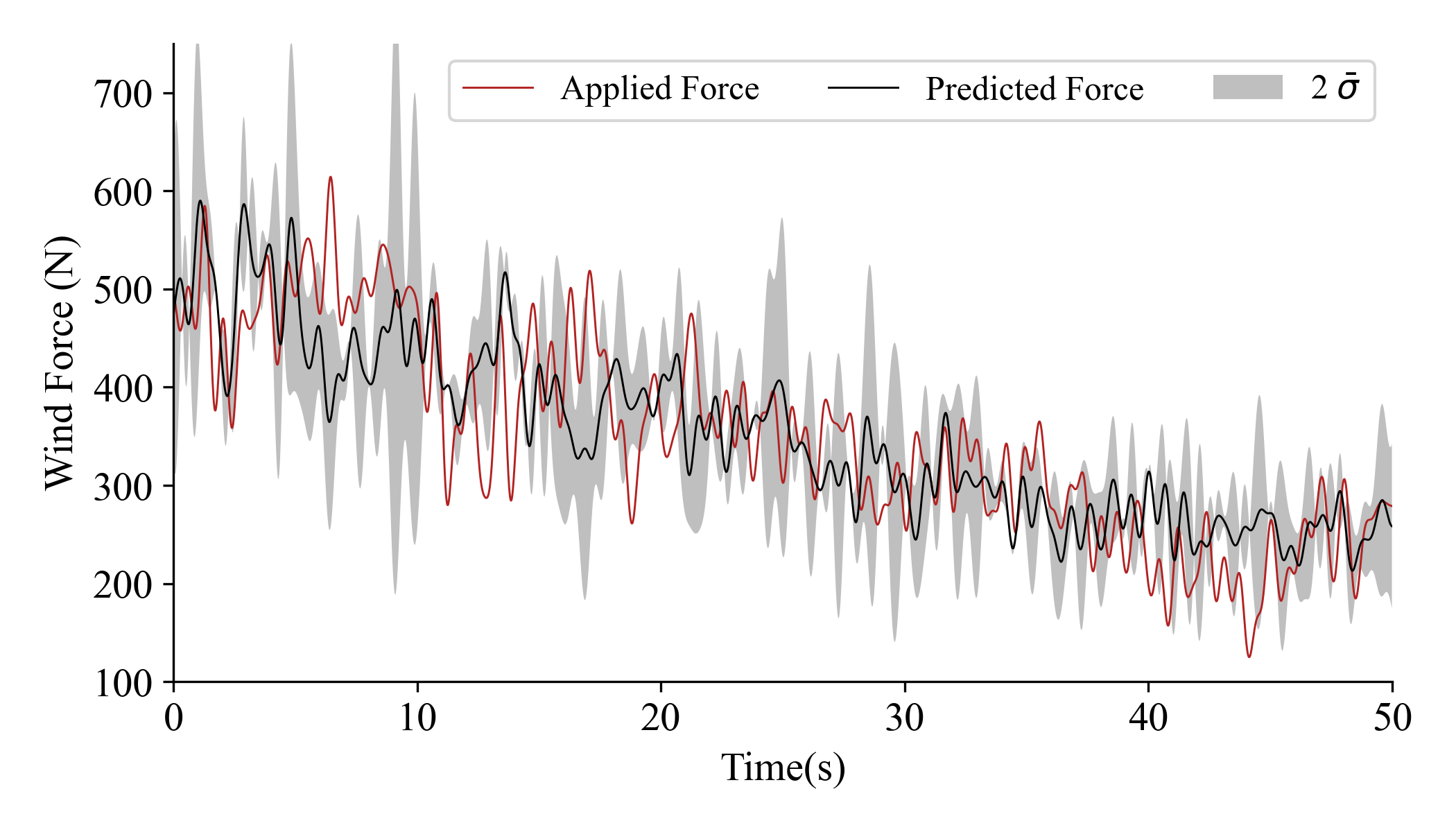}
        \captionsetup{justification=centering}
        \caption{ }
        \label{fig:plot3}
    \end{subfigure}
    \begin{subfigure}[b]{0.49\textwidth}
        \centering
        \includegraphics[width=\textwidth]{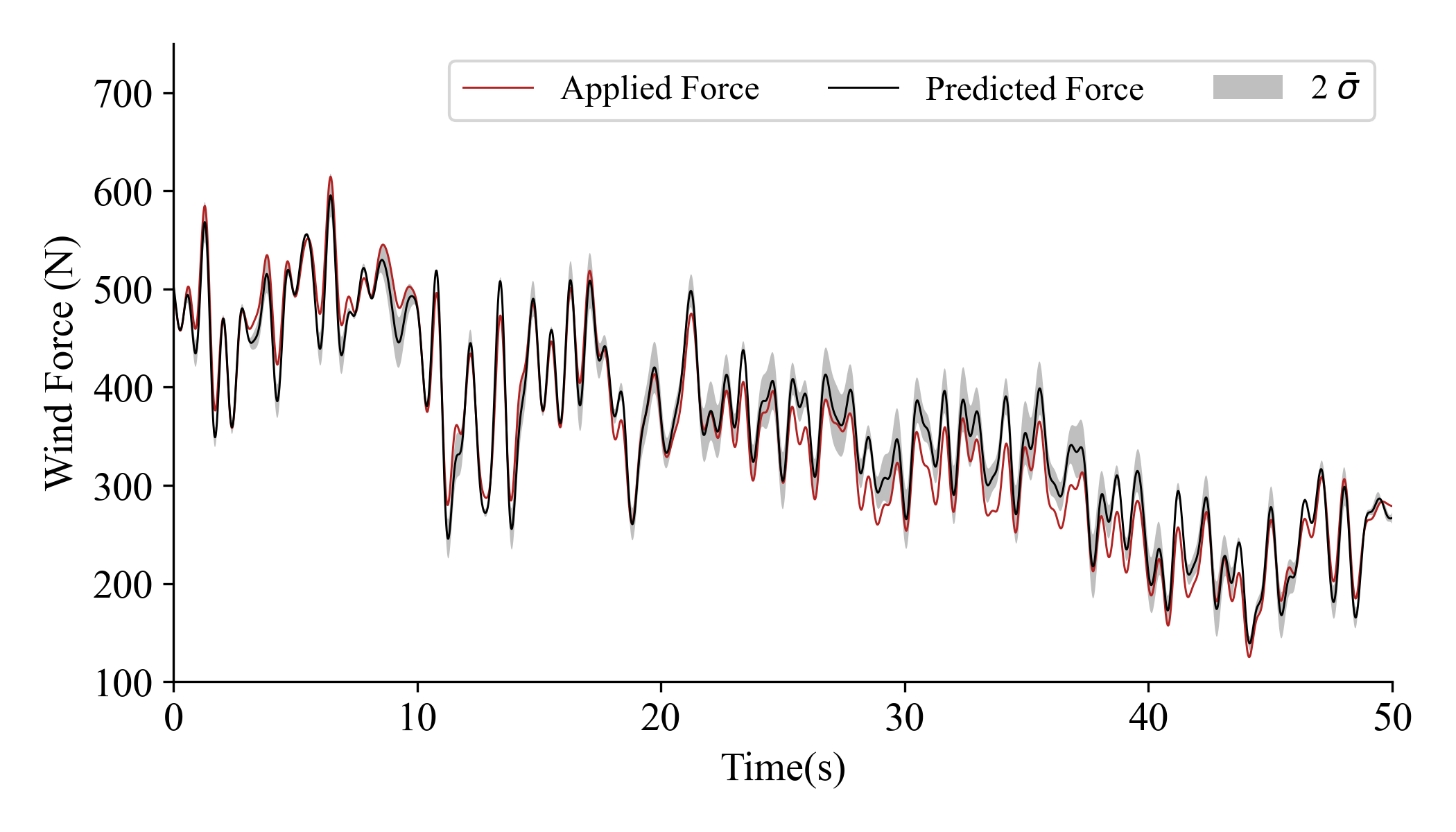}
        \captionsetup{justification=centering}
        \caption{}
        \label{fig:plot4}
    \end{subfigure}
    \caption{Comparison of thunderstorm excitation force reconstruction using PINNs and SaPINNs under four scenarios: (a) PINNs with known force, (b) PINNs when no excitation profile is provided, (c) SaPINNs with random phases, and (d) SaPINNs with trained phases. Each plot shows the true force, predicted force, and $2\sigma$ credible intervals.}
    \label{fig:Fig6}
\end{figure}

\begin{figure}[!htbp]
    \centering
    \begin{subfigure}[b]{0.45\textwidth}
        \centering
        \includegraphics[width=\textwidth]{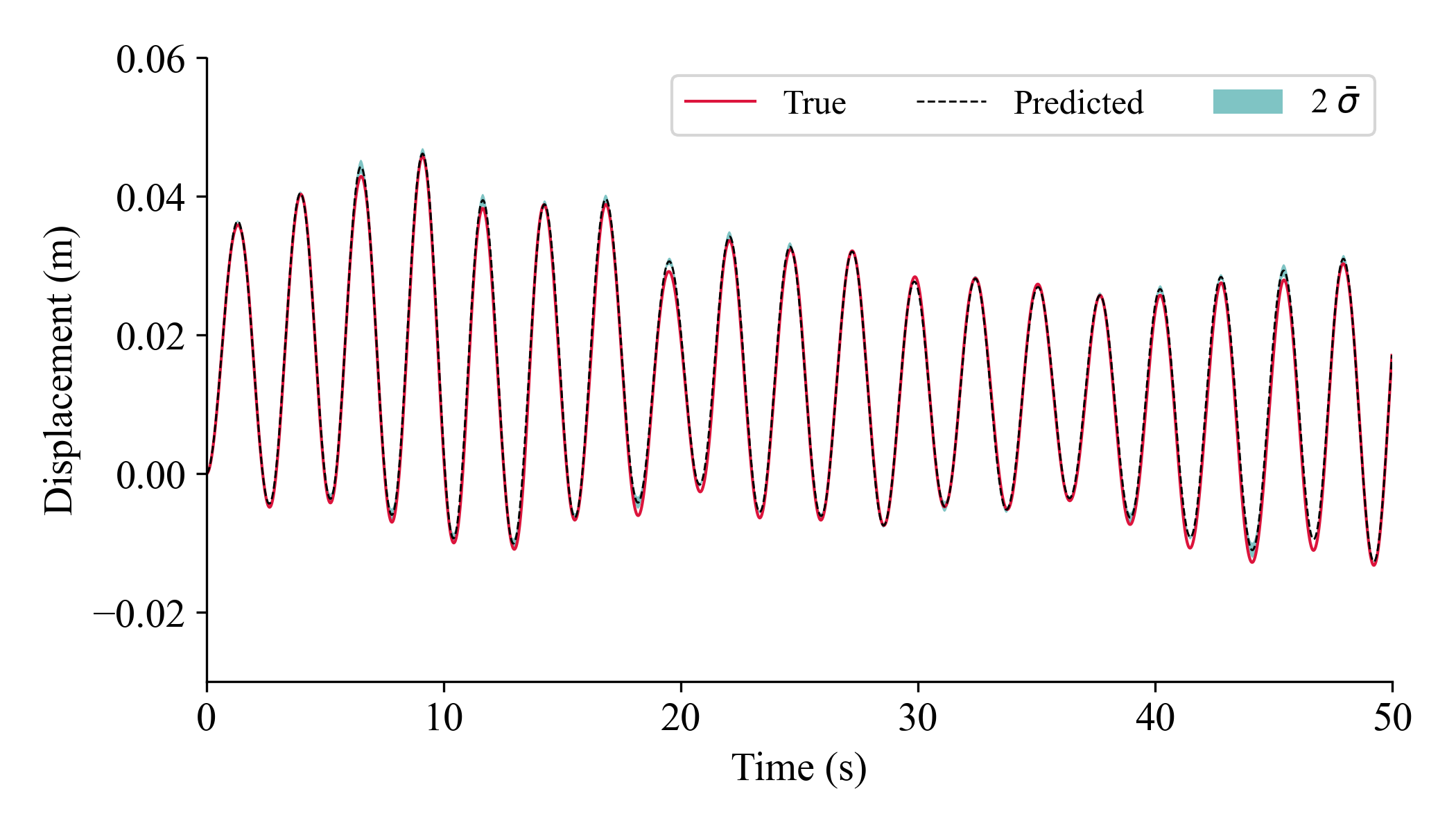}
        \captionsetup{justification=centering}
        \caption{}
        \label{fig:plot1}
    \end{subfigure}
    \hspace{0em}
    \begin{subfigure}[b]{0.45\textwidth}
        \centering
        \includegraphics[width=\textwidth]{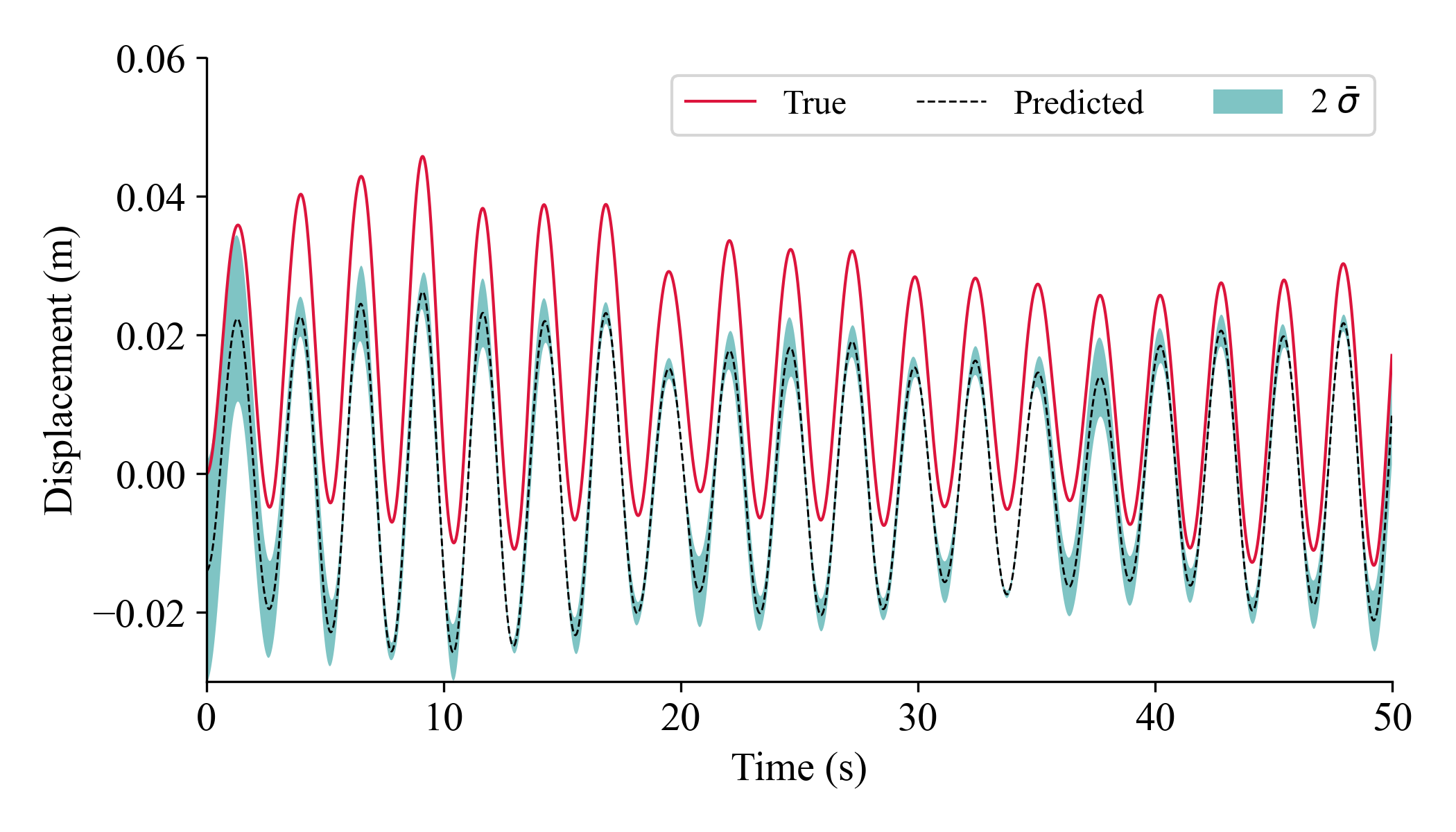}
        \captionsetup{justification=centering}
        \caption{}
        \label{fig:plot2}
    \end{subfigure}
    \vspace{0em} 
    \begin{subfigure}[b]{0.45\textwidth}
        \centering
        \includegraphics[width=\textwidth]{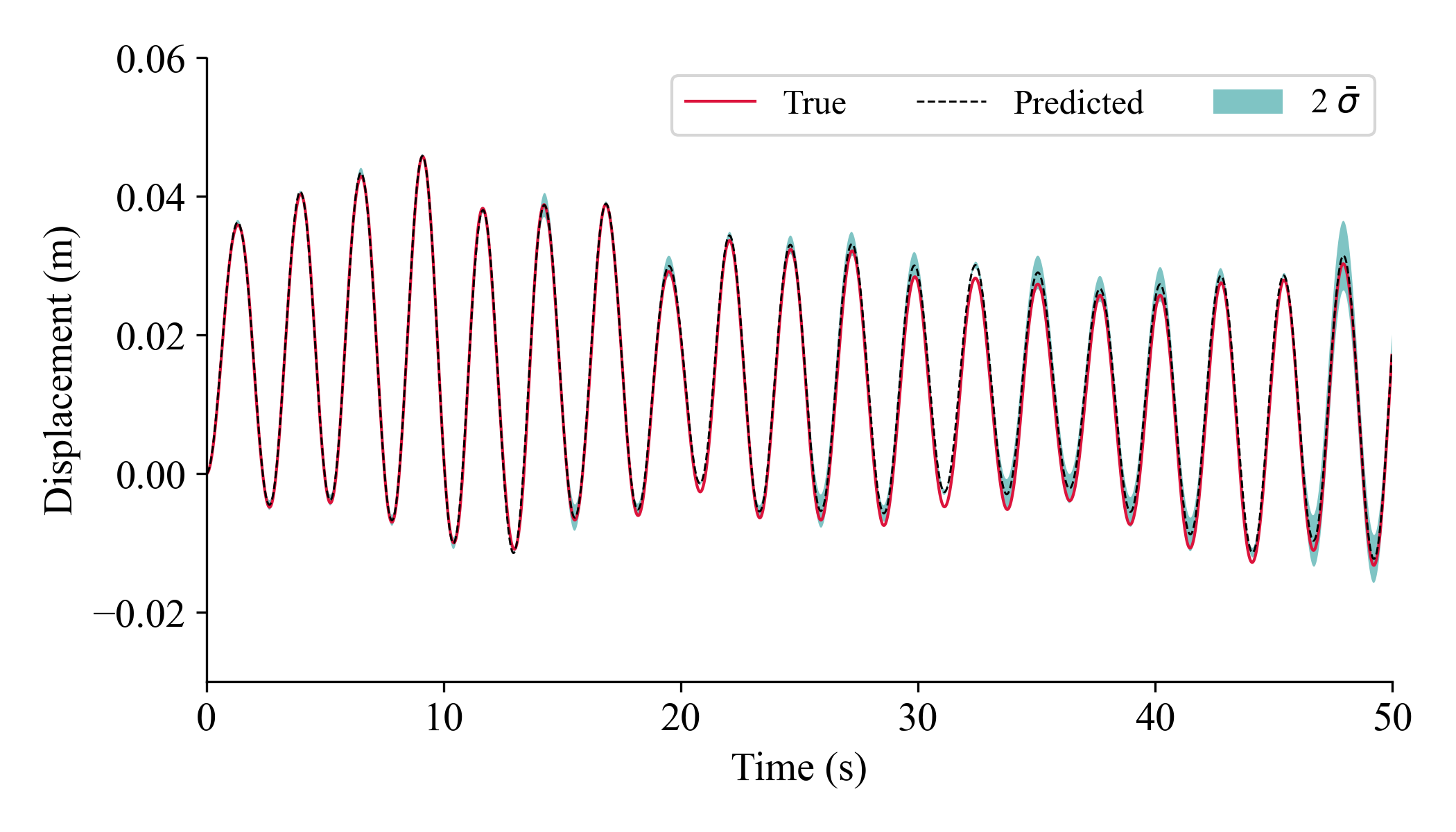}
        \captionsetup{justification=centering}
        \caption{}
        \label{fig:plot3}
    \end{subfigure}
    \hspace{0em}
    \begin{subfigure}[b]{0.45\textwidth}
        \centering
        \includegraphics[width=\textwidth]{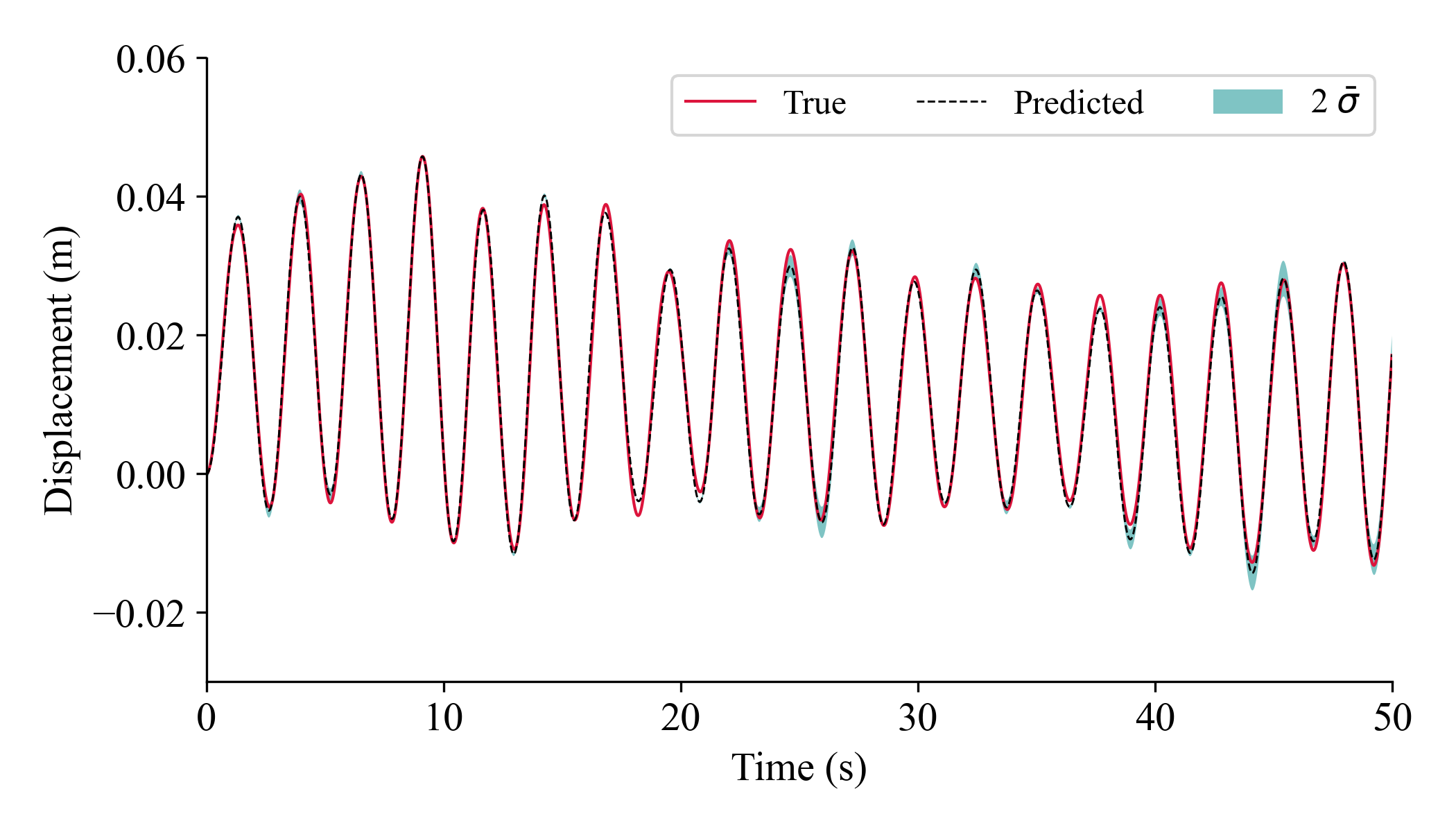}
        \captionsetup{justification=centering}
        \caption{ }
        \label{fig:plot4}
    \end{subfigure}
    \caption{True and predicted displacements with their respective $2\sigma$ credible intervals estimated under four scenarios: (a) PINNs with known force, (b) PINNs when no excitation profile is provided, (c) SaPINNs with random phases, and (d) SaPINNs with trained phases.}
    \label{fig:Fig7}
\end{figure}

\begin{figure}[H]
    \centering
    \begin{subfigure}[b]{0.35\textwidth}
        \centering
        \includegraphics[width=\textwidth]{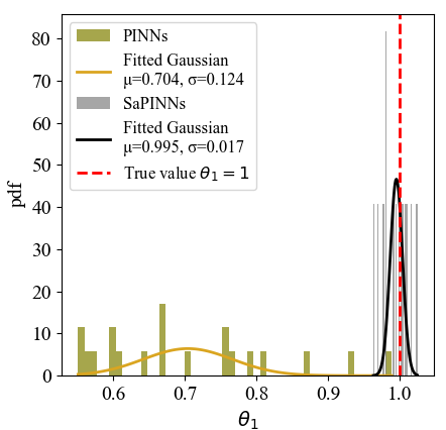}
        \label{subfig:label_for_first_image}
    \end{subfigure}
    \begin{subfigure}[b]{0.355\textwidth}
        \centering
        \includegraphics[width=\textwidth]{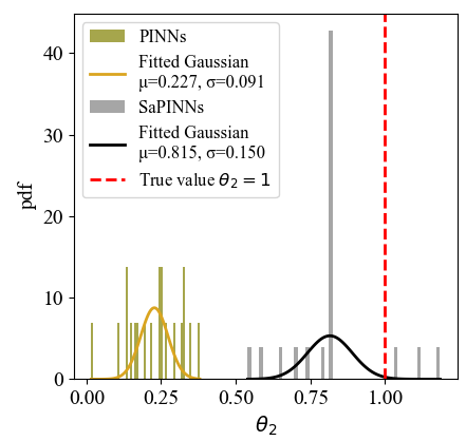}
        \label{subfig:label_for_second_image}
    \end{subfigure}
        \vspace{-2ex} 
    \captionsetup{justification=justified}
  \caption{SDoF stiffness $\theta_1$ and damping ratio $\theta_2$ scaling factors estimates from ensemble of PINNs (unknown force) and SaPINNs (trained phase angles), along with associated uncertainties. Values are obtained from 20 independently trained instances of each model. }
    \label{fig:Fig8}
\end{figure}

\begin{figure}[H]
    \centering
    \begin{subfigure}[b]{0.4\textwidth}
        \centering
        \includegraphics[width=\textwidth]{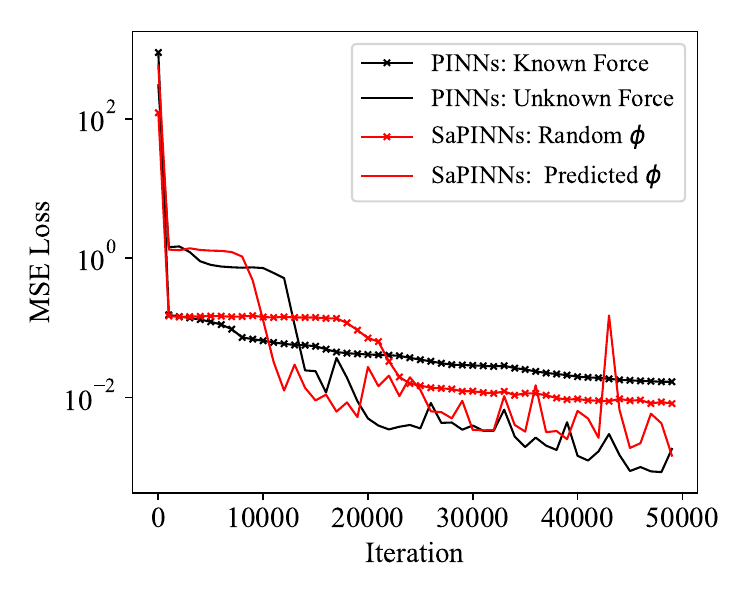}
        \captionsetup{justification=centering}
            \vspace{-5ex} 
        \caption{} 
        \label{subfig:Fig9a}
    \end{subfigure}
    \begin{subfigure}[b]{0.48\textwidth}
        \centering
        \includegraphics[width=\textwidth]{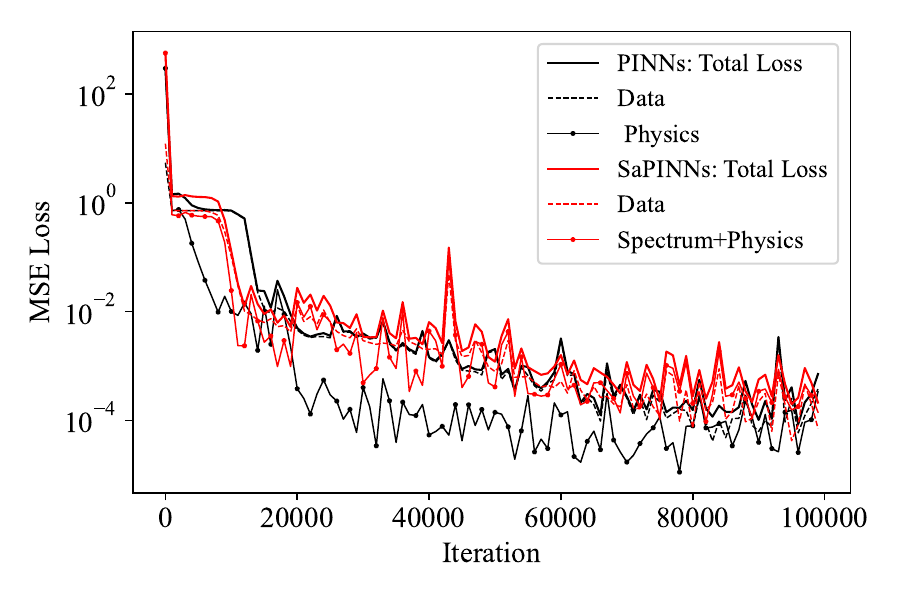}
        \captionsetup{justification=centering}
            \vspace{-5ex} 
        \caption{} 
        \label{subfig:Fig9b}
    \end{subfigure}
    
    \captionsetup{justification=justified}
    \caption{Comparison of loss convergence: (a) across four different scenarios; (b) between PINN, when no input force provided, and SaPINN, where the phase angles are incorporated into the training.}
    \label{fig:Fig9}
\end{figure}

To evaluate the performance of the proposed model under thunderstorm excitation, we consider four scenarios. First, we test PINNs in two settings: (i) when the input force time history is known, and (ii) when the input is entirely unknown. Next, to examine the sensitivity of the input reconstruction to the identified phase angles, we compare the SaPINNs using (iii) randomly assigned phases against the model (iv) where the phases are incorporated into the training .  While classical identification methods remain more robust and computationally efficient for state–parameter estimation in the presence of known input force, we include  these comparisons to illustrate that PINNs can solve the problem effectively when provided with sufficient input information. 
For the training, we employed an ensemble of 20 parallel networks, each consisting of five hidden layers with 128, 64, 64, 64, and 128 nodes, respectively.  The results for the input force reconstruction are presented in Fig.\ref{fig:Fig6}, while the predicted unknown displacements are shown in Fig.\ref{fig:Fig7}.
As shown in Fig. \ref{fig:Fig6}a,  the network parameters are fitted well into an excitation time-history,  and the model achieves high accuracy with negligible uncertainty, as could be expected.  The second scenario requires the network to infer a completely unknown excitation based on the observed acceleration signal and the governing model constraints, at which the PINNs fail dramatically, as demonstrated in Fig. \ref{fig:Fig6}b. Further,  Fig. \ref{fig:Fig6}c shows the performance of SaPINNs for the unknown input force, demonstrating improved estimation accuracy achieved by incorporating spectral information into the training process. However, in the absence of explicitly incorporated phase information, the model can only capture the overall trend of the excitation and is unable to reconstruct its exact, unique force profile. Finally, Fig. \ref{fig:Fig6}d shows an improved performance in the input force reconstruction when phase angles are directly embedded into the learning process. A detailed comparison of the models'  performances for the input–state–parameter estimation task is presented in Table \ref{table:1}, and the distribution over the system parameters estimates is shown in Fig. \ref{fig:Fig8}.
\begin{table}[!htbp]
\centering
\caption{Performance evaluation across models: PINNs and SaPINNs for input, state, and parameter estimation.}
\begin{tabular}{lccccccc}
\toprule
Model & $\theta_1$ & $\sigma(\theta_1)$ & $\theta_2$ & $\sigma(\theta_2)$ & MSE($x(t)$) & MSE($U(t)$)  \\
\midrule
{PINNs} \\
\quad Known Force        & 1.000 & 0.005 & 0.905 & 0.225 & $1.2 \times 10^{-6}$ & -- \\
\quad Unknown Force      & 0.704 & 0.124 & 0.227 & 0.091 & $5.9 \times 10^{-4}$ & $6.2 \times 10^{-2}$ \\
{SaPINNs} \\
\quad Random $\phi$      & 0.971 & 0.037 & 0.587 & 0.360 & $1.9 \times 10^{-5}$ & $1.7 \times 10^{-5}$  \\
\quad Predicted $\phi$     & 0.995 & 0.017 & 0.815 & 0.150 & $6.4 \times 10^{-6}$ & $6.4 \times 10^{-5}$ \\
\bottomrule
\end{tabular}
\label{table:1}
\end{table}
Based on the inference results, the stiffness scaling factor was estimated with an error of  0.05\% using SaPINNs with trained phases, whereas the damping coefficient was not reliably identified in all cases with both PINNs and SaPINNs.  The poor performance of the models can be attributed to the low sensitivity of the system response to small variations in damping, especially when the system is lightly damped as in the present case. Additionally, measurement noise in the observed system response, combined with the system mass misspecification, further obscures the already subtle damping-related effects. Despite this inaccuracy,  SaPINNs significantly outperformed PINNs in reconstructing the excitation force and in estimating the displacements and stiffness scaling factor. \\
The convergence behavior of the models in each scenario is illustrated by the evolution of their respective loss functions, shown in  Fig. \ref{subfig:Fig9a} for the first 50,000 iterations. A detailed comparison of the individual loss components for cases (ii) and (iv) is demonstrated in Fig.\ref{subfig:Fig9b}.  For illustrative purposes, the evolution of the loss functions is shown for the first network in the ensemble (i.e., the first of 20). 
Naturally, when the time history of the external excitation is fully known, a PINN exhibits fast and stable convergence. However, in a case of jointly estimating the input excitation, latent states, and the system parameters, an inherently ill-posed problem, PINN encounter significant difficulty, resulting in instability in convergence behavior,  and inaccurate force and parameters estimates. On the other hand, a SaPINN with randomly specified phase angles demonstrate slow convergence toward a lower loss, but since the model lacks the parametrization required to lead the training to a unique solution, it can never reach the desired minimum. This limitation is also reflected in the unknowns' estimates, as the absence of uniquely defined phase angles during the training process is expected to lead to inaccurate parameter estimates and hinder the model’s ability to reconstruct the applied wind force. 
In contrast, a SaPINN that learns the phase angles during training converge to a lower loss, compared to the SaPINN with random phases, and although the loss trajectories of a PINN in the unknown force scenario and SaPINN with embedded phase angles both appear similarly unstable (Fig. \ref{fig:Fig9}), the latter delivers markedly superior predictions across every evaluation metric. These findings underscore the importance of appropriate model parametrization in enabling accurate joint input–state–parameter estimation when using neural network-based frameworks. 

\subsection{Nonlinear systems: Seismic excitation}\label{ex.nonlin}
The identification of nonlinear dissipative systems remains a topic of active interest in structural dynamics, owing to their widespread use in environments subject to earthquakes, sea waves, and other forms of natural excitation. To demonstrate the performance of SaPINNs for input–state–parameter estimation in nonlinear systems, synthetic data were generated by simulating the dynamic response of a hysteretic 3-DoF Bouc–Wen model \cite{https://doi.org/10.1002/stc.290}, with nonlinearity introduced at the base element. The hysteretic component was modeled as:
\begin{equation}
\dot{r}_1(t) = \dot{x}_1(t) - \beta |\dot{x}_1(t)| |r_1(t)|^{n-1} r (t)- \alpha_r(\dot{x}_1(t)) |r_1(t)|^n,
\end{equation}
where $r_1(t)$ denotes the dimensionless restoring force at the base, $\dot{x}_1(t)$ is the velocity of the 1st DoF, $\alpha_r$, $\beta$, are the parameters that regulate the post‑yield stiffness ratio and the hysteretic loop width, respectively, and $n$ governs the smoothness of the transition from elastic to plastic system response .
For the synthetic examples (Section \ref{sec.synthetic}), the applied excitation is a scaled ground acceleration generated from an earthquake spectrum defined by the time-modulated Kanai-Tajimi EPSD function:
\begin{equation}
S_p(\omega, t)=\left|\gamma(t)\right|^2 S(\omega).
\end{equation}
With the modulating function $\gamma(t)$ defined as:
\begin{equation}
\gamma(t)=g_k(e^{-at}-e^{-bt}).
\label{eq:mod_function}
\end{equation}

\begin{figure}[!htbp]
    \centering
    \begin{subfigure}[b]{0.4\textwidth}
        \centering
        \includegraphics[width=\textwidth]{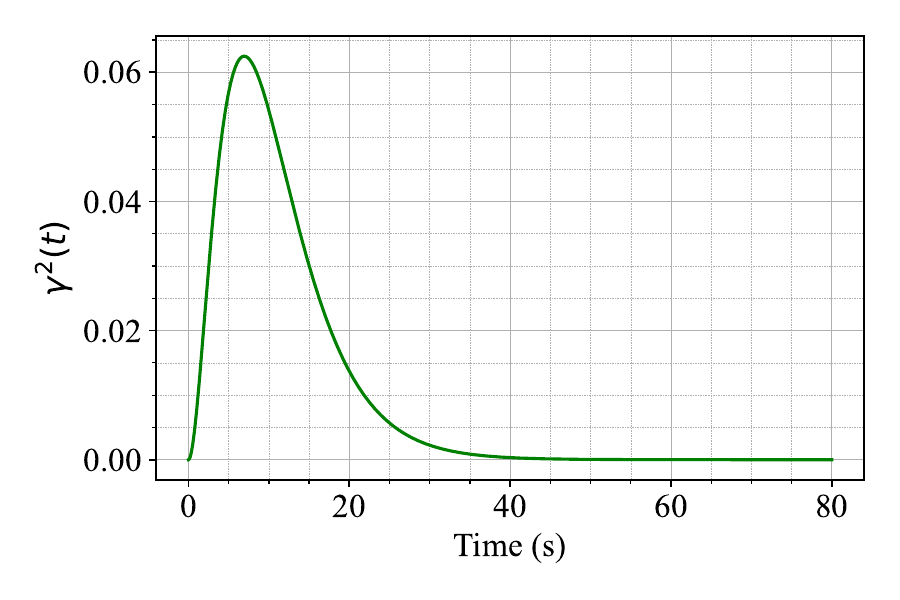}
        \captionsetup{justification=centering}
        \caption{} 
        \label{subfig:label_for_first_image}
    \end{subfigure}
    \begin{subfigure}[b]{0.4\textwidth}
        \centering
        \includegraphics[width=\textwidth]{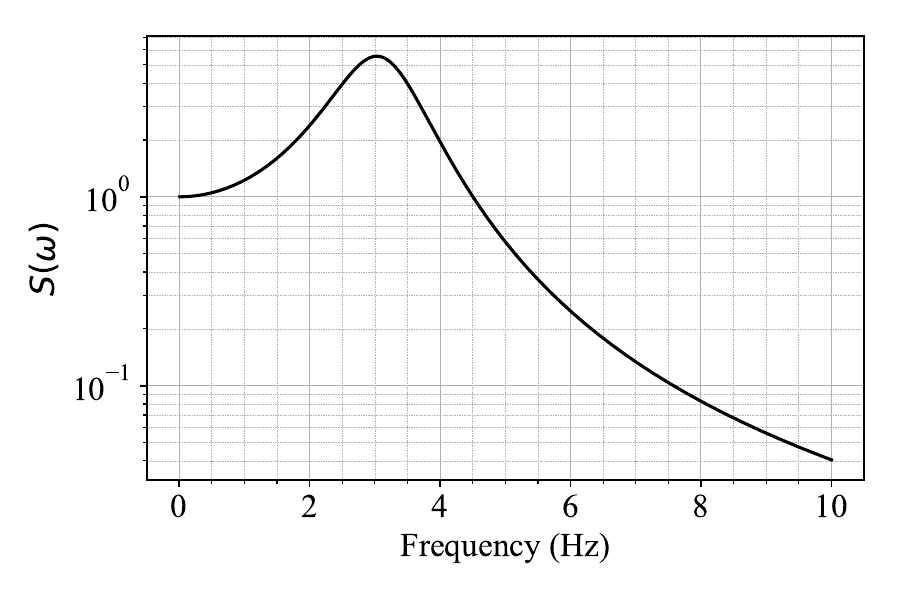}
        \captionsetup{justification=centering}
        \caption{} 
        \label{subfig:label_for_second_image}
    \end{subfigure}
    \captionsetup{justification=justified}
    \caption{(a) The modulating function  $\gamma(t) ^2$, which represents the time-dependent amplitude variation of the ground acceleration. (b) The Kanai–Tajimi power spectrum \( S(\omega) \), shown as a function of frequency \( \omega \) in Hz, highlighting the energy content, with a peak indicating the dominant frequency of the ground motion $\omega_g$.}
    \label{fig:kanai_tajimi}
\end{figure}

Here, $g_k$ is a normalizing constant, and the parameters $a=0.1$ and $b=0.2$ represent empirically derived parameters based on analyzed earthquake data \cite{SPANOS201257}. The Kanai–Tajimi spectrum is described as the following:  
\begin{equation}
S(\omega)=S_1\frac{\omega_g^4+4 \zeta_g^2 \omega_g^2 \omega^2}{\left(\omega_g^2-\omega^2\right)^2+4 \zeta_g^2 \omega_g^2 \omega^2},
\end{equation}
where $S_1$ is the intensity scaling factor, $\omega_g$ is the natural frequency of the ground motion, and $\xi_g$ is the damping ratio that reflects the energy dissipation characteristics of the soil.
For the simulation, a spectrum is considered over the frequency range  $\omega  \sim[0, 10] $ Hz, with the $\omega_g=3.5$  Hz,  intensity $S_1=0.1$, and a damping characteristic assigned as $\zeta_g=0.24$ based on the original Kanai-Tajimi model \cite{kanai1957semi}. It is important to note that these parameters are site-specific and require calibration for each location based on geotechnical investigations or previously recorded ground motion data \cite{47f10642b6fa413986dc1d34dee28aa5}. Fig. \ref{fig:kanai_tajimi} shows the modulating function $\gamma(t)^2$, where the $g_k=1$,  and the spectrum $ S(\omega)$, presented in logarithmic scale.\\
\\
Using the notation introduced in Section \ref{section.4}, the nonlinear system state vector for the SaPINNs can be formulated as the following:
 \begin{equation}
\dot{\mathbf{z}}_{\boldsymbol{\Theta}}(t)=
\begin{bmatrix}
\dot{\boldsymbol{x}}_{\boldsymbol{\xi}}(t)\\[6pt]
 M^{-1} \left( \boldsymbol{U}_{\boldsymbol{\phi}}(t) 
- C \dot{\boldsymbol{x}}_{\boldsymbol{\xi}}(t) 
- K(\boldsymbol{\theta}) \boldsymbol{x}_{\boldsymbol{\xi}}(t) 
- G \boldsymbol{r}_{\boldsymbol{\xi}}(t) \right)\\[10pt]
 \boldsymbol{\dot{r}}_{\boldsymbol{\xi}}(t)
\end{bmatrix},
\end{equation}

where $G$ is the vector capturing the Bouc–Wen hysteresis contribution, which is equal to $[1 \ 0 \ 0]^T$ in the example, as the nonlinearity was introduced at the 1st DoF only, and the input force $\boldsymbol{U}_{\boldsymbol{\phi}}$ is defined using the time-modulated Kanai-Tajimi EPSD function. In both the PINNs and SaPINNs training setups, the restoring force and its derivative were not assigned explicit physics‑based parameters; but instead, they were treated as purely latent processes and inferred solely through the network parameters as $\boldsymbol{r}_{\boldsymbol{\xi}}$, and $ \boldsymbol{\dot{r}}_{\boldsymbol{\xi}}(t)$, respectively. As part of the input force parametrization for the SaPINNs, the time-modulated function $h(t) =\left|\gamma(t)\right|^2$ to reflect the influence of earthquake-related parameters.


In the present examples, the system response is simulated using a fourth-order Runge–Kutta time integration method, and noisy acceleration data are assumed to be available for all DoF.  The noise ratio in the measurements was set to 15\%, consistent with the examples of linear systems. The objective is to reconstruct the force applied to a system, while simultaneously recovering the latent states $[{x}_1,{x}_2,  {x}_3,   {r}_1]$, which represent the displacements at each DoF and the restoring force at the 1st DoF,  as well as the scaling factors for the stiffness ratios at the 2nd and 3rd DoF defined as $ \boldsymbol\theta=[\theta_1,\theta_2]$. The system is modeled as a 3-DoF lumped-mass structure subjected to a base excitation, with the system response being measured over a duration of 80 seconds. The mass of each DoF was considered as  $m_1=m_2=m_3=50$ kg, and the damping coefficients were taken as $ c_1=c_2=c_3=0.25$. The modeling error was introduced by intentionally misspecifying the damping ratios as 1 instead of the true values of 0.25 for all DoF.  \\
Since the majority of the work on PINNs for the SHM  relies on slowly varying responses \cite{lai2022neural,buildings13030650,haywood2025response}, the next sections examine the performance of the proposed framework on two systems with different stiffness and hysteretic model parameters–here referred to as low-frequency and high-frequency systems–subjected to a scaled excitation at the base. For these illustrative cases, the ground motion is simulated using a Kanai–Tajimi spectrum. Subsequently, we present an example for the high-frequency system subjected to the recorded El Centro earthquake, with intentionally missspecified spectrum.

\subsubsection{3-DoF system subjected to synthetic earthquake}\label{sec.synthetic}
\begin{figure}[H]
    \centering
    \begin{subfigure}[b]{0.47\textwidth}
        \centering
        \includegraphics[width=\textwidth]{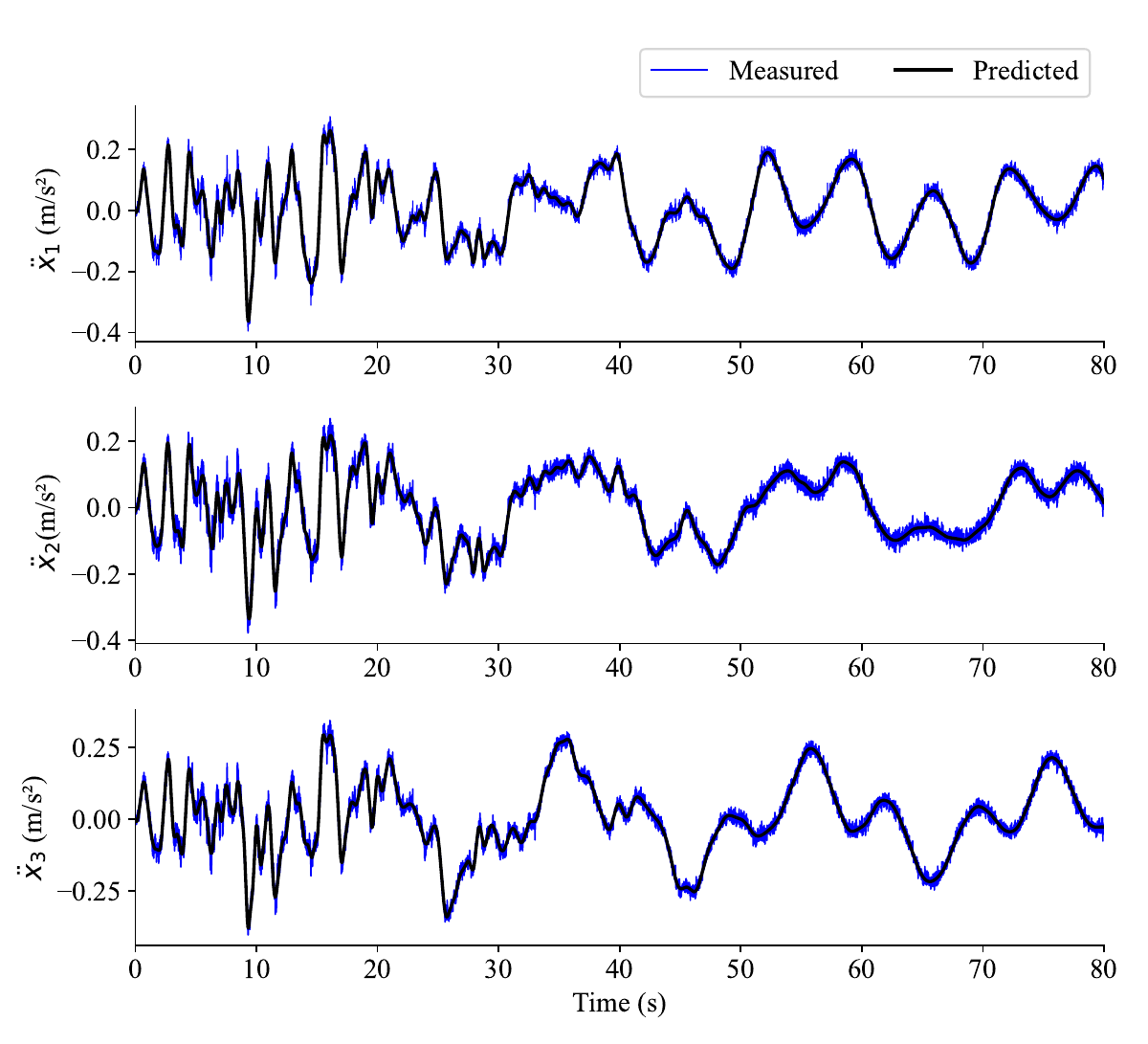}
            \vspace{-3ex}
         \captionsetup{justification=centering}
         \caption{} 
    \end{subfigure}
    \begin{subfigure}[b]{0.52\textwidth}
        \centering
        \includegraphics[width=\textwidth]{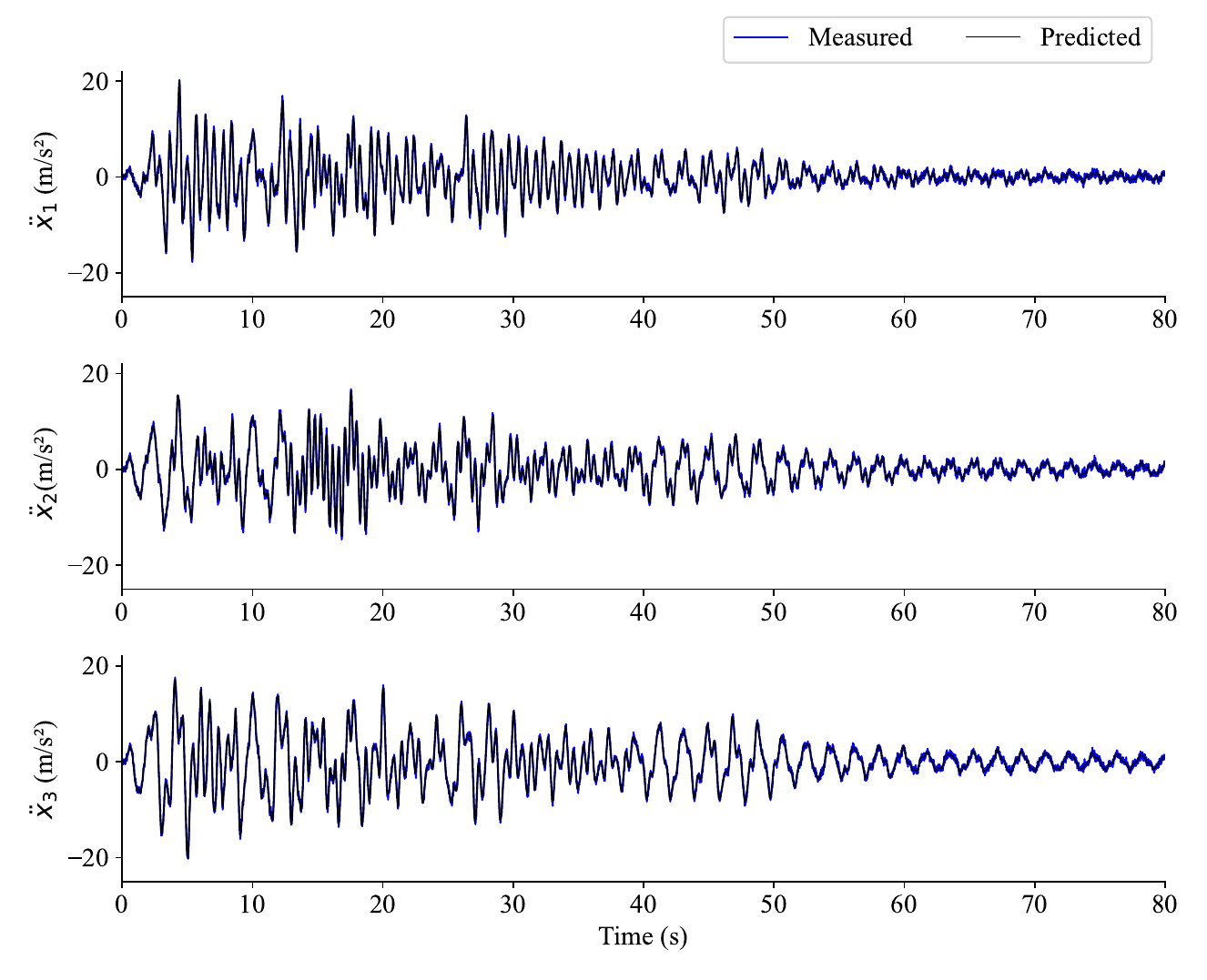}
         \vspace{-3ex}
         \captionsetup{justification=centering}
         \caption{} 
    \end{subfigure}
    \label{subfig:Fig11b}
     \vspace{-4ex} 
    \caption{Measured and predicted system response with the SaPINNs for (a) low-frequency, (b) high-frequency systems.}
    \label{fig:Fig11}
\end{figure}

\begin{figure}[H]
    \centering
    \begin{subfigure}[b]{0.5\textwidth}
        \centering
        \includegraphics[width=\textwidth]{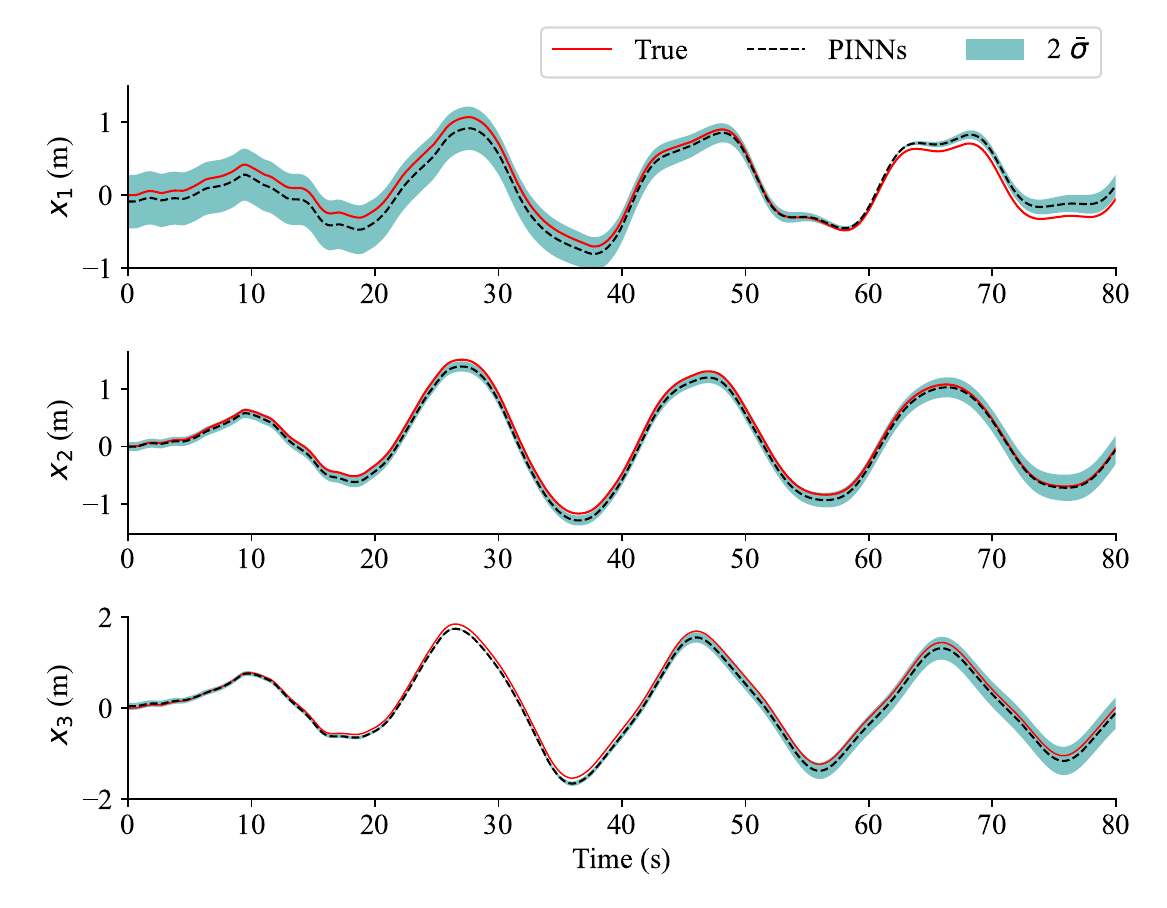}
        \label{subfig:label_for_first_image}
    \end{subfigure}
    \begin{subfigure}[b]{0.47\textwidth}
        \centering
        \includegraphics[width=\textwidth]{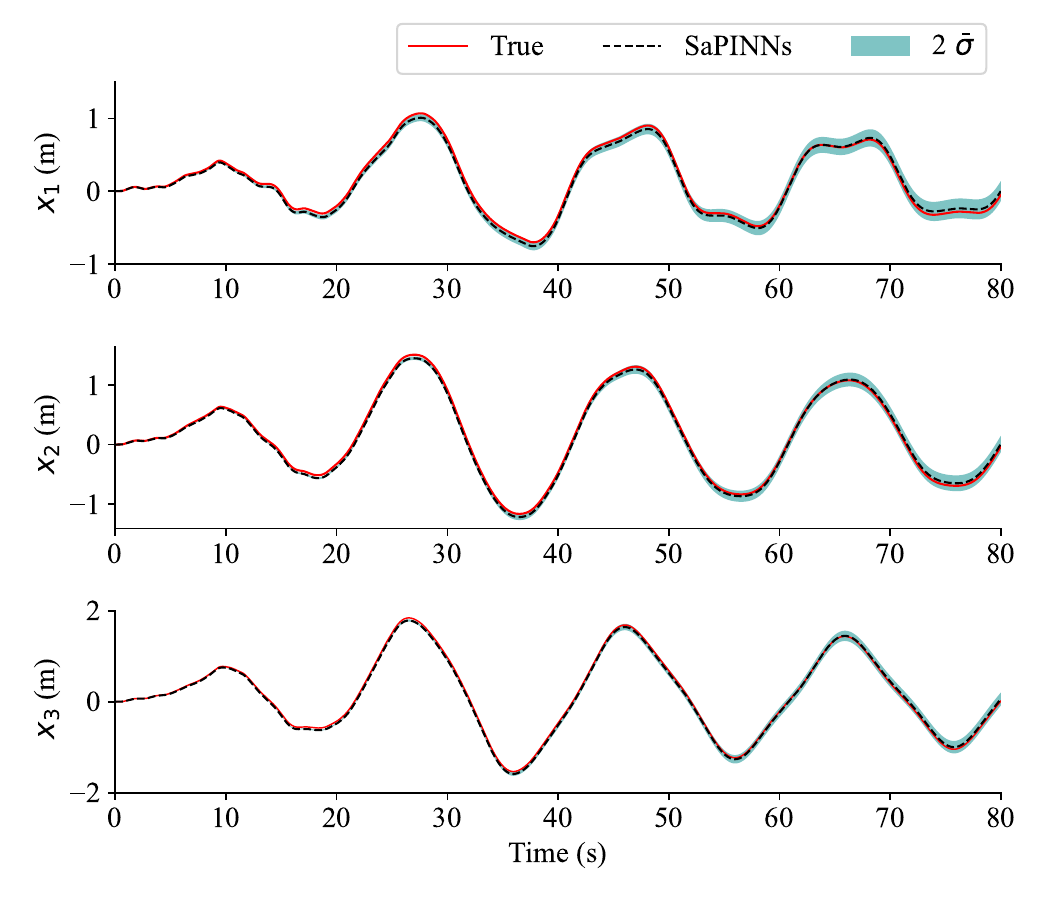}
        \label{subfig:label_for_second_image}
    \end{subfigure}
    \captionsetup{justification=centering}
     \vspace{-5ex} 
   \caption{ Low-frequency system: predicted displacements using  PINNs and SaPINNs.}
    \label{fig:Fig12}
\end{figure}

\begin{figure}[H]
    \centering
    \begin{subfigure}[b]{0.495\textwidth}
        \centering
        \includegraphics[width=\textwidth]{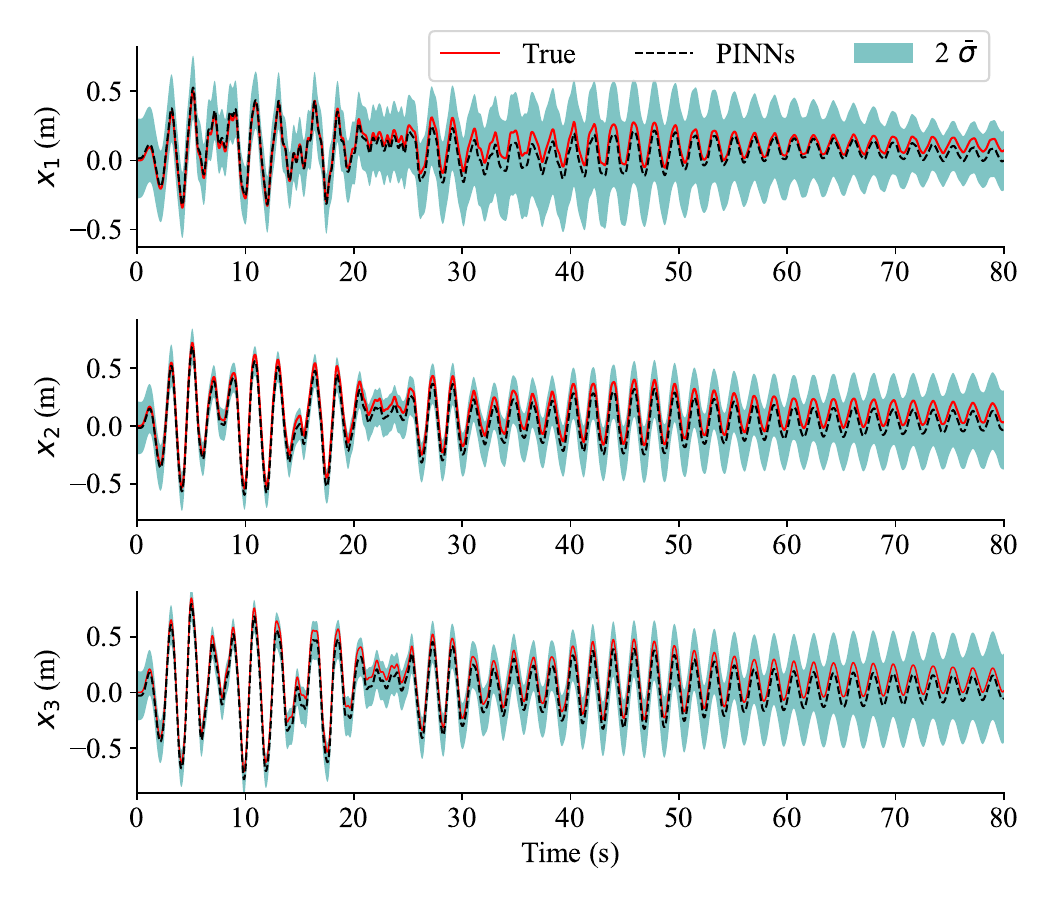}
        \label{subfig:label_for_first_image}
    \end{subfigure}
    \hspace{-1ex} 
    \begin{subfigure}[b]{0.495\textwidth}
        \centering
        \includegraphics[width=\textwidth]{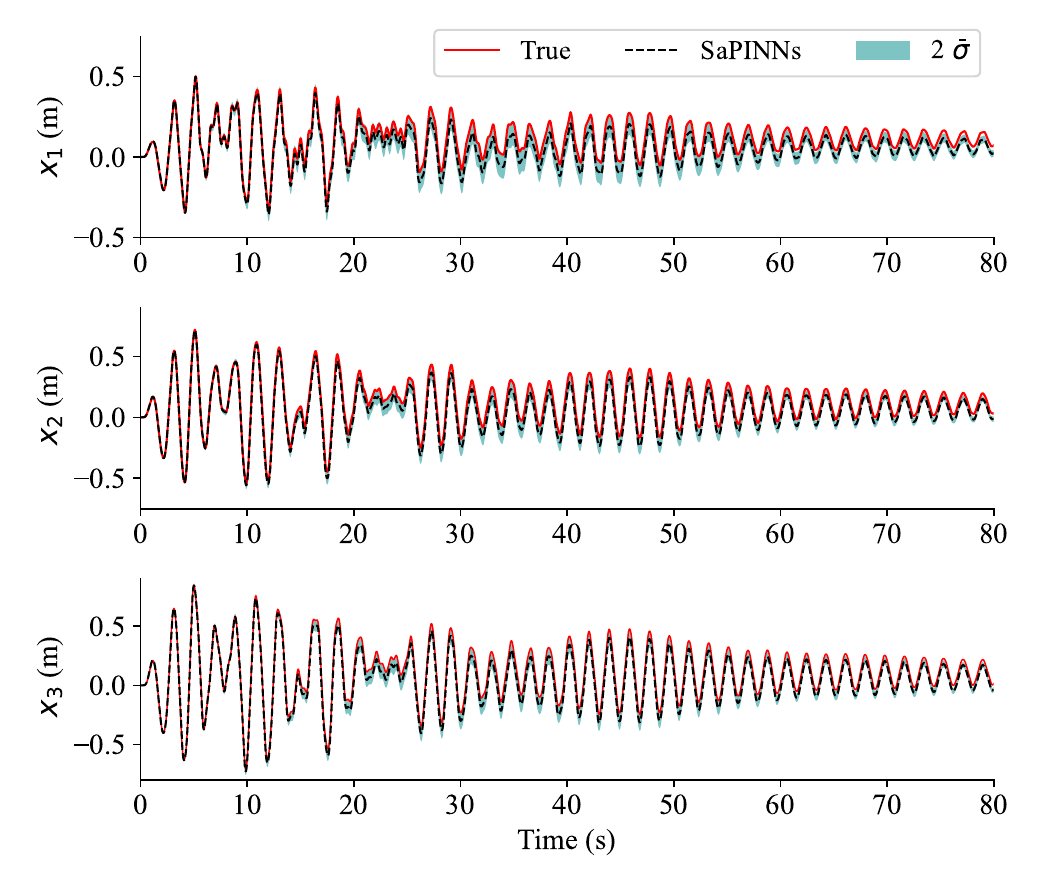}
        \label{subfig:label_for_second_image}
    \end{subfigure}
    \captionsetup{justification=centering}
    \vspace{-3ex} 
    \caption{High-frequency system: predicted displacements using PINNs and SaPINNs.}
    \label{fig:Fig13}
\end{figure}

For the simulation of the low-frequency system response, the properties of the hysteretic 3-DoF Bouc–Wen model were taken as the following: $k_1=k_2=k_3=30$ N/m,  $\alpha = 1$, $\beta = 0.5$, and $n = 1$. For the high-frequency system, the model was considered with parameters:  $k_1=k_2=k_3=3$000 N/m, $\alpha=1$ $\beta=0.5$,  $n=1$.
Within the SaPINNs framework, the nonlinear system response at the 1st DoF can be defined explicitly as:
\begin{equation}
f_{\boldsymbol{\Theta},1}(t) = \frac{1}{m_1} \left( 
- k_1 r_{\boldsymbol{\xi},1} 
- \theta_1 k_2 \left( x_{\boldsymbol{\xi},1}(t) + x_{\boldsymbol{\xi},2}(t_i) \right) 
- (c_1 + c_2) \dot{x}_{\boldsymbol{\xi},1}(t) 
+ c_2 \dot{x}_{\boldsymbol{\xi},2}(t) 
- U_{\boldsymbol{\phi},1}(t)\right). 
\end{equation}
Following the comparative assessment of PINNs and SaPINNs presented in Section \ref{sec.thunder}, we demonstrate here the performance of two representative configurations: (i) the baseline PINNs trained without prescribing the input force and (ii) SaPINNs in which the phase angles are included in the training.  For each configuration, an ensemble of 20 networks was trained in parallel, where every network consisted of five hidden layers with a topology of 64–64–32–64–64 nodes. A side‑by‑side comparison of the measurements data $  \boldsymbol{ y}^{obs}(t)$ is shown in Fig. \ref{fig:Fig11}. The outputs of the trained PINNs and SaPINNs ensembles for the latent states prediction, specifically, the displacement responses, are shown in Fig.\ref{fig:Fig12} and Fig.\ref{fig:Fig13} for the low- and high-frequency systems, respectively. 
\begin{figure}[H]
    \centering
    \begin{subfigure}[b]{0.49\textwidth}
        \centering
        \includegraphics[width=\textwidth]{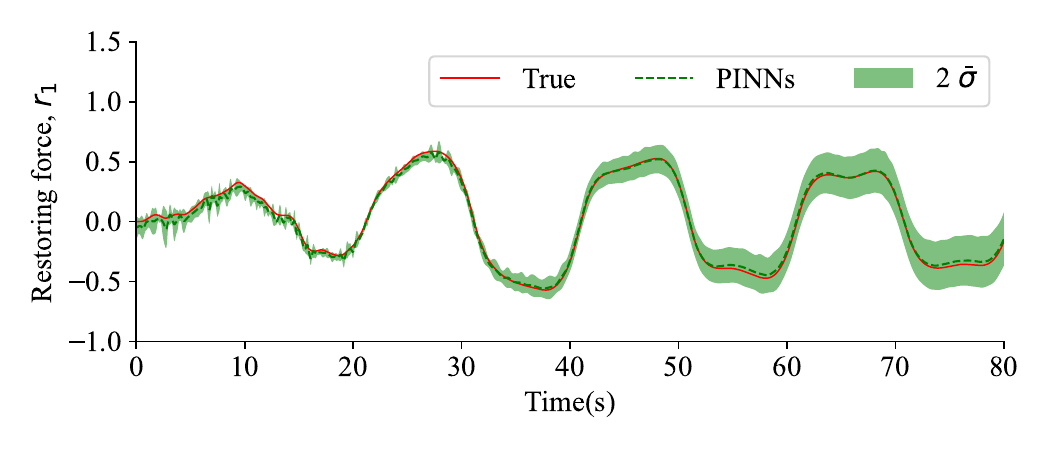}
    \end{subfigure}
    \begin{subfigure}[b]{0.49\textwidth}
        \centering
        \includegraphics[width=\textwidth]{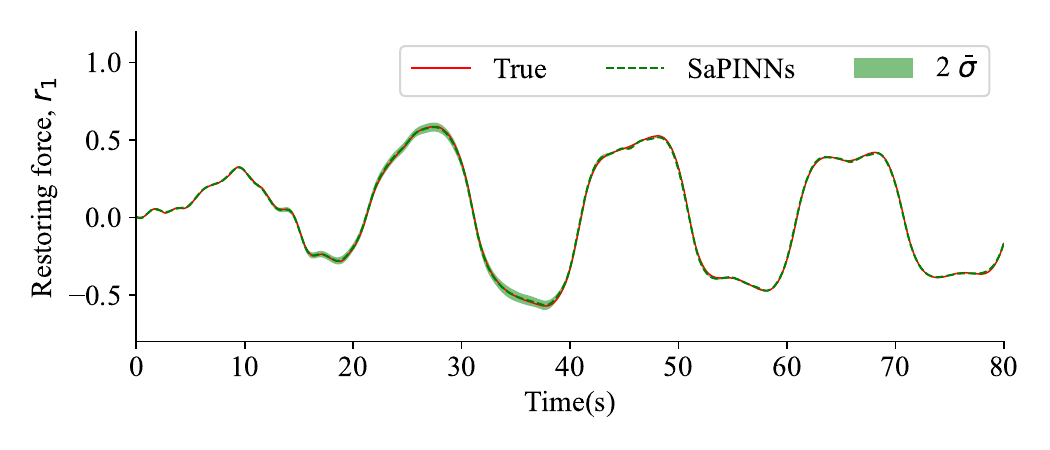}
    \end{subfigure}
     \vspace{-3ex} 
    \caption{Low-frequency system: comparison of restoring force time-history predictions from  acceleration data using PINNs and  SaPINNs. }
       \label{fig:Fig14}
\end{figure} 
\begin{figure}[H]
    \centering
    \begin{subfigure}[b]{0.49\textwidth}
        \centering
        \includegraphics[width=\textwidth]{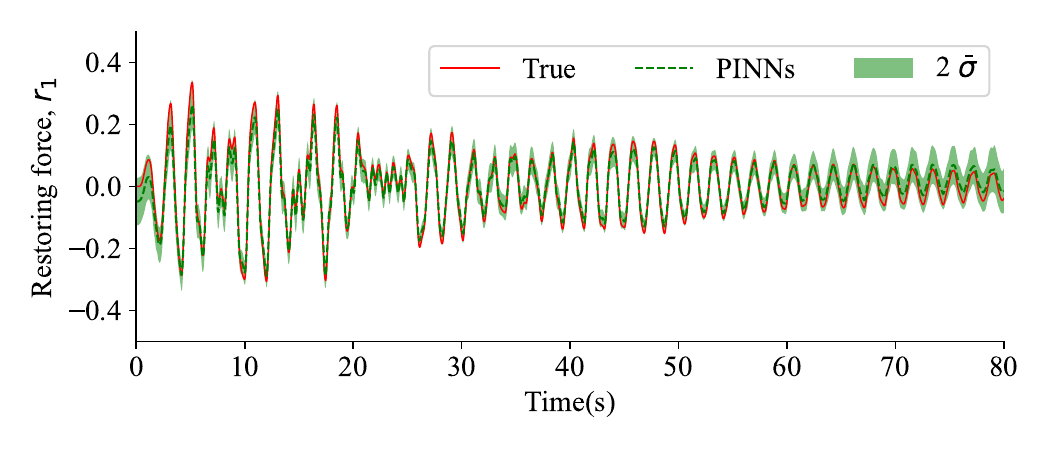}
    \end{subfigure}
    \begin{subfigure}[b]{0.49\textwidth}
        \centering
        \includegraphics[width=\textwidth]{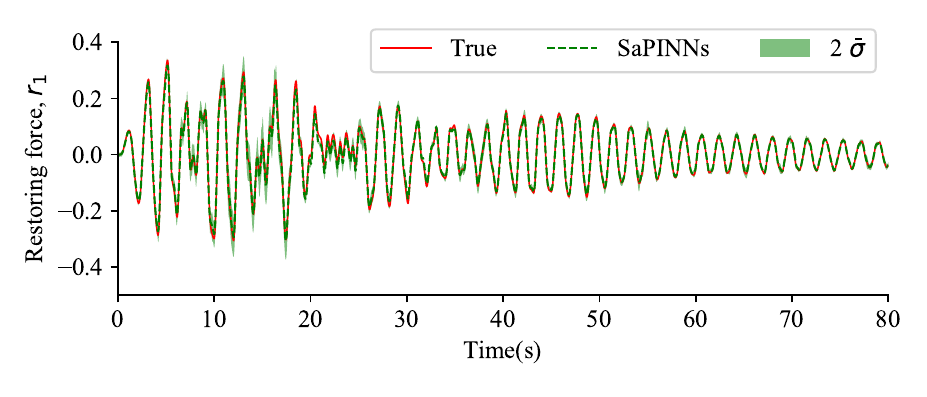}
    \end{subfigure}
      \vspace{-1ex} 
    \caption{High-frequency system: comparison of restoring force time-history predictions from  acceleration data using PINNs and  SaPINNs. }
    \label{fig:Fig15}
\end{figure}

\begin{figure}[H]
    \centering
    \begin{subfigure}[b]{\textwidth}
        \centering
        \begin{minipage}[b]{0.32\textwidth}
            \centering
            \includegraphics[width=\textwidth]{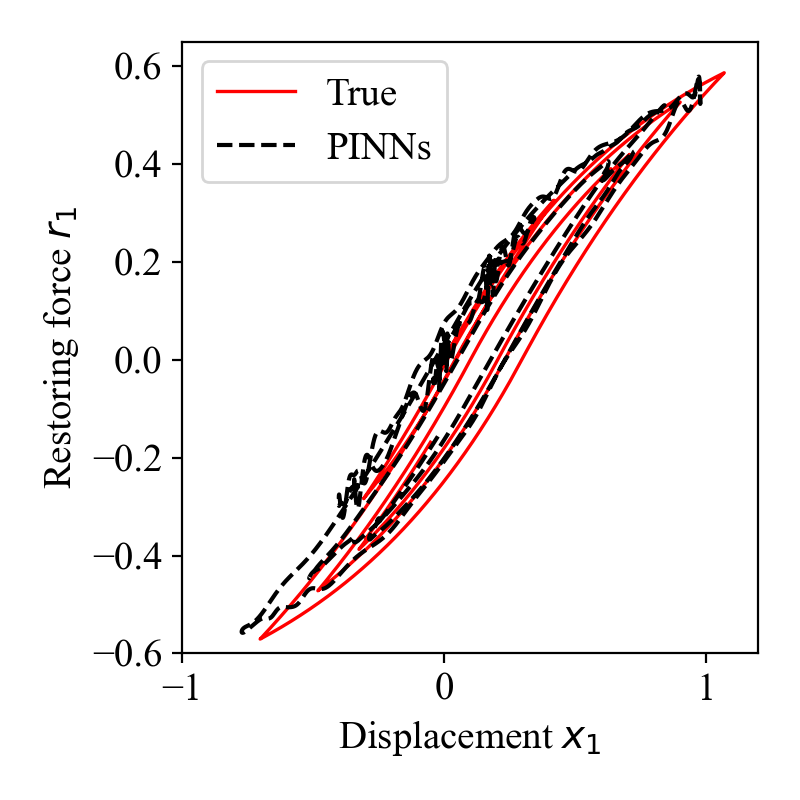}
        \end{minipage}
        \begin{minipage}[b]{0.32\textwidth}
            \centering
            \includegraphics[width=\textwidth]{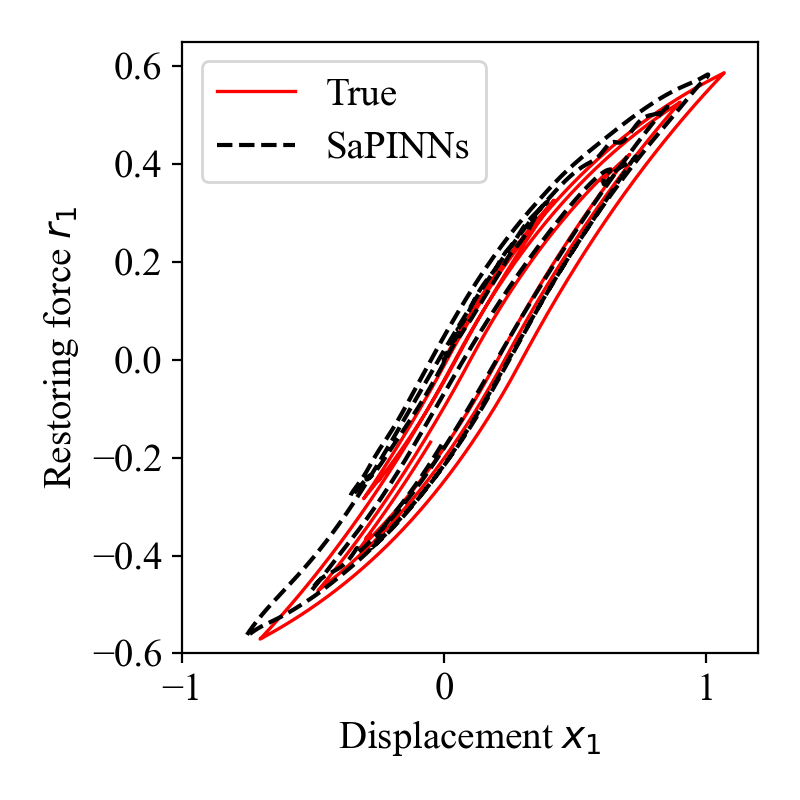}
        \end{minipage}
        \vspace{-3ex}
        \captionsetup{justification=centering}
        \caption*{(a)}
    \end{subfigure}
    \vspace{1ex}
    \begin{subfigure}[b]{\textwidth}
        \centering
        \begin{minipage}[b]{0.32\textwidth}
            \centering
            \includegraphics[width=\textwidth]{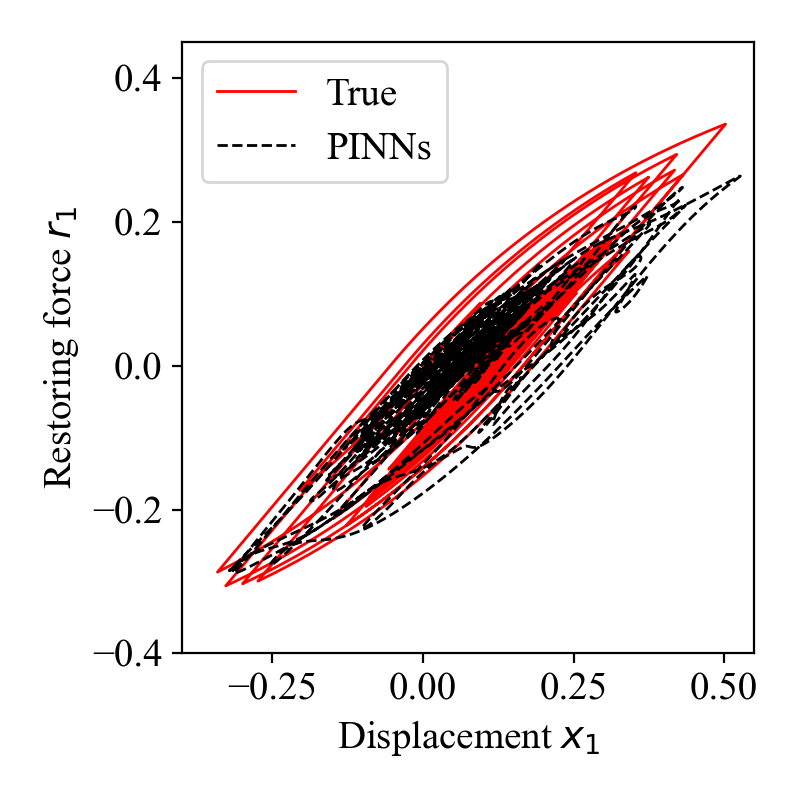}
        \end{minipage}
         \begin{minipage}[b]{0.32\textwidth}
            \centering
            \includegraphics[width=\textwidth]{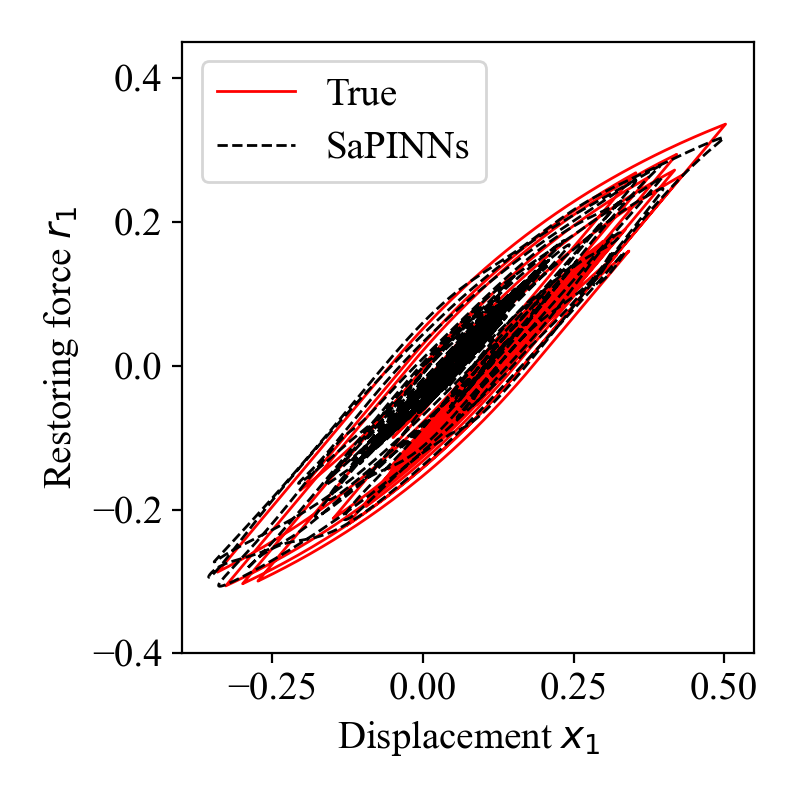}
        \end{minipage}
        \vspace{-3ex}
        \captionsetup{justification=centering}
        \caption*{(b)}
    \end{subfigure}
    \vspace{-3ex}
    \caption{Reconstructed hysteretic response of a 3-DoF Bouc–Wen system using  PINNs and SaPINNs for (a) low-frequency, (b) high-frequency systems.}
    \label{fig:Fig16}
\end{figure}

\begin{figure}[!htbp]
    \centering
    \begin{subfigure}[b]{0.495\textwidth}
        \centering
        \includegraphics[width=\textwidth]{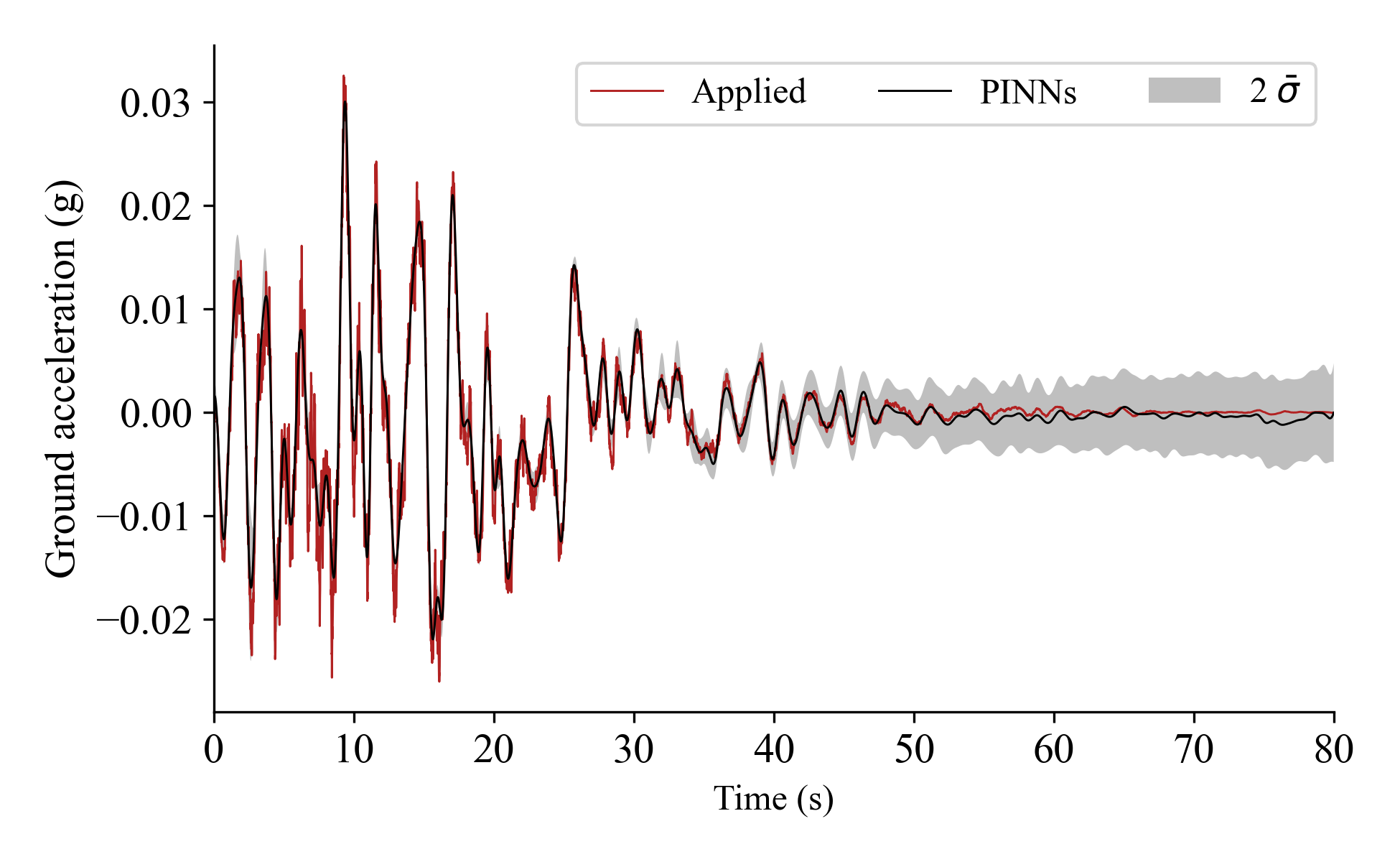}
    \end{subfigure}
    \begin{subfigure}[b]{0.495\textwidth}
        \centering
        \includegraphics[width=\textwidth]{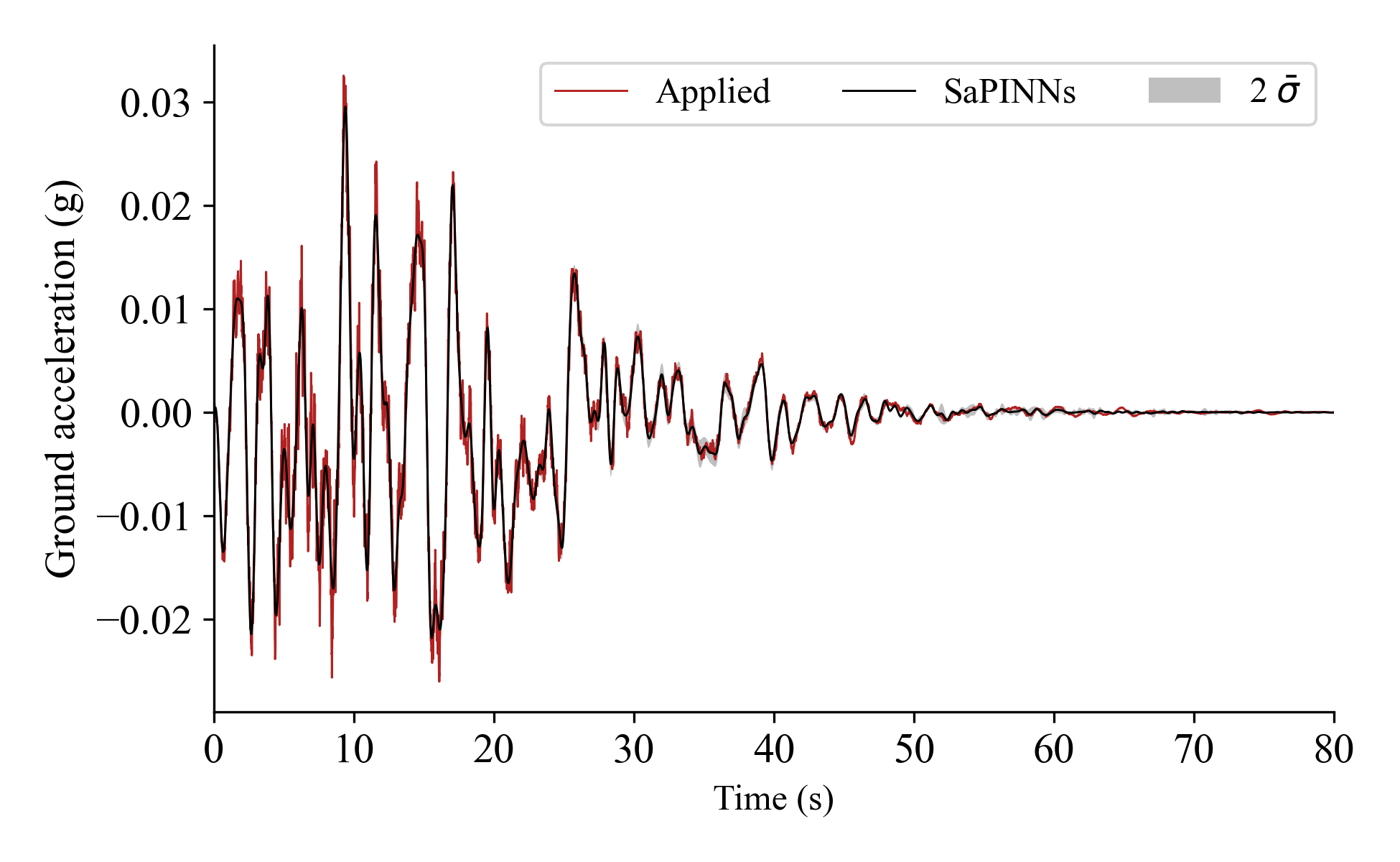}
    \end{subfigure}
     \vspace{-3ex} 
    \caption{Low-frequency system: predicted time-history of the applied synthetic ground acceleration using PINNs and SaPINNs ensemble.}
    \label{fig:Fig17}
\end{figure}

\begin{figure}
    \centering
    \begin{subfigure}[b]{0.495\textwidth}
        \centering
        \includegraphics[width=\textwidth]{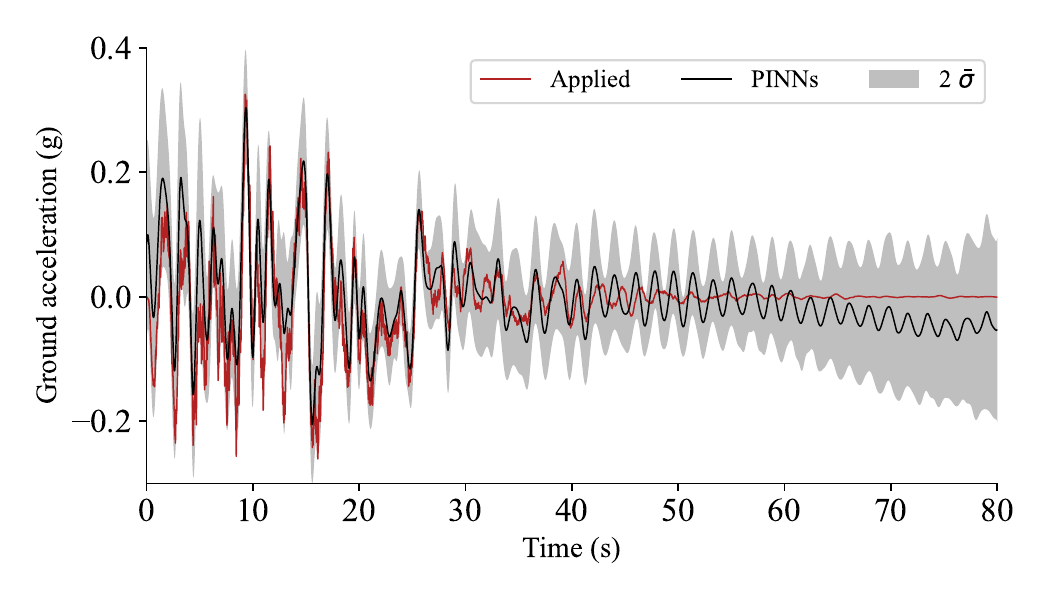}
    \end{subfigure}
    \begin{subfigure}[b]{0.495\textwidth}
        \centering
        \includegraphics[width=\textwidth]{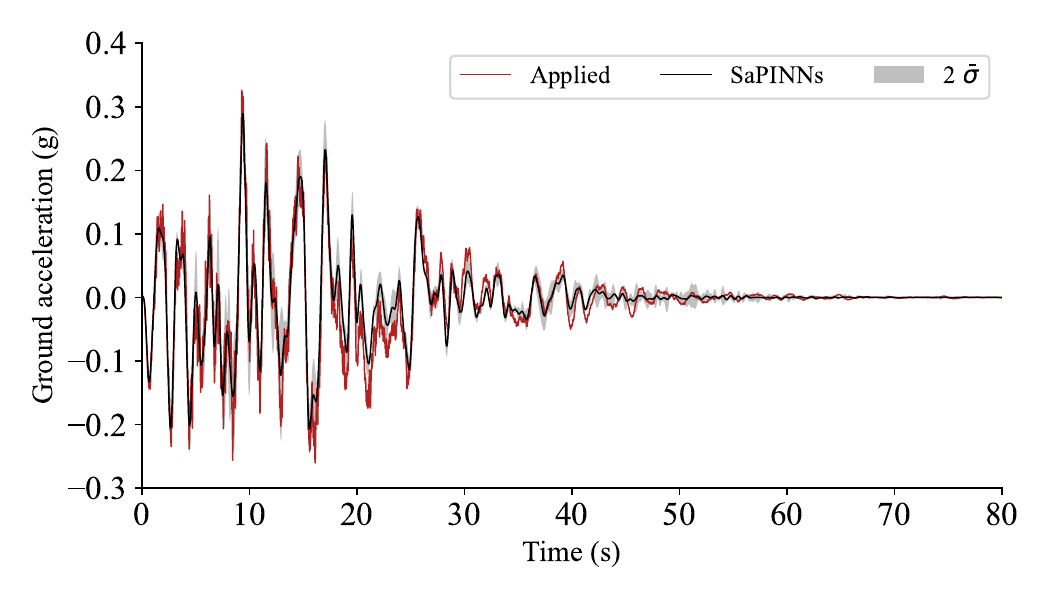}
    \end{subfigure}
     \vspace{-3ex} 
    \caption{High-frequency system: predicted time-history of the applied synthetic ground acceleration using PINNs and SaPINNs ensemble.}
    \label{fig:Fig18}
\end{figure}

As shown in Figs.\ref{fig:Fig12} -\ref{fig:Fig13}, for both the low- and high-frequency systems, the deviations from the true displacement trajectories are more pronounced in the outputs of the PINNs model. Additionally, the uncertainty associated with the predicted displacements is significantly higher compared to the SaPINNs model output. A similar trend is observed in the reconstruction of the restoring forces, as illustrated in Figs.\ref{fig:Fig14}-\ref{fig:Fig15}. The hysteresis loops inferred by each model,  shown in Fig.\ref{fig:Fig16},  demonstrates that the SaPINNs ensemble reproduces the characteristic shape and energy‑dissipation pattern with minimal bias, whereas, the PINNs exhibit greater deviation and reduced physical consistency, failing to fully represent the underlying nonlinear dynamics.
Further, both models recovered the input force in the low-frequency system with similar accuracy, but the PINNs exhibited higher uncertainty in its inference, as shown in Fig.\ref{fig:Fig17}.  A key factor contributing to the success of PINNs along the SaPINNs for the input force reconstruction is the nature of the excitation applied in the form of base excitation, which simplifies the inverse problem by inducing a spatially coherent forcing structure across all DoF.  In the case of base excitation, the external force enters the system as a ground motion input, resulting in equivalent inertial forces at each DoF that are proportional to the respective masses. This generates a global excitation profile that simultaneously influences all DoFs, thereby increasing the amount of information encoded in the acceleration measurements. From a system identification perspective, this form of excitation enhances the observability of the system, as each state variable becomes directly affected by the input through the mass-proportional forcing term. However, despite the success in the low-frequency system, the accuracy of the PINNs deteriorates markedly in the high-frequency case, while the SaPINNs managed to get a reasonable prediction of the applied force (Fig.\ref{fig:Fig18}). 

\begin{figure}[!htbp]
    \centering
    \begin{subfigure}[b]{0.4\textwidth}
        \centering
        \includegraphics[width=\textwidth]{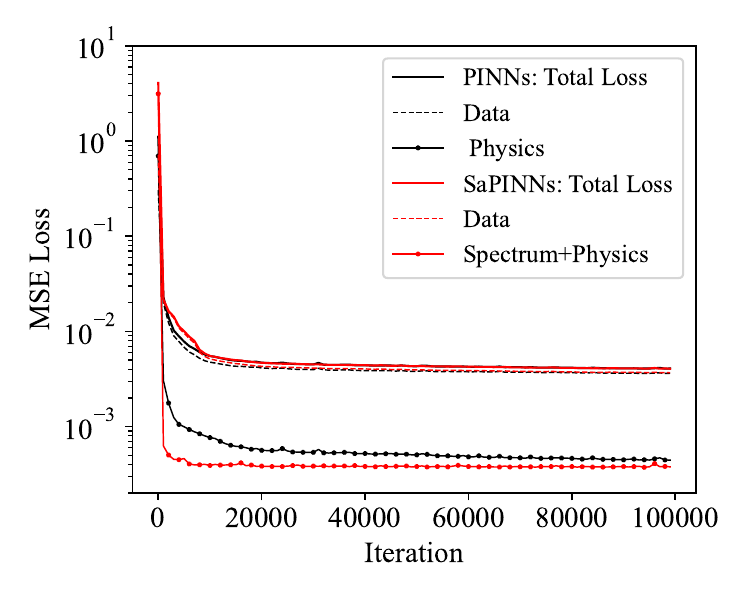}
        \vspace{-3ex} 
        \captionsetup{justification=centering}
        \caption{} 
        \label{subfig:label_for_first_image}
    \end{subfigure}
    \begin{subfigure}[b]{0.49\textwidth}
        \centering
        \includegraphics[width=\textwidth]{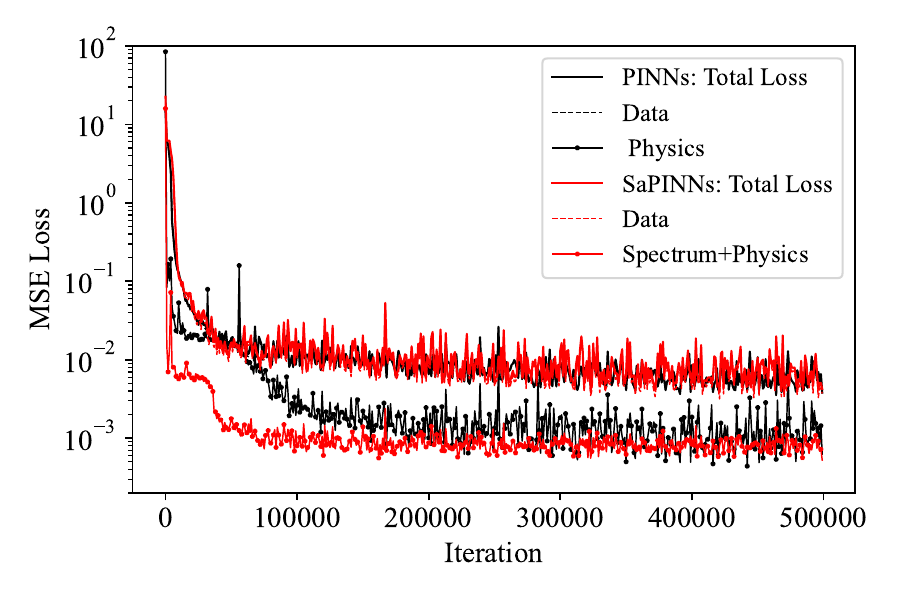}
        \captionsetup{justification=centering}
         \vspace{-3ex} 
        \caption{} 
        \label{subfig:label_for_second_image}
    \end{subfigure}
    \captionsetup{justification=centering}
    \caption{Loss convergence behavior for (a) low-frequency system and (b) high-frequency system for the first component of the ensemble. }
    \label{fig:Fig.19}
\end{figure}

The performance gap observed between PINNs and SaPINNs for the high-frequency system, despite both models being trained under the same base excitation setting, reflects fundamental differences in how the models handle information in highly oscillatory inputs. The deterioration in PINNs performance is primarily due to the increased complexity of the observed data, the acceleration signal (Fig.\ref{fig:Fig11}b), that introduces rapidly varying features that are difficult to infer without additional  guidance. While SaPINNs are able to regularize this complexity through their spectral parameterization, guiding the model toward physically meaningful solutions, the baseline PINNs lack the necessary conditions to constrain the inference, resulting in reduced accuracy in force estimation. \\
\begin{table}[H] 
\centering
\caption{ Performance metrics of the PINNs and SaPINNs for low- and high-frequency systems excited by a synthetic earthquake simulated from a Kanai-Tajimi EPSD. Uncertainties in the system parameters are inferred from a 20-member ensemble. }
\begin{tabular}{lcccccccc}
\toprule
     Model & \(\theta_1\) & \(\sigma(\theta_1)\) & \(\theta_2\) & \(\sigma(\theta_2)\) & \(MSE(x_1)\) & \(MSE(x_2)\) & \(MSE(x_3)\)& \(MSE(r_1)\) \\ \midrule
\multicolumn{9}{c}{Low-Frequency} \\ \midrule
{PINNs}& 0.969& 0.004& 0.988& 8.6e-4& 0.426 &1.041 &1.528 & 7.9e-5\\
{SaPINNs}  &0.992& 0.002& 0.997& 5.9e-4& 0.421 &1.033&1.507& 4.5e-5\\ \midrule
\multicolumn{9}{c}{High-Frequency} \\ \midrule
{PINNs}  & 0.879 & 0.022 & 0.905 & 0.017 & 0.031&0.081&0.119& 4.8e-4\\
{SaPINNs} & 1.009& 0.060& 0.996&0.037&0.030& 0.077&0.114 & 1.9e-4\\ \bottomrule
\end{tabular}
\label{table:2}
\end{table}
The models’ performance in the parameter estimation demonstrates additional differences. As summarized in Table \ref{table:2}, the SaPINNs yield stiffness estimates that are closer to the ground truth than those inferred by the PINNs.  Further, for the high-frequency scenario,  the system's parameter uncertainty  increases, which indicates that the larger standard deviations inferred by the ensemble for $\theta_1$ and $\theta_2$ are driven by the training data complexity rather than the learning architecture. Given the same record duration and sampling rate, the high-frequency response contains a greater number of oscillation cycles, resulting in fewer data points per cycle.  Such a signal exhibits greater variability and sharper transitions relative to the samples, which make the mapping between the observed responses and underlying parameters more sensitive and nonlinear. This added complexity increases the likelihood of convergence to different local optima across ensemble members, leading to greater spread in the recovered parameter distributions. The differences in convergence behavior between the two systems are reflected in their respective loss functions,  shown in Fig. \ref{fig:Fig.19}.

For the low-frequency system, both PINNs (black lines) and SaPINNs (red lines) show stable and monotonic convergence of their loss functions. However, the SaPINNs converge faster and to a slightly lower total loss compared to the PINNs over the same training period.  In contrast, the high-frequency system presents a more challenging learning scenario for both models.  As the networks try to accommodate the sharp fluctuations present in the training data, gradient updates occasionally drag the weights into regions that strongly violate the physics and similarly the spectrum-physics residuals. The subsequent regularization penalty then forces a corrective step, resulting in the pronounced saw-tooth pattern. Additionally, the losses cannot decrease indefinitely as they are ultimately bounded by the measurement noise and the modeling error floor (as discussed in Section \ref{section.4.1}). Once the residual reaches its minimum, every attempt to squeeze out another fraction of the data misfit is countered by an increase in the physics and the spectrum-physics residual, making the optimizer to oscillate around a quasi-equilibrium.
  \vspace{1em} 
\subsubsection{3-DoF system subjected to an El Centro earthquake}\label{section.elcentro}
\begin{figure}[H]
    \centering
    \begin{subfigure}[b]{0.495\textwidth}
        \centering
        \includegraphics[width=\textwidth]{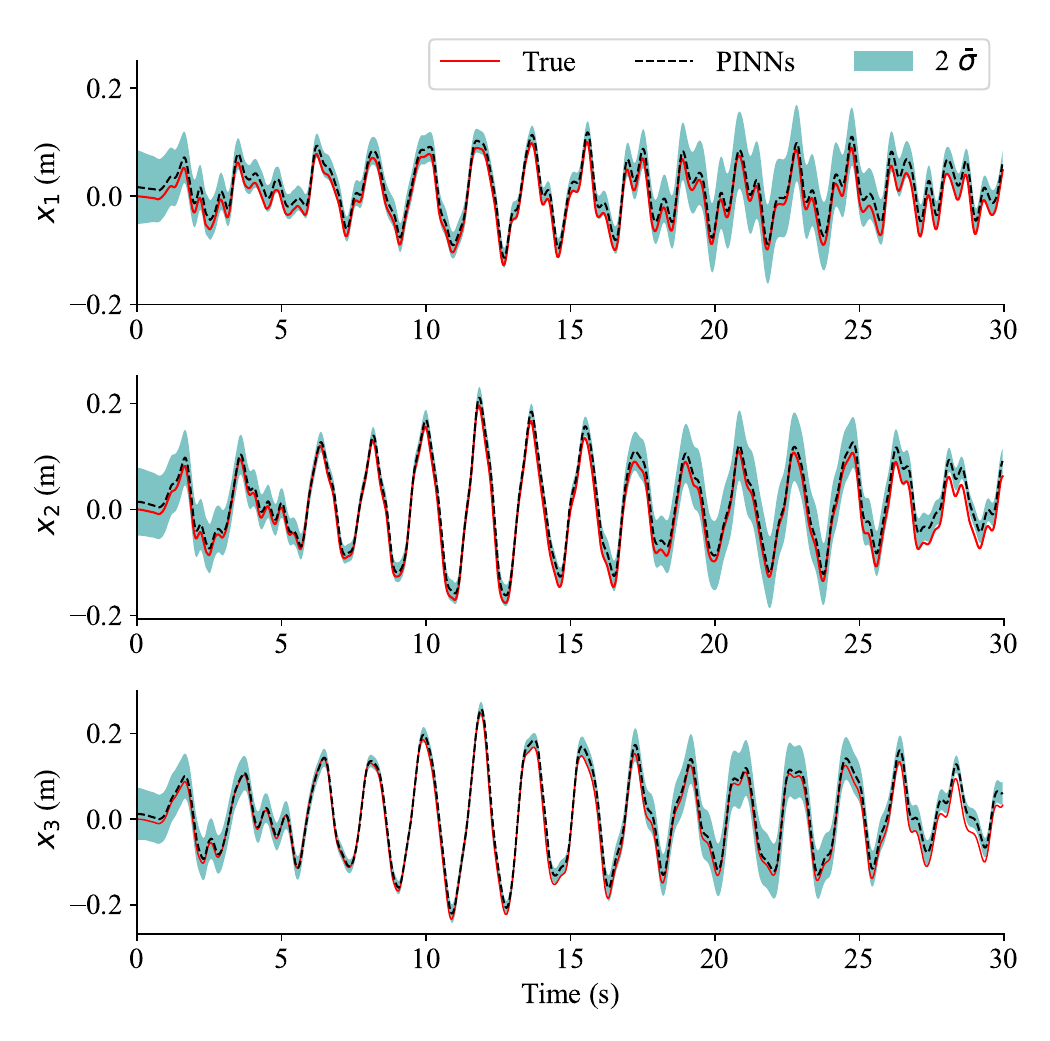}
    \end{subfigure}
    \begin{subfigure}[b]{0.495\textwidth}
        \centering
        \includegraphics[width=\textwidth]{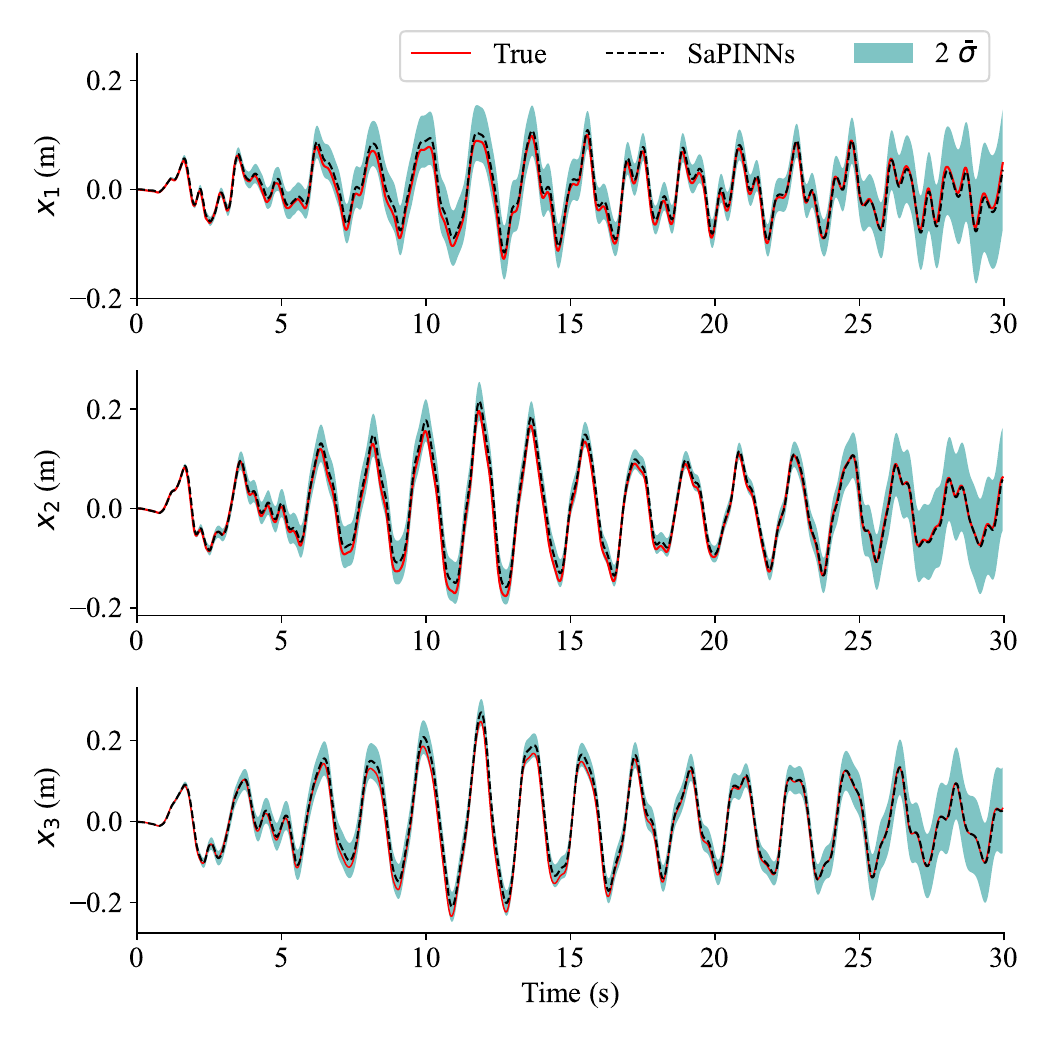}
    \end{subfigure}
     \captionsetup{justification=centering}
    \vspace{-5ex} 
    \caption{Predicted displacements using PINNs and SaPINNs ensemble under the unknown El Centro base excitation. }
    \label{fig:Fig.20}
\end{figure}

\begin{figure}[H]
    \centering
    \begin{subfigure}[b]{0.495\textwidth}
        \centering
        \includegraphics[width=\textwidth]{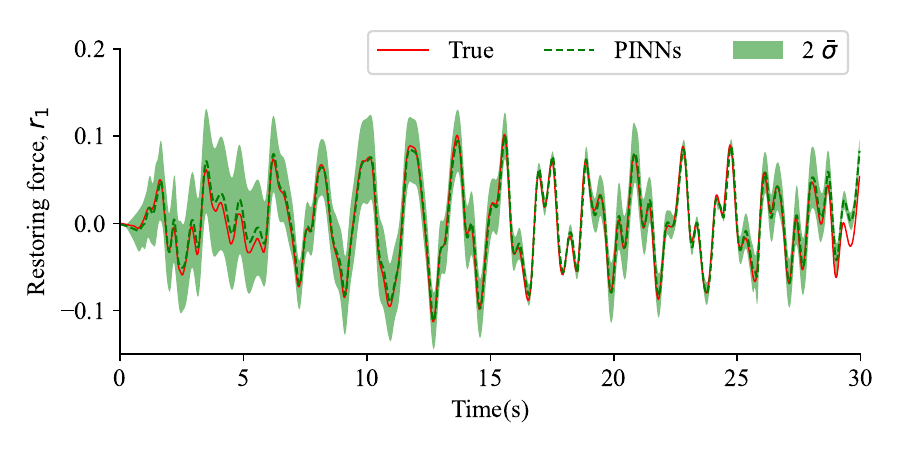}
    \end{subfigure}
    \begin{subfigure}[b]{0.495\textwidth}
        \centering
        \includegraphics[width=\textwidth]{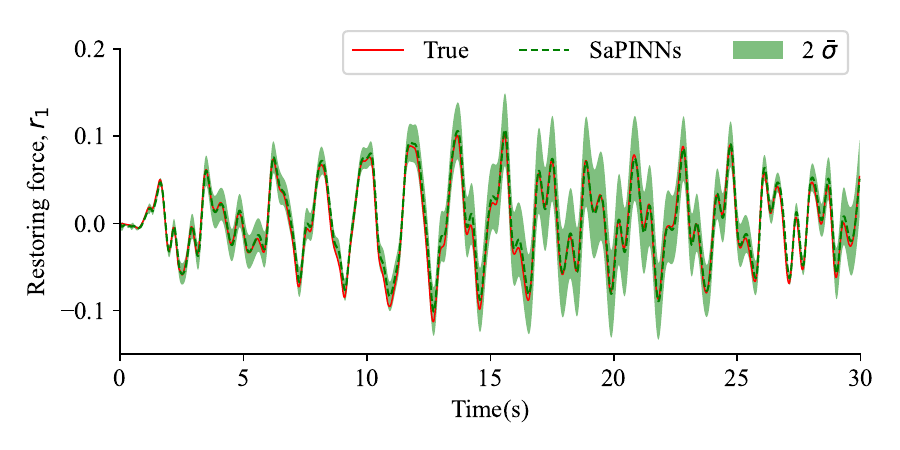}
    \end{subfigure}
    \vspace{-5ex} 
    \caption{Predicted restoring force at 1st DoF using PINNs and SaPINNs under the unknown El Centro base excitation.}
    \label{fig:Fig21}
\end{figure}

\begin{figure}[H]
    \centering
    \begin{subfigure}[b]{0.495\textwidth}
        \centering
        \includegraphics[width=\textwidth]{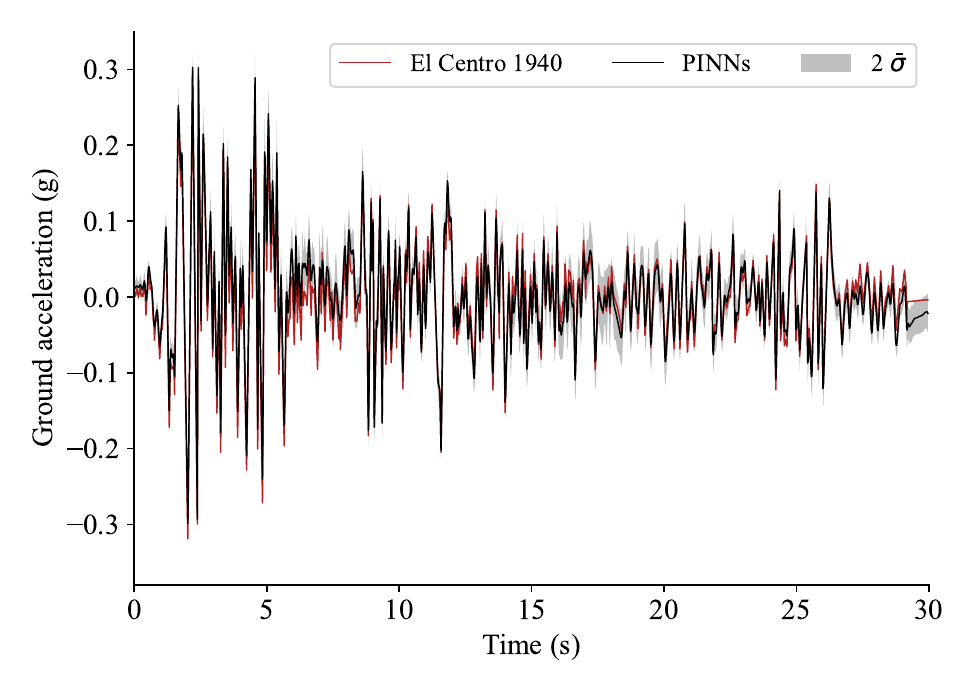}
        \captionsetup{justification=centering}
        \label{subfig:label_for_first_image}
    \end{subfigure}
    \begin{subfigure}[b]{0.495\textwidth}
        \centering
        \includegraphics[width=\textwidth]{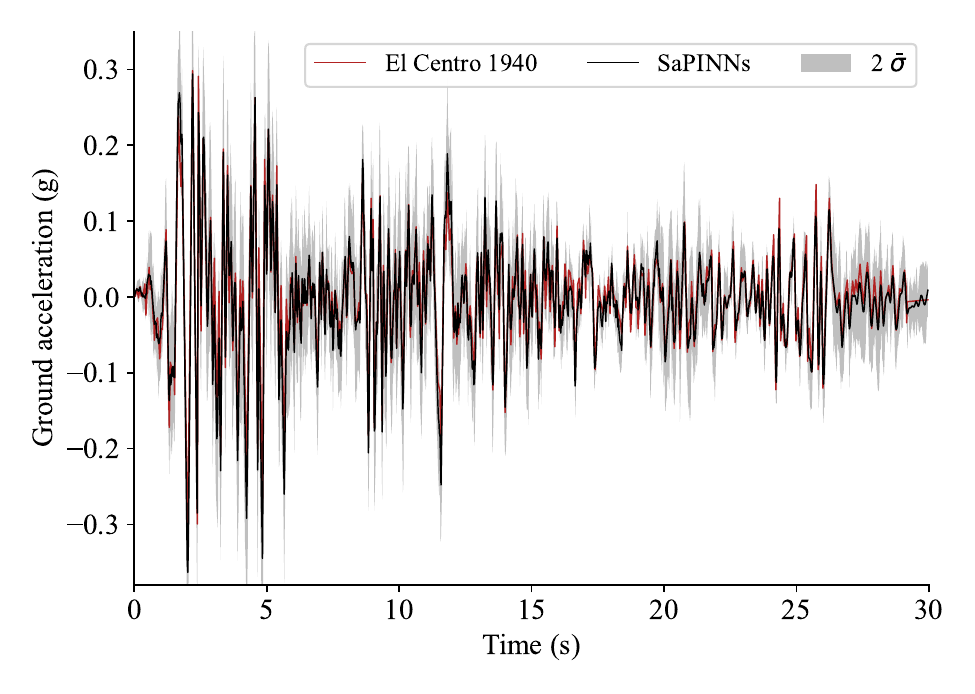}
        \captionsetup{justification=centering}
           \label{subfig:label_for_second_image}
    \end{subfigure}
    \captionsetup{justification=centering}
    \vspace{-6ex} 
    \caption{Predicted time-history of applied El Centro ground acceleration with  PINNs and SaPINNs ensemble.}
    \label{fig:Fig.22}
\end{figure}

\begin{figure}[H]
    \centering
    \begin{subfigure}[t]{0.495\textwidth}
        \centering
        \includegraphics[scale=0.5]{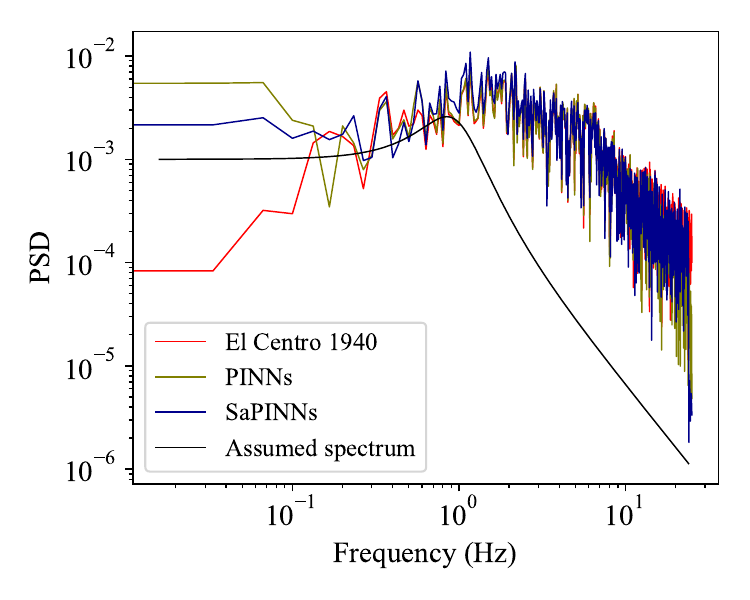}
        \captionsetup{justification=centering}
                 \caption{} 
    \end{subfigure}
    \begin{subfigure}[t]{0.495\textwidth}
        \centering
        \includegraphics[scale=0.5]{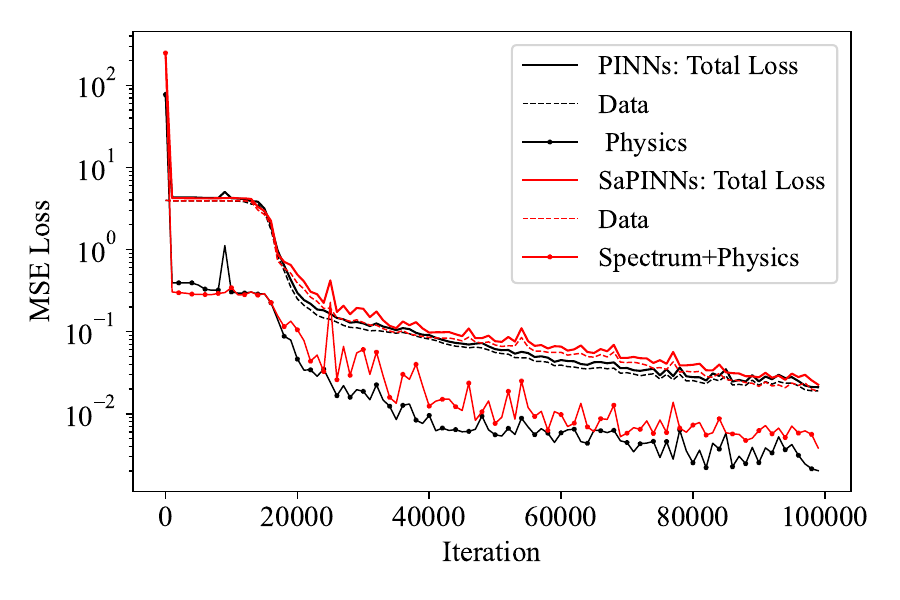}
        \captionsetup{justification=centering}
         \caption{} 
    \end{subfigure}
    \caption{(a) PSD comparison between the applied El Centro excitation, predicted excitations from PINNs and SaPINNs, and the assumed target spectrum for the SaPINNs. (b) Evolution of  the loss functions over the training horizon for the PINNs and SaPINNs across loss components. }
    \label{fig:spec_elcentro}
\end{figure}

To further illustrate the performance of the proposed framework in a more realistic scenario, we examine a high-frequency system subjected to a seismic excitation whose spectral parameters are incorrectly characterized, reflecting the limited a priori information available in practice.  The seismic excitation used for the example is the North–South component of the 1940 El Centro earthquake in California \cite{ElCentro}. The spectrum is modeled using the Kanai–Tajimi formulation with a scaling factor as a function of the ground motion intensity $S_1=0.1$, a predominant ground frequency $\omega_g=1.59$ Hz, and a damping ratio $\xi_g=0.6$.  In the example, a spectrum is considered in the range of $\omega  \sim[0,24] $ Hz,  with a frequency discretization of $\Delta \omega=0.015$ Hz. The modulating function was adopted in the same form as in the synthetic earthquake example  (Eq. \ref{eq:mod_function}) , with parameters set to $a=0.1$ and $b=0.2$. 
Fig. \ref{fig:Fig.20} shows the ground acceleration time histories reconstructed by the PINNs and SaPINNs ensembles against the recorded earthquake motion, Fig.\ref{fig:Fig21} displays the estimated displacements with their associated uncertainties, and Fig.\ref{fig:Fig.22} demonstrates the restoring forces at the 1st DoF inferred by the models.
As shown in the prediction of the ground motion and latent states, both the PINNs and SaPINNs have sufficient expressivity and constraints to learn meaningful solutions. Similarly to the synthetic earthquake excitation, the better accuracy achieved here by PINNs, relative to the linear examples, stems largely from the use of base excitation, which renders the inverse problem considerably more tractable. However, the SaPINNs' ensemble exhibits noticeably wider predictive uncertainty over the predicted quantities compared to the synthetic examples. This increase is expected due to the mismatch between the assumed excitation spectrum and the true ground motion characteristics (Fig.\ref{fig:spec_elcentro}a).  Nonetheless, despite this bias, the ensemble still converges to a credible solution, and the actual base excitation is within the $2\sigma$ credibility interval of the SaPINNs' predictions. This outcome is a direct consequence of the model formulation: the phases control the temporal alignment of the harmonic components, allowing the optimizer to fine-tune their interference patterns to a more appropriate waveform shape. By calibrating the phases, the model enables adjacent frequencies to cancel out at certain time instants and reinforce at others, thereby reshaping the instantaneous force history to be consistent with the observed measurements.
The loss functions evolution is shown in Fig.\ref{fig:spec_elcentro}b, where the losses' trajectories reveal a a similar trend between the two architectures: the PINNs and SaPINNs losses descend steeply and exhibit oscillations during the training. 

\begin{table}[H]
\centering
\caption{Comparison of PINNs and SaPINNs performance metrics for input–state–parameter estimation for a 3-DoF Bouc–Wen system excited by 1940 El Centro earthquake base excitation. Uncertainties in the system parameters are inferred from a 20-member ensemble.}
\begin{tabular}{lcccccccc}
\toprule
    Model & \(\theta_1\) & \(\sigma(\theta_1)\) & \(\theta_2\) & \(\sigma(\theta_2)\) & \(MSE(x_1)\) & \(MSE(x_2)\) & \(MSE(x_3)\)& \(MSE(r_1)\) \\ \midrule
PINNs & 0.956 & 0.002 & 0.981 & 3.0e-3 & 0.005&0.012 &0.019 & 2.1e-4\\
SaPINNs &0.957& 0.002& 0.992 &8.7e-4&0.005 & 0.012&0.018 & 7.7e-5\\ \bottomrule
\end{tabular}
\label{table:3}
\end{table}

The performance metrics summarized in Table \ref{table:3} show that both models estimate $\theta_1$ with comparable accuracy;  yet, the SaPINNs ensemble achieves a more accurate estimate of $\theta_2$, which is closer to the true value of 1.  The two architectures reconstruct the displacements almost equally, but SaPINNs exhibit the reduced error for the reconstructed restoring force. 

\section{Conclusions}\label{conclusion}
This work introduced Spectrum and Physics-Informed Neural Networks (SaPINNs), which incorporate natural-hazard excitation spectrum parametrization into the loss function to address the problem of input–state–parameter estimation. The performance of the proposed framework was compared with baseline PINNs to assess its efficiency for structural health monitoring of systems subjected to unmeasured input force. Across both linear and nonlinear examples, including thunderstorm-induced wind loading and seismic excitation, SaPINNs consistently produced more accurate estimates of the unknowns than PINNs, most notably in reconstructing the excitation force time history and in parameter estimation, while maintaining comparable accuracy in the reconstruction of latent states. Deep Ensembles (DEns) were employed to quantify uncertainty, and the results demonstrate that incorporating spectral characteristics into the loss function effectively reduces the uncertainty of the estimates. In the absence of explicit force information, conventional PINNs often converge to non-unique or physically inconsistent solutions, whereas the spectral information incorporated in SaPINNs act as constraints, guiding the network toward more physically plausible solutions. \\
 While SaPINNs demonstrated clear advantages over conventional PINNs, certain limitations remain. The performance of the SaPINNs is influenced by the quality of the available measurements, as well as by the spectral content of the excitation. As illustrated by the comparative loss trajectories across the presented examples, problems with slowly varying observations exhibit smooth and steadily declining losses. Conversely, scenarios involving high-frequency content or significant model misspecification result in  non-monotonic loss behavior and increased sensitivity to the network parameters tuning. These cases demand greater model capacity, i.e., deeper architectures, wider layers, and more trainable parameters, to adequately capture the underlying system dynamics, which in turn increases training time and the risk of overfitting without proper regularization. To enhance accuracy of the predictions, we recommend a pre-calibration phase using baseline measurements under known operational conditions that can help to improve the balance between expressivity and generalization of the neural network-based framework, thereby strengthening the reliability of the estimates.\\
 Despite these challenges, SaPINNs offer promising applications in domains where excitation spectra are well studied. The proposed framework could be extended to more complex multi-degree-of-freedom (MDoF) systems. However, in such cases, a higher level of discrepancy between the model and the real monitored system is expected due to oversimplified modeling assumptions. To account for this, we recommend incorporating an auxiliary process that explicitly represents the modeling error.  This process could be represented as Gaussian Process or a predefined function, potentially estimated from the system’s free-vibration response or derived from prior system identification experiments where the excitation force was measured or controlled.  Explicitly accounting for the modeling error narrows the solution space and reduces bias in parameter estimation, even when the excitation spectrum and modulation function are not accurately characterized. \\
Beyond system identification in civil structures, the proposed framework also holds practical promise in aerospace engineering. Aircraft structures are routinely subjected to turbulence, gust loads, and acoustic excitations that have been extensively characterized through flight testing and wind-tunnel studies. Embedding such spectral priors within the SaPINNs framework could enable more accurate input–state–parameter estimation for structural health monitoring, flutter detection, and fatigue-life assessment.  Similarly, spacecraft applications may benefit from this approach in modeling structural responses under launch vibration spectra or micro-meteorite impact loads, where excitation characteristics are relatively well defined. 
\section*{Acknowledgments}
This work was partially supported by the Center for Smart Streetscapes, an NSF Engineering Research Center,  under Grant No. EEC-2133516.

\bibliography{REFERENCES}

@article{lai2021structural,
  title={Structural identification with physics-informed neural ordinary differential equations},
  author={Lai, Zhilu and Mylonas, Charilaos and Nagarajaiah, Satish and Chatzi, Eleni},
  journal={Journal of Sound and Vibration},
  volume={508},
  pages={116196},
  year={2021},
  publisher={Elsevier}
}

@article{yang2021b,
  title={B-PINNs: Bayesian physics-informed neural networks for forward and inverse PDE problems with noisy data},
  author={Yang, Liu and Meng, Xuhui and Karniadakis, George Em},
  journal={Journal of Computational Physics},
  volume={425},
  pages={109913},
  year={2021},
  publisher={Elsevier}
}

@Article{buildings13030650,
AUTHOR = {Moradi, Sarvin and Duran, Burak and Eftekhar Azam, Saeed and Mofid, Massood},
TITLE = {Novel Physics-Informed Artificial Neural Network Architectures for System and Input Identification of Structural Dynamics PDEs},
JOURNAL = {Buildings},
VOLUME = {13},
YEAR = {2023},
NUMBER = {3},
ARTICLE-NUMBER = {650},
ISSN = {2075-5309},
DOI = {10.3390/buildings13030650}
}

@article{sitzmann2020implicit,
  title={Implicit neural representations with periodic activation functions},
  author={Sitzmann, Vincent and Martel, Julien and Bergman, Alexander and Lindell, David and Wetzstein, Gordon},
  journal={Advances in neural information processing systems},
  volume={33},
  pages={7462--7473},
  year={2020}
}

@article{kingma2014adam,
  title={Adam: A method for stochastic optimization},
  author={Kingma, Diederik P and Ba, Jimmy},
  journal={arXiv preprint arXiv:1412.6980},
  year={2014}
}

@article{mustajab2024physics,
  title={Physics-informed neural networks for high-frequency and multi-scale problems using transfer learning},
  author={Mustajab, Abdul Hannan and Lyu, Hao and Rizvi, Zarghaam and Wuttke, Frank},
  journal={Applied Sciences},
  volume={14},
  number={8},
  pages={3204},
  year={2024},
  publisher={MDPI}
}

@article{roncallo2020evolutionary,
  title={An evolutionary power spectral density model of thunderstorm outflows consistent with real-scale time-history records},
  author={Roncallo, Luca and Solari, Giovanni},
  journal={Journal of Wind Engineering and Industrial Aerodynamics},
  volume={203},
  pages={104204},
  year={2020},
  publisher={Elsevier}
}

@article{RONCALLO2022104978,
title = {Maximum dynamic response of linear elastic SDOF systems based on an evolutionary spectral model for thunderstorm outflows},
journal = {Journal of Wind Engineering and Industrial Aerodynamics},
volume = {224},
pages = {104978},
year = {2022},
issn = {0167-6105},
doi = {https://doi.org/10.1016/j.jweia.2022.104978},
author = {Luca Roncallo and Giovanni Solari and Giuseppe Muscolino and Federica Tubino}
}

@article{raissi2018hidden,
  title={Hidden physics models: Machine learning of nonlinear partial differential equations},
  author={Raissi, Maziar and Karniadakis, George Em},
  journal={Nature Communications},
  volume={9},
  pages={1--10},
  year={2018}
}

@article{SOLARI200173,
title = {Probabilistic 3-D turbulence modeling for gust buffeting of structures},
journal = {Probabilistic Engineering Mechanics},
volume = {16},
number = {1},
pages = {73-86},
year = {2001},
issn = {0266-8920},
doi = {https://doi.org/10.1016/S0266-8920(00)00010-2},
author = {G. Solari and G. Piccardo}
}

@article{rogers2020application,
  title={On the application of Gaussian process latent force models for joint input-state-parameter estimation: With a view to Bayesian operational identification},
  author={Rogers, TJ and Worden, K and Cross, EJ},
  journal={Mechanical Systems and Signal Processing},
  volume={140},
  pages={106580},
  year={2020},
  publisher={Elsevier}
}

@article{liu2024neural,
  title={Neural extended Kalman filters for learning and predicting dynamics of structural systems},
  author={Liu, Wei and Lai, Zhilu and Bacsa, Kiran and Chatzi, Eleni},
  journal={Structural Health Monitoring},
  volume={23},
  number={2},
  pages={1037--1052},
  year={2024},
  publisher={SAGE Publications Sage UK: London, England}
}

@article{azam2015dual,
  title={A dual Kalman filter approach for state estimation via output-only acceleration measurements},
  author={Azam, Saeed Eftekhar and Chatzi, Eleni and Papadimitriou, Costas},
  journal={Mechanical systems and signal processing},
  volume={60},
  pages={866--886},
  year={2015},
  publisher={Elsevier}
}

@article{raissi2019physics,
  title={Physics-informed neural networks: A deep learning framework for solving forward and inverse problems involving nonlinear partial differential equations},
  author={Raissi, Maziar and Perdikaris, Paris and Karniadakis, George E},
  journal={Journal of Computational physics},
  volume={378},
  pages={686--707},
  year={2019},
  publisher={Elsevier}
}

@article{dertimanis2019input,
  title={Input-state-parameter estimation of structural systems from limited output information},
  author={Dertimanis, Vasilis K and Chatzi, EN and Azam, S Eftekhar and Papadimitriou, Costas},
  journal={Mechanical Systems and Signal Processing},
  volume={126},
  pages={711--746},
  year={2019},
  publisher={Elsevier}
}

@article{psaros2022meta,
  title={Meta-learning PINN loss functions},
  author={Psaros, Apostolos F and Kawaguchi, Kenji and Karniadakis, George Em},
  journal={Journal of computational physics},
  volume={458},
  pages={111121},
  year={2022},
  publisher={Elsevier}
}

@article{vlachas2022multiscale,
  title={Multiscale simulations of complex systems by learning their effective dynamics},
  author={Vlachas, Pantelis R and Arampatzis, Georgios and Uhler, Caroline and Koumoutsakos, Petros},
  journal={Nature Machine Intelligence},
  volume={4},
  number={4},
  pages={359--366},
  year={2022},
  publisher={Nature Publishing Group UK London}
}

@article{simpson2021machine,
  title={Machine learning approach to model order reduction of nonlinear systems via autoencoder and LSTM networks},
  author={Simpson, Thomas and Dervilis, Nikolaos and Chatzi, Eleni},
  journal={Journal of Engineering Mechanics},
  volume={147},
  number={10},
  pages={04021061},
  year={2021},
  publisher={American Society of Civil Engineers}
}

@article{lai2022neural,
  title={Neural modal ordinary differential equations: Integrating physics-based modeling with neural ordinary differential equations for modeling high-dimensional monitored structures},
  author={Lai, Zhilu and Liu, Wei and Jian, Xudong and Bacsa, Kiran and Sun, Limin and Chatzi, Eleni},
  journal={Data-Centric Engineering},
  volume={3},
  pages={e34},
  year={2022},
  publisher={Cambridge University Press}
}

@article{alimonti2024number,
  title={Is the number of global natural disasters increasing?},
  author={Alimonti, Gianluca and Mariani, Luigi},
  journal={Environmental Hazards},
  volume={23},
  number={2},
  pages={186--202},
  year={2024},
  publisher={Taylor \& Francis}
}

@article{SPANOS201257,
title = {Harmonic wavelets based statistical linearization for response evolutionary power spectrum determination},
journal = {Probabilistic Engineering Mechanics},
volume = {27},
number = {1},
pages = {57-68},
year = {2012},
note = {The IUTAM Symposium on Nonlinear Stochastic Dynamics and Control},
issn = {0266-8920},
doi = {https://doi.org/10.1016/j.probengmech.2011.05.008},
author = {P.D. Spanos and I.A. Kougioumtzoglou}
}

@article{DEODATIS1996149,
title = {Non-stationary stochastic vector processes: seismic ground motion applications},
journal = {Probabilistic Engineering Mechanics},
volume = {11},
number = {3},
pages = {149-167},
year = {1996},
issn = {0266-8920},
doi = {https://doi.org/10.1016/0266-8920(96)00007-0},
author = {George Deodatis}
}

@article{47f10642b6fa413986dc1d34dee28aa5,
title = "Response of a frame structure on a canyon site to spatially varying ground motions",
author = "Kaiming Bi and Hong Hao and W. Ren",
year = "2010",
doi = "10.12989/sem.2010.36.1.111",
language = "English",
volume = "36",
pages = "111--127",
journal = "Structural Engineering and Mechanics",
issn = "1225-4568",
publisher = "Techno-Press",
number = "1",
}

@inproceedings{NIPS2017_9ef2ed4b,
 author = {Lakshminarayanan, Balaji and Pritzel, Alexander and Blundell, Charles},
 booktitle = {Advances in Neural Information Processing Systems},
 editor = {I. Guyon and U. Von Luxburg and S. Bengio and H. Wallach and R. Fergus and S. Vishwanathan and R. Garnett},
 pages = {},
 publisher = {Curran Associates, Inc.},
 title = {Simple and Scalable Predictive Uncertainty Estimation using Deep Ensembles},
 volume = {30},
 year = {2017}
}

@article{maes2019tracking,
  title={Tracking of inputs, states and parameters of linear structural dynamic systems},
  author={Maes, K and Karlsson, Freddie and Lombaert, G},
  journal={Mechanical Systems and Signal Processing},
  volume={130},
  pages={755--775},
  year={2019},
  publisher={Elsevier}
}

@article{nayek2019gaussian,
  title={A Gaussian process latent force model for joint input-state estimation in linear structural systems},
  author={Nayek, Rajdip and Chakraborty, Souvik and Narasimhan, Sriram},
  journal={Mechanical Systems and Signal Processing},
  volume={128},
  pages={497--530},
  year={2019},
  publisher={Elsevier}
}

@inproceedings{alvarez2009latent,
  title={Latent force models},
  author={Alvarez, Mauricio and Luengo, David and Lawrence, Neil D},
  booktitle={Artificial Intelligence and Statistics},
  pages={9--16},
  year={2009},
  organization={PMLR}
}

@inproceedings{vlachas2022coupling,
  title={On the coupling of reduced order modeling with substructuring of structural systems with component nonlinearities},
  author={Vlachas, Konstantinos and Tatsis, Konstantinos and Agathos, Konstantinos and Brink, Adam R and Quinn, Dane and Chatzi, Eleni},
  booktitle={Dynamic Substructures, Volume 4: Proceedings of the 39th IMAC, A Conference and Exposition on Structural Dynamics 2021},
  pages={35--43},
  year={2022},
  organization={Springer}
}

@article{lourens2012augmented,
  title={An augmented Kalman filter for force identification in structural dynamics},
  author={Lourens, E and Reynders, Edwin and De Roeck, Guido and Degrande, Geert and Lombaert, Geert},
  journal={Mechanical systems and signal processing},
  volume={27},
  pages={446--460},
  year={2012},
  publisher={Elsevier}
}

@article{https://doi.org/10.1002/stc.290,
author = {Chatzi, Eleni N. and Smyth, Andrew W.},
title = {The unscented Kalman filter and particle filter methods for nonlinear structural system identification with non-collocated heterogeneous sensing},
journal = {Structural Control and Health Monitoring},
volume = {16},
number = {1},
pages = {99-123},
doi = {https://doi.org/10.1002/stc.290},
year = {2009}
}

@article{VETTORI2023109654,
title = {An adaptive-noise Augmented Kalman Filter approach for input-state estimation in structural dynamics},
journal = {Mechanical Systems and Signal Processing},
volume = {184},
pages = {109654},
year = {2023},
issn = {0888-3270},
doi = {https://doi.org/10.1016/j.ymssp.2022.109654},
author = {S. Vettori and E. {Di Lorenzo} and B. Peeters and M.M. Luczak and E. Chatzi}
}

@article{haywood2025response,
  title={Response estimation and system identification of dynamical systems via physics-informed neural networks},
  author={Haywood-Alexander, Marcus and Arcieri, Giacomo and Kamariotis, Antonios and Chatzi, Eleni},
  journal={Advanced Modeling and Simulation in Engineering Sciences},
  volume={12},
  number={1},
  pages={8},
  year={2025},
  publisher={Springer}
}

@article{Haywood-Alexander2025,
  author       = {Haywood-Alexander, Marcus and Arcieri, Giacomo and Kamariotis, Antonios and Chatzi, Eleni},
  title        = {Response estimation and system identification of dynamical systems via physics-informed neural networks},
  journal      = {Advanced Modeling and Simulation in Engineering Sciences},
  year         = {2025},
  volume       = {12},
  number       = {1},
  pages        = {8},
  doi          = {10.1186/s40323-025-00291-9},
  issn         = {2213-7467}
}

@inproceedings{gal2016dropout,
  title={Dropout as a bayesian approximation: Representing model uncertainty in deep learning},
  author={Gal, Yarin and Ghahramani, Zoubin},
  booktitle={international conference on machine learning},
  pages={1050--1059},
  year={2016},
  organization={PMLR}
}

@article{doi:10.1061/JSDEAG.0003465,
author = {Rimas Vaicaitis  and Masanobu Shinozuka  and Masaru Takeno },
title = {Parametric Study of Wind Loading on Structures},
journal = {Journal of the Structural Division},
volume = {99},
number = {3},
pages = {453-468},
year = {1973},
doi = {10.1061/JSDEAG.0003465}
}

@article{observability,
author = {Chatzis, Manolis N. and Chatzi, Eleni N. and Smyth, Andrew W.},
title = {On the observability and identifiability of nonlinear structural and mechanical systems},
journal = {Structural Control and Health Monitoring},
volume = {22},
number = {3},
pages = {574-593},
doi = {https://doi.org/10.1002/stc.1690},
year = {2015}
}

@book{rezaeian2010stochastic,
  title={Stochastic modeling and simulation of ground motions for performance-based earthquake engineering},
  author={Rezaeian, Sanaz},
  year={2010},
  publisher={University of California, Berkeley}
}

@article{castiglione2020auto,
  title={Auto-regressive model based input and parameter estimation for nonlinear finite element models},
  author={Castiglione, Juan and Astroza, Rodrigo and Azam, Saeed Eftekhar and Linzell, Daniel},
  journal={Mechanical Systems and Signal Processing},
  volume={143},
  pages={106779},
  year={2020},
  publisher={Elsevier}
}

@article{gustfactor,
author = {Dae-Kun Kwon  and Ahsan Kareem },
title = {Gust-Front Factor: New Framework for Wind Load Effects on Structures},
journal = {Journal of Structural Engineering},
volume = {135},
number = {6},
pages = {717-732},
year = {2009},
doi = {10.1061/(ASCE)0733-9445(2009)135:6(717)}
}

@article{STAFFORD20091123,
title = {An energy-based envelope function for the stochastic simulation of earthquake accelerograms},
journal = {Soil Dynamics and Earthquake Engineering},
volume = {29},
number = {7},
pages = {1123-1133},
year = {2009},
issn = {0267-7261},
doi = {https://doi.org/10.1016/j.soildyn.2009.01.003},
author = {P.J. Stafford and S. Sgobba and G.C. Marano}
}

@article{solari2016thunderstorm,
  title={Thunderstorm response spectrum technique: Theory and applications},
  author={Solari, Giovanni},
  journal={Engineering Structures},
  volume={108},
  pages={28--46},
  year={2016},
  publisher={Elsevier}
}

@article{psaros2023uncertainty,
  title={Uncertainty quantification in scientific machine learning: Methods, metrics, and comparisons},
  author={Psaros, Apostolos F and Meng, Xuhui and Zou, Zongren and Guo, Ling and Karniadakis, George Em},
  journal={Journal of Computational Physics},
  volume={477},
  pages={111902},
  year={2023},
  publisher={Elsevier}
}

@article{MAES2019378,
title = {Observability of nonlinear systems with unmeasured inputs},
journal = {Mechanical Systems and Signal Processing},
volume = {130},
pages = {378-394},
year = {2019},
issn = {0888-3270},
doi = {https://doi.org/10.1016/j.ymssp.2019.05.010},
author = {K. Maes and M.N. Chatzis and G. Lombaert}
}

@article{kanai1957semi,
  title={Semi-empirical formula for the seismic characteristics of the ground},
  author={Kanai, Kiyoshi},
  journal={Bulletin of the Earthquake Research Institute},
  volume={35},
  pages={309--325},
  year={1957}
}

@article{ElCentro,
title = {Nonstationary Kanai-Tajimi models for El Centro 1940 and Mexico City 1985 earthquakes},
journal = {Probabilistic Engineering Mechanics},
volume = {5},
number = {4},
pages = {171-181},
year = {1990},
issn = {0266-8920},
doi = {https://doi.org/10.1016/0266-8920(90)90018-F},
author = {F-G. Fan and G. Ahmadi}
}

\pagebreak

\section*{Appendix A. Observability Rank Condition}
 
The nonlinear observability matrix is constructed based on the gradients of the Lie derivatives of a function $g_s(\boldsymbol{z}_a)$
Using the functional form $\boldsymbol{z}_a=[x_1,x_2, x_3, x_4, x_5]$,  the observability matrix can be constructed as:
\begin{align*}
\mathcal{O}(\mathbf{z_a},t) =
&\scalebox{0.9}{$
\begin{bmatrix}
- x_4 \\
x_3 x_4 \\
- x_4 (x_3^2 - x_4) \\
- x_4 (x_3 x_4 - x_3 (x_3^2 - x_4)) \\
- x_4 (- x_3 (x_3 x_4 - x_3 (x_3^2 - x_4)) - x_4 (x_3^2 - x_4))
\end{bmatrix}
$}
\\[1em]
\phantom{\mathcal{O}(\mathbf{z_a},t) =}
&\scalebox{0.8}{$
\begin{bmatrix}
- x_3 \\
x_3^2 - x_4 \\
x_3 x_4 - x_3 (x_3^2 - x_4) \\
- x_3 (x_3 x_4 - x_3 (x_3^2 - x_4)) - x_4 (x_3^2 - x_4) \\
- x_3 (- x_3 (x_3 x_4 - x_3 (x_3^2 - x_4)) - x_4 (x_3^2 - x_4)) - x_4 (x_3 x_4 - x_3 (x_3^2 - x_4))
\end{bmatrix}
$}
\\[1em]
\phantom{\mathcal{O}(\mathbf{z_a},t) =}
&\scalebox{0.8}{$
\begin{bmatrix}
- x_2 \\
x_1 x_4 + 2 x_2 x_3 - A \cos(\omega t + x_5) \\
x_2 x_4 - x_2 (x_3^2 - x_4) + 2 x_3 (A \cos(\omega t + x_5) - x_1 x_4 - 2 x_2 x_3) \\
- 2 x_2 x_3 x_4 - x_2 (x_3 x_4 - x_3 (x_3^2 - x_4)) + (x_3^2 - x_4) (A \cos(\omega t + x_5) - x_1 x_4 - 2 x_2 x_3) \\
- x_2 x_4 (- 3 x_3^2 + 2 x_4) - x_2 (- x_3 (x_3 x_4 - x_3 (x_3^2 - x_4)) - x_4 (x_3^2 - x_4)) + (x_3 x_4 - x_3 (x_3^2 - x_4)) (A \cos(\omega t + x_5) - x_1 x_4 - 2 x_2 x_3)
\end{bmatrix}
$}
\\[1em]
\phantom{\mathcal{O}(\mathbf{z_a},t) =}
&\scalebox{0.8}{$
\begin{bmatrix}
- x_1 \\
x_1 x_3 - x_2 \\
- A \cos(\omega t + x_5) + x_1 x_4 - x_1 (x_3^2 - x_4) + x_2 x_3 \\
- x_1 (x_3 x_4 - x_3 (x_3^2 - x_4)) + x_2 x_4 - x_2 (x_3^2 - x_4) + x_3 (A \cos(\omega t + x_5) - x_1 x_4 - 2 x_2 x_3) \\
- x_1 (- x_3 (x_3 x_4 - x_3 (x_3^2 - x_4)) - x_4 (x_3^2 - x_4)) + x_2 x_4 - x_2 (x_3 x_4 - x_3 (x_3^2 - x_4)) + x_3 (- A \cos(\omega t + x_5) + x_1 x_4 - x_1 (x_3^2 - x_4) + x_2 x_3)
\end{bmatrix}
$}
\\[1em]
\phantom{\mathcal{O}(\mathbf{z_a},t) =}
&\scalebox{0.8}{$
\begin{bmatrix}
- A \sin(\omega t + x_5) \\
A x_3 \sin(\omega t + x_5) \\
- A (x_3^2 - x_4) \sin(\omega t + x_5) \\
- A (x_3 x_4 - x_3 (x_3^2 - x_4)) \sin(\omega t + x_5) \\
- A (- x_3 (x_3 x_4 - x_3 (x_3^2 - x_4)) - x_4 (x_3^2 - x_4)) \sin(\omega t + x_5)
\end{bmatrix}
$}
\end{align*}

Based on the rank, the observability matrix is degenerate (rank 4) at one time, however, the system is \textit{globally observable} (or observable over an interval of the measurement) .

\vspace{1em} 
\section*{Appendix B. Observability under Uncertain Excitation Amplitude}
Considering unit mass (\(m=1\)) the equation of motion for a linear SDoF is described as:
\begin{equation}
\ddot{x}+c\,\dot{x}+k\,x = A\cos\!\bigl(\omega_0 t+\phi\bigr).
\end{equation}
Considering in this case that the damping coefficient is known  and the unknown states and parameters are embedded in the augmented state vector:
\begin{equation}
\boldsymbol{z}_a(t)=\begin{bmatrix}
x_1 \\
x_2 \\
x_3 \\
x_4 \\
x_5
\end{bmatrix}
=
\begin{bmatrix}
x(t) \\
\dot{x}(t) \\
k \\
A\\
\phi
\end{bmatrix},
\qquad
\dot{\boldsymbol{z}}_a(t)=
\mathbf f_s(\boldsymbol{z}_a)=
\begin{bmatrix}
x_2\\[4pt]
-\;c\,x_2-\;x_3\,x_1+\;x_4\cos\!\bigl(\omega_0 t+x_5\bigr)\\[4pt]
0\\[2pt]
0\\[2pt]
0
\end{bmatrix}.
\end{equation}
The parameters \(A,k,\phi\) are treated as time‐invariant states (\(\dot{x}_3=0\), \(\dot{x}_4=0\), \(\dot{x}_5=0\)).
Considering the system output as the measured acceleration:
\begin{equation}
y(t)=g_s(\boldsymbol{z}_a)=
-\;c\,x_2-\;x_3\,x_1+\;x_4\cos\!\bigl(\omega_0 t+x_5\bigr).
\end{equation}
The gradients for the ORC‐DF matrix can be estimated as:
\begin{align}
\nabla g_s &=
\bigl[-x_3,\; -c,\; -x_1,\;\cos(\omega_0 t+x_5),\;  -x_4\sin(\omega_0 t+x_5)\bigr], 
\\[6pt]
\mathcal L_{\mathbf f}^{1}g_s &=
\frac{\mathrm d}{\mathrm dt}g_s
=
-c\,x_2-x_3x_2
-\;x_4\omega_0\sin(\omega_0 t+x_5), \quad
\end{align}
The gradients of successive Lie derivatives yields the augmented observability matrix:
\begin{equation}
\mathcal{O}(\boldsymbol{z}_a)=
\begin{bmatrix}
\nabla\mathcal L_{\mathbf f}^0 g_s\\
\nabla\mathcal L_{\mathbf f}^1 g_s\\
\nabla\mathcal L_{\mathbf f}^2 g_s\\
\nabla\mathcal L_{\mathbf f}^3 g_s\\
\nabla\mathcal L_{\mathbf f}^4 g_s
\end{bmatrix}\in\mathbb{R}^{5\times5}.
\end{equation}
Provided that
\begin{equation}
A\neq0,\quad\omega_0\neq0,\quad
\text{and}\;x(t),\dot{x}(t)\not\equiv0
\end{equation}
over the observation window, the rows are algebraically independent and
\begin{equation}
\operatorname{rank}\bigl(\mathcal{O}(\boldsymbol{z}_a)\bigr)=5,
\end{equation}
satisfying the Observability Rank Condition for all augmented states. Hence, all five augmented states \((x,\,\dot{x},\,k,\,A,\,\phi)\) are uniquely recoverable from a single continuous acceleration record, rendering the joint input–state–parameter estimation problem well‐posed.
\end{document}